\documentclass[a4paper,11pt]{article}
\usepackage{amssymb}
\usepackage{amsmath}
\usepackage{bm}
\usepackage{boxedminipage}
\usepackage[pdftex]{graphicx}
\usepackage{graphicx}
\usepackage{enumerate}

\makeatletter \@addtoreset{equation}{section}

\makeatletter\renewcommand\section{\@startsection {section}{1}{\z@}
{-3.5ex \@plus -1ex \@minus -.2ex} {2.3ex \@plus.2ex}
{\normalfont\large\bfseries}}
\renewcommand\subsection{\@startsection{subsection}{2}{\z@}                                     {-3.25ex\@plus -1ex \@minus -.2ex}                                     {1.5ex \@plus .2ex}                                     {\normalfont\bfseries}}
\def\Label#1{\label{#1}  \smash{\hbox to0pt{\raise1ex\hbox{\tiny[#1]}\hss}}}
\def\noLabels{\let\Label=\label}
\def\nobbibitem{\let\bbibitem=\bibitem}

\begin{document}

%\ifpdf\DeclareGraphicsExtensions{.pdf, .jpg, .tif} \else%
%\DeclareGraphicsExtensions{.eps, .jpg} \fi
\begin{titlepage}

    \thispagestyle{empty}
    \begin{flushright}
        %\hfill{LBNL-...}\\
        \hfill{CERN-PH-TH/2010-130}\\
        \hfill{SU-ITP-10/08}\\
        \hfill{UCB-PTH-10/12}\\
    \end{flushright}

    %\vspace{5pt}
    \begin{center}
        { \huge{\textbf{Charge Orbits of\\\vspace{5pt}Extremal Black Holes\\ \vspace{10pt}in Five Dimensional Supergravity}}}

        \vspace{10pt}

        {\large{{\bf Bianca L. Cerchiai$^{\sharp}$, \ Sergio Ferrara$^{\clubsuit,\flat}$,\\\ Alessio Marrani$^{\heartsuit}$ and \ Bruno Zumino$^{\spadesuit,\diamondsuit}$
        }}}

        \vspace{20pt}

        {$\sharp$ \it Universit\`{a} degli Studi di Milano, \\
        Dipartimento di Matematica ``Federigo Enriques"\\
        Via Cesare Saldini 50, I-20133 Milano, Italy\\
        \texttt{Bianca.Cerchiai@unimi.it}}

        \vspace{2pt}

        {$\clubsuit$ \it Theory division, CERN, Geneva, Switzerland \\
        CH 1211, Geneva 23, Switzerland\\
        \texttt{sergio.ferrara@cern.ch}}

        \vspace{2pt}

         {$\flat$ \it INFN - LNF, \\
          Via Enrico Fermi 40, I-00044 Frascati, Italy}

          \vspace{2pt}

        {$\heartsuit$ \it Stanford Institute for Theoretical Physics\\
        %Department of Physics, 382 Via Pueblo Mall, Varian Lab,\\
        Stanford University, Stanford, CA 94305-4060, USA\\
        \texttt{marrani@lnf.infn.it}}

         \vspace{2pt}

        {$\spadesuit$ \it Lawrence Berkeley National Laboratory, \\
        Theory Group, Bldg 50A5104\\
        1 Cyclotron Rd, Berkeley, CA 94720-8162, USA\\
        \texttt{zumino@thsrv.lbl.gov}}

        \vspace{2pt}

        {$\diamondsuit$ \it Department of Physics, University of California,\\
        Berkeley, CA 94720-8162, USA}

        \vspace{15pt}

        %\vspace{80pt}
 \vspace{5pt}
        {ABSTRACT}
\end{center}

We derive the $U$-duality charge orbits, as well as the related \textit{%
moduli spaces}, of ``large'' and ``small'' extremal black holes in
non-maximal \textit{ungauged} Maxwell-Einstein supergravities with
symmetric scalar manifolds in $d=5$ space-time dimensions.

The stabilizer groups of the various classes of orbits are obtained
by determining and solving suitable $U$-invariant sets of
constraints, both in ``bare'' and ``dressed'' charges bases, with
various methods.

After a general treatment of attractors in \textit{real special
geometry} (also considering non-symmetric cases), the
$\mathcal{N}=2$ ``magic'' theories, as well as the $\mathcal{N}=2$
Jordan symmetric sequence, are analyzed in detail.\ Finally, the
half-maximal ($\mathcal{N}=4$) matter-coupled supergravity is also
studied in this context.

\end{titlepage}
\newpage\tableofcontents%\newpage

\section{\label{Intro}Introduction}

Five-dimensional supergravity theories with non-maximal supersymmetry ($%
2\leqslant \mathcal{N}<8$), emerging from Calabi-Yau compactifications of $M$%
-theory, admit extremal black $p$-brane solutions in their spectrum \cite
{Gibbons-Townsend-1}. In particular, \textit{ungauged} theories admit
extremal black holes ($p=0$) and black strings ($p=1$) which are
asymptotically flat, and reciprocally related through $U$-duality\footnote{%
Here $U$-duality is referred to as the ``continuous'' version, valid for
large values of the charges, of the $U$-duality groups introduced by Hull
and Townsend \cite{HT-1}.}. These objects have been intensely studied along
the years, due to the wide range of classical and quantum aspects they
exhibit.

For asymptotically flat, spherically symmetric and stationary solutions, the
\textit{Attractor Mechanism} \cite{FKS,Strom1,FK1-FK2,FGK} proved to be a
crucial phenomenon, determining, in a universal fashion, the stabilization
of scalar fields in the near-horizon geometry in terms of the fluxes of the
two-form field strengths of the Abelian vector fields coupled to the system.
Moreover, the \textit{Attractor Mechanism} turned out to be important also
to unravel dynamical properties such as split attractor flows \cite
{split-attr-flow} and wall crossing \cite{WC}, and to gain insights in the
microstate counting analysis (see \textit{e.g.} \cite{microstate-counting},
and Refs. therein), also in relation to string topological partition
functions \cite{topological} (see also \cite{Pioline-Lectures-06} for a
recent account and list of Refs.). In $d=5$ space-time dimensions, progress
has been achieved also with the discovery of new attractor solutions (see
\textit{e.g.} \cite{KL}), as well as with the formulation of a first-order
formalism governing the evolution dynamics of non-supersymmetric scalar
flows \cite{FO-d=5}.

For supergravity theories with scalar manifolds which are symmetric cosets,
the extremal solutions of the \textit{ungauged} theory can be classified
through the orbits of the relevant representation space of the $U$-duality
group, in which the corresponding supporting charges sit. The relation
between $U$-invariant BPS conditions and charge orbits in $d=5$
supergravities has been the subject of various studies along the years \cite
{Ferrara-Maldacena,FG1,LPS,ADF-U-duality-d=5,DFL-0-brane,FG2,AFMT1}.

The present paper extends to $d=5$ space-time dimensions the $4$-dimensional
investigation of \cite{CFMZ1}.

We derive the $U$-duality charge orbits, as well as the related \textit{%
moduli spaces}, of ``large'' and ``small'' extremal black holes and black
strings in \textit{ungauged} Maxwell-Einstein supergravities with symmetric
scalar manifolds. The stabilizer groups of the various classes of orbits are
obtained by determining and solving suitable $U$-invariant sets of
constraints, both in ``bare'' and ``dressed'' charges bases, as well by
exploiting \.{I}n\"{o}n\"{u}-Wigner contractions and $SO\left( 1,1\right) $%
-gradings.

It is here worth pointing out that in this paper we will not deal with
maximal $\mathcal{N}=8$, $d=5$ supergravity, because a complete analysis of
extremal black hole attractors and their ``large'' and ``small'' charge
orbits is already present in literature, see \textit{e.g.} \cite
{ADF-U-duality-d=5,Ferrara-Maldacena,FG1,LPS,DFL-0-brane,Ferrara-Marrani-1,Ferrara-Marrani-2,AFMT1,CFGM1,CFG1,ICL-1}%
. We will just mention such a theory shortly below Eq. (\ref{jjjjj-1}).{%
\medskip }

The plan of the paper is as follows.

We start and give a \textit{r\'{e}sum\'{e}} of \textit{real special geometry}%
$\mathbb{\ }$in Sect. \ref{RSG}, setting up notation and presenting all
formul\ae\ needed for the subsequent treatment of charge orbits and
attractors.

In Sect. \ref{Attractors-RSG} extremal black hole (black string) attractors
are studied in full generality within real special geometry. Starting from
the treatment of \cite{FG2}, various refinements and generalizations are
performed, in particular addressing the issue of generic, \textit{%
non-symmetric} vector multiplets' scalar manifolds. In Subsect. \ref
{Critical-Points-V} we analyze the various classes of critical points of the
effective potential $V$, also within the so-called ``new attractor''
approach (see Subsubsect. \ref{New-Attr}). Then, in Subsect. \ref
{Higher-Order-T} we compute the higher order covariant derivatives of the
previously introduced rank-$3$ invariant tensor $T_{xyz}$, which will play a
key role in the subsequent developments and results, exposed in Subsects.
\ref{Generic-RSG} and \ref{Symmetric-RSG}, respectively dealing with generic
and homogeneous symmetric real special manifolds. A general analysis of the
Hessian matrix of $V$, crucial in order to establish the stability of
considered attractor points, is then performed in Subsect. \ref
{Hessian-Matrix}.

In Sect. \ref{Small-Orbits-Symmetric-RSG} all ``small'' charge orbits of
\textit{symmetric} ``magic'' real special geometries are explicitly
determined and classified, by exploiting the properties of the functional $%
\widehat{\mathcal{I}}_{3}$ introduced in Subsubsect. \ref{I_3-hat-d}. Note
that ``small'' charge orbits support non-attractor solutions, which have
vanishing Bekenstein-Hawking \cite{BH1} entropy in the Einsteinian
approximation. Nevertheless, they can be treated by exploiting their
symmetry properties under $U$-duality.

Sect. \ref{N=2-magic-H-d=5---N=6-d=5} analyzes the ``duality'' relating the $%
\mathcal{N}=2$ ``magic'' theory coupled to $14$ Abelian vector multiplets
and the $\mathcal{N}=6$ \textit{``pure''} supergravity, both based on the
rank-$3$ Euclidean Jordan algebra $J_{3}^{\mathbb{H}}$ and thus sharing the
very same bosonic sector.

Then, Sect. \ref{N=2,d=5-Jordan-Symm-Seq} is devoted to the analysis of the
``large'' (Subsect. \ref{N=2-d=5-Jordan-Symm-Seq-Large}) and ``small''
(Subsect. \ref{N=2-d=5-Jordan-Symm-Seq-Small}) charge orbits of $\mathcal{N}%
=2$ Jordan symmetric sequence. Similarly, Sect. \ref{N=4,d=5} provides a
detailed treatment of the ``large'' (Subsect. \ref{N=4-d=5-Large}) and
``small'' (Subsect. \ref{N=4-d=5-Small}) charge orbits of the half-maximal ($%
\mathcal{N}=4$) matter coupled supergravity. The analysis of both Sects. \ref
{N=2,d=5-Jordan-Symm-Seq} and \ref{N=4,d=5} is made in the ``bare'' charges
basis, and various subtleties, related to the \textit{reducible} nature of
the $d=5$ $U$-duality group and disconnectedness of orbits in these two
theories, are elucidated.

Some Appendices conclude the paper, containing various details concerning
the determination of the ``small'' orbits in \textit{symmetric} ``magic''
real special geometries.

The resolution of $U$-invariant defining (differential) constraints, both in
``bare'' and ``dressed'' charges bases, is performed in App. \ref
{Solutions-Constraints-Small-Orbits}.

Then, in App. \ref{Group-Theory-Small-Orbits} we give an equivalent
derivation of all ``small'' charge orbits of \textit{symmetric} ``magic''
real special geometries, relying on group theoretical procedures, namely
\.{I}n\"{o}n\"{u}-Wigner contractions (Sub-App. \ref{IW-Contrs.}) and $%
SO\left( 1,1\right) $-three-grading (Sub-App. \ref{SO(1,1)-Three-Grading}%
).\medskip

Finally, we point out that all results on charge orbits can actually be
obtained in various other ways, including the analysis of cubic norm forms
of the relevant Jordan systems in $d=5$; this will be investigated elsewhere.

\section{\label{RSG}\textit{R\'{e}sum\'{e}} of Real Special Geometry (RSG)}

\textit{Real special geometry} (RSG) (\cite{GST1,GST2,GST3,GST4,dWVVP,CD},
and Refs. therein) is the geometry underlying the scalar manifold $M_{5}$
(with Euclidean signature) of Abelian vector multiplets coupled to the
\textit{minimal} supergravity in $d=5$ space-time dimensions, namely to $%
\mathcal{N}=2$, $d=5$ theory.

In the present Section, we recall some basic facts about RSG, setting up
notation and presenting all formul\ae\ needed for the subsequent treatment
of charge orbits and attractors. Apart from a slight changes in notation, we
will adopt the conventions of \cite{FG2}, which are slightly different from
the ones used in \cite{CFM1} (see the observations in \cite{CFM1}
itself).\bigskip

We start by specifying the kind and range of indices being used. $%
i=0,1,...n_{V}$ is the index in the \textit{``ambient space'' }(in which $%
M_{5}$ is defined through a cubic constraint; see Eq. (\ref{constraint-RSG})
below). ``$0$'' is the index pertaining to the (\textit{``bare''}) $d=5$
graviphoton, and $n_{V}$ stands for the number of Abelian vector multiplets
coupled to the supergravity multiplet. On the other hand, $x=1,...,n_{V}$,
and $a=1,...,n_{V}$ respectively denote \textit{``curved''} and (local)
\textit{``flat''} coordinates in $M_{5}$.

The metric $a_{ij}$ in the \textit{``ambient space''} (named $g_{ij}$ in
\cite{CFM1}) can be defined as follows:
\begin{equation}
a_{ij}=-\frac{1}{3}\frac{\partial ^{2}\log \mathcal{V}\left( \lambda \right)
}{\partial \lambda ^{i}\partial \lambda ^{j}},
\end{equation}
where
\begin{equation}
\mathcal{V}\left( \lambda \right) \equiv d_{ijk}\lambda ^{i}\lambda
^{j}\lambda ^{k}>0  \label{Vol}
\end{equation}
is the volume of $M_{5}$ itself, and $d_{ijk}=d_{\left( ijk\right) }$ is a
rank-$3$ completely symmetric invariant tensor (see further below). In turn,
the $\lambda ^{i}$'s are some real functions (with suitable features of
smoothness and regularity) of the set of scalars $\phi ^{x}$ of the theory,
coordinatizing $M_{5}$:
\begin{equation}
\lambda ^{i}=\lambda ^{i}\left( \phi ^{x}\right) .
\end{equation}
They do satisfy the inequality (\ref{Vol}). As elucidated \textit{e.g.} in
\cite{CFM1}, the $\lambda ^{i}$'s are nothing but the (opposite of the)
imaginary (\textit{``dilatonic''}) part of the complex scalar fields of the
\textit{special K\"{a}hler geometry} (SKG) based on a cubic holomorphic
prepotential (usually named $d$-SKG; see \textit{e.g.} \cite{dWVVP,dWVP-2}),
endowing the Abelian vector multiplets' scalar manifold of $\mathcal{N}=2$
Maxwell-Einstein supergravity in $4$ space-time dimensions. In this respect,
the \textit{``ambient space''} in $5$ dimensions is nothing but the \textit{%
``dilatonic sector''} of the $d$-SKG in $4$ dimensions.

It is now convenient to introduce rescaled variables as follows:
\begin{equation}
\widehat{\lambda }^{i}\equiv \lambda ^{i}\mathcal{V}^{-1/3}\Leftrightarrow
d_{ijk}\widehat{\lambda }^{i}\widehat{\lambda }^{j}\widehat{\lambda }^{k}=%
\mathcal{V}\left( \widehat{\lambda }\right) =1.  \label{cubic-constr-1}
\end{equation}
Thus, the metric of $M_{5}$ is the pull-back of $a_{ij}$ on the hypersurface
\begin{equation}
\mathcal{V}\left( \lambda \right) \equiv 1  \label{constraint-RSG}
\end{equation}
in the \textit{``ambient space''}, namely:
\begin{equation}
g_{xy}\equiv \widehat{\lambda }_{~x}^{i}\widehat{\lambda }_{~y}^{j}\left.
a_{ij}\right| _{\mathcal{V}\left( \lambda \right) \equiv 1}=-\frac{1}{3}%
\widehat{\lambda }_{~x}^{i}\widehat{\lambda }_{~y}^{j}\left. \frac{\partial
^{2}\log \mathcal{V}\left( \lambda \right) }{\partial \lambda ^{i}\partial
\lambda ^{j}}\right| _{\mathcal{V}\left( \lambda \right) \equiv
1}=g_{xy}\left( \widehat{\lambda }\left( \phi \right) \right) =\overset{%
\left( \sim \right) }{g}_{xy}\left( \phi \right) ,
\end{equation}
where (the semicolon denotes Riemann-covariant differentiation throughout)
\begin{equation}
\widehat{\lambda }_{~x}^{i}\equiv -\sqrt{\frac{3}{2}}\frac{\partial \widehat{%
\lambda }^{i}}{\partial \phi ^{x}}\equiv -\sqrt{\frac{3}{2}}\widehat{\lambda
}_{,x}^{i}=-\sqrt{\frac{3}{2}}\widehat{\lambda }_{;x}^{i}.  \label{pre-diff}
\end{equation}
Notice that the constraint (\ref{cubic-constr-1}) implies
\begin{equation}
\frac{\partial \mathcal{V}\left( \widehat{\lambda }\right) }{\partial \phi
^{x}}=3d_{ijk}\widehat{\lambda }_{,x}^{i}\widehat{\lambda }^{j}\widehat{%
\lambda }^{k}-\sqrt{6}d_{ijk}\widehat{\lambda }_{~x}^{i}\widehat{\lambda }%
^{j}\widehat{\lambda }^{k}=0.
\end{equation}

Let us now introduce $T_{xyz}$, a rank-$3$ completely symmetric invariant
tensor, related to $d_{ijk}$ through the definition
\begin{equation}
T_{xyz}\equiv \widehat{\lambda }_{~x}^{i}\widehat{\lambda }_{~y}^{j}\widehat{%
\lambda }_{~z}^{k}d_{ijk}=-\left( \frac{3}{2}\right) ^{3/2}\widehat{\lambda }%
_{,x}^{i}\widehat{\lambda }_{,y}^{j}\widehat{\lambda }_{,z}^{k}d_{ijk}=T_{%
\left( xyz\right) },  \label{T}
\end{equation}
whose inversion reads
\begin{equation}
d_{ijk}=\frac{5}{2}\widehat{\lambda }_{i}\widehat{\lambda }_{j}\widehat{%
\lambda }_{k}-\frac{3}{2}\overset{\circ }{a}_{(ij}\widehat{\lambda }%
_{k)}+T_{xyz}\widehat{\lambda }_{i}^{~x}\widehat{\lambda }_{j}^{~y}\widehat{%
\lambda }_{k}^{~z},  \label{d}
\end{equation}
where
\begin{equation}
\overset{\circ }{a}_{ij}\left( \widehat{\lambda }\right) \equiv \left.
a_{ij}\right| _{\mathcal{V}\left( \lambda \right) \equiv 1}.
\end{equation}
In other words, $T_{xyz}$ is the $\phi $-dependent \textit{``dressing''}
(through $\widehat{\lambda }_{,x}^{i}\left( \phi \right) $'s) of the
constant ($\phi $-independent) tensor $d_{ijk}$. It is here worth
anticipating that Eqs. (\ref{T}) and (\ref{d}) play the key role to relate
the formalism based on \textit{``bare''} charges with the formalism based on
the \textit{``dressed''} charges (see further below).

$T_{xyz}$ enters the so-called \textit{``RSG constraints''}, relating in $%
M_{5}$ the Riemann tensor $R_{xyzu}$ to the metric tensor $g_{xy}$, as
follows:
\begin{equation}
R_{xyzu}=\frac{4}{3}\left( g_{x[u}g_{z]y}+T_{x[u}^{~~~w}T_{z]yw}\right) =%
\frac{4}{3}\left( g_{x[u}g_{z]y}+T_{xw^{\prime }[u}T_{z]yw}g^{ww^{\prime
}}\right) .  \label{RSG-constraints}
\end{equation}
It is worth noticing a direct consequence of such \textit{``RSG constraints''%
}: the sectional curvature (see \textit{e.g.} \cite{Riemann-Finsler} and
\cite{Differential-Geometry}) of matter charges in RSG globally vanishes:
\begin{equation}
\mathcal{R}\left( Z\right) \equiv R_{xyzw}g^{xx^{\prime }}g^{yy^{\prime
}}g^{zz^{\prime }}g^{ww^{\prime }}Z_{x^{\prime }}Z_{y^{\prime }}Z_{z^{\prime
}}Z_{w^{\prime }}=0.
\end{equation}
This is trivially due to the symmetry properties of the Riemann tensor $%
R_{xyzw}$ (which are the ones for a generic Riemann geometry), and it is a
feature discriminating RSG from SKG (in which $\mathcal{R}\left( Z\right) $
generally does not vanish; see \textit{e.g.} \cite{Kallosh-rev,Raju-1}).

As a consequence of the constraints (\ref{RSG-constraints}) (within the
\textit{metric postulate}), the definition of $M_{5}$ to be an \textit{%
homogeneous} symmetric manifold
\begin{equation}
R_{xyzu;t}=0
\end{equation}
yields
\begin{equation}
\left( T_{xw^{\prime }[u;t}T_{z]yw}+T_{xw^{\prime }[u}T_{z]yw;t}\right)
g^{ww^{\prime }}=0\Leftrightarrow T_{xw^{\prime }[u}T_{z]yw;t}g^{ww^{\prime
}}=T_{xw[u}T_{z]y~;t}^{~~w}=0,  \label{TTg=0}
\end{equation}
solved by
\begin{equation}
T_{xyz;u}=0.  \label{T-symm}
\end{equation}
Through Eqs. (\ref{T}) and (\ref{d}), and exploiting the constraints imposed
by local $\mathcal{N}=2$ supersymmetry, it can be shown that Eq. (\ref
{T-symm}) implies the following relation between the tensors $d_{ijk}$:
\begin{equation}
d^{ijk}d_{j(mn}d_{pq)k}=\delta _{(m}^{i}d_{npq)}\Leftrightarrow
d_{j(mn}d_{pq)k}d_{rst}\overset{\circ }{a}^{sj}\overset{\circ }{a}^{tk}%
\overset{\circ }{a}^{ri}=\delta _{(m}^{i}d_{npq)},  \label{d-symm}
\end{equation}
where the index-raising through the contravariant metric $\overset{\circ }{a}%
^{ij}$ has been explicited.

\section{\label{Attractors-RSG}Attractors in RSG}

The present Section is dedicated to the study of attractors in RSG. This has
been firstly treated in \cite{FG2} (and then reconsidered in \cite{AFMT1},
in connection to $d=6$).

Starting from the treatment of \cite{FG2}, we will generalize and elaborate
further various results obtained therein.

It is worth recalling that no asymptotically-flat \textit{dyonic} solutions
of Einstein Eqs. exist in $d=5$. Thus, the $d=5$ asymptotically flat black
holes (BHs) can only carry \textit{electric }charges $q_{i}$. Their magnetic
\textit{duals} are the $d=5$ asymptotically flat black strings, which can
only carry \textit{magnetic} charges $p^{i}$.

We will perform all our treatment within the electric charge configuration.
Due to the mentioned BH/black string duality, this does not imply any loss
of generality. Furthermore, we will study attractors within the \textit{%
Ans\"{a}tze} of asymptotical (Minkowski) flatness, staticity, spherical
symmetry and extremality of the BH space-time metric (if no scalars are
coupled, this is nothing but the so-called Tangherlini extremal $d=5$ BH).
The near-horizon geometry of extremal \textit{electric} BHs and extremal
\textit{magnetic} black strings respectively is $AdS_{2}\times S^{3}$ and $%
AdS_{3}\times S^{2}$.

\subsection{\label{Critical-Points-V}Classes of Critical Points of $V$}

From the general theory of \textit{Attractor Mechanism }\cite
{FKS,Strom1,FK1-FK2,FGK}, the stabilization of scalar fields in proximity of
the (unique) event horizon of a static, spherically symmetric and
asymptotically flat extremal BH in $\mathcal{N}=2$, $d=5$ Maxwell-Einstein
supergravity is described by the critical points of the positive-definite
\textit{effective potential} function
\begin{equation}
V\equiv \overset{\circ }{a}^{ij}q_{i}q_{j}=\left( \widehat{\lambda }%
^{i}q_{i}\right) ^{2}+\frac{3}{2}g^{xy}\widehat{\lambda }_{,x}^{i}q_{i}%
\widehat{\lambda }_{,y}^{j}q_{j}=Z^{2}+\frac{3}{2}g^{xy}Z_{x}Z_{y},
\label{V}
\end{equation}
where the $\mathcal{N}=2$, $d=5$ central charge function $Z$ and its
Riemann-covariant derivatives (\textit{``matter charges''}) have been
defined as follows:
\begin{eqnarray}
Z &\equiv &\widehat{\lambda }^{i}q_{i};  \label{Z} \\
Z_{x} &\equiv &\widehat{\lambda }_{,x}^{i}q_{i}=Z_{,x}=Z_{;x}.  \label{Zx}
\end{eqnarray}

The definitions (\ref{Z}) and (\ref{Zx}) can be inverted, obtaining the
fundamental identities of RSG (in \textit{electric} formulation) \cite{FG2}:
\begin{equation}
q_{i}=\widehat{\lambda }_{i}Z-\frac{3}{2}g^{xy}\widehat{\lambda }_{i,x}Z_{y}.
\label{RSG-electric-Ids}
\end{equation}
The identities (\ref{RSG-electric-Ids}) relate the basis of \textit{``bare''}
($\phi $-independent) electric charges $q_{i}$ to the basis of \textit{%
``dressed''} (central and matter) charges $\left\{ Z,Z_{x}\right\} $, which
do depend on the scalars $\phi ^{x}$, as yielded by definitions (\ref{Z})
and (\ref{Zx}).

By recalling definitions (\ref{Z}) and (\ref{Zx}), one obtains that
\begin{equation}
Z_{xy}\equiv Z_{x;y}=Z_{,x;y}=Z_{;x;y}=\widehat{\lambda }_{,x;y}^{i}q_{i}=%
\frac{2}{3}g_{xy}Z-\sqrt{\frac{2}{3}}T_{xyz}g^{zw}Z_{w}.  \label{DDZ}
\end{equation}

Therefore, by using Eq. (\ref{DDZ}) the criticality conditions (\textit{alias%
} \textit{Attractor Eqs.}) for the \textit{effective potential} $V$ can be
easily computed to be \cite{FG2}:
\begin{equation}
V_{x}\equiv V_{,x}=V_{;x}=2\left( 2ZZ_{x}-\sqrt{\frac{3}{2}}%
T_{xyz}g^{ys}g^{zt}Z_{s}Z_{t}\right) =0.  \label{RSG-AEs}
\end{equation}
\textit{A} \textit{priori}, the classes of critical points of $V$ which are
\textit{non-degenerate} (\textit{i.e.} with $\left. V\right| _{V_{x}=0}\neq
0 $) are three:

\subsubsection{($\frac{1}{2}$-)BPS}

This class is defined by the sufficient (but not necessary) criticality
constraint
\begin{equation}
Z_{x}=0,~\forall x,  \label{1/2-BPS}
\end{equation}
implying
\begin{equation}
V=Z^{2}.\label{pppp-1}
\end{equation}

\subsubsection{Non-BPS}

This class is defined by the constraints
\begin{equation}
\left\{
\begin{array}{l}
Z\neq 0; \\
Z_{x}\neq 0\text{ for~\textit{at~least} some~}x\text{'s},
\end{array}
\right.
\end{equation}
which are critical provided the following algebraic constraint among $Z$ and
$Z_{x}$'s hold:
\begin{equation}
Z_{x}=\frac{1}{2Z}\sqrt{\frac{3}{2}}T_{xyz}g^{ys}g^{zt}Z_{s}Z_{t}.
\label{nBPS-Z<>0-large}
\end{equation}
\textit{At least} in symmetric RSG, this implies \cite{FG2}
\begin{equation}
V=9Z^{2}.\label{pppp-2}
\end{equation}

\subsubsection{Remark}

It is here worth recalling the Bekenstein-Hawking entropy-area formula \cite
{BH1}, implemented for critical points of $V$:
\begin{equation}
\frac{S_{BH,d=5}}{\pi }=\frac{A_{H}}{4\pi }\equiv R_{H}^{2}=\left( \left.
V\right| _{\partial V=0}\right) ^{3/4}.  \label{Bek-Haw}
\end{equation}

The \textit{Attractor Mechanism} \cite{FKS,Strom1,FK1-FK2,FGK} is known to
hold only for the so-called ``large'' BHs, which, through Eq. (\ref{Bek-Haw}%
), have a non-vanishing Bekenstein-Hawking entropy.

Therefore, attractors \textit{in strict sense} are given by \textit{%
non-degenerate} critical points of $V$. On the other hand, degenerate
critical points of $V$, namely critical points such that $\left. V\right|
_{\partial V=0}=0$ are trivial. Indeed, by virtue of the positive
definiteness of $V$ (inherited from the strictly positive definiteness of $%
\overset{\circ }{a}^{ij}$ throughout all its domain of definition), it holds
that
\begin{equation}
V=0\Leftrightarrow q_{i}=0~\forall i,  \label{V=0}
\end{equation}
which is the trivial limit of the theory with \textit{all} (electric)
charges switched off.

The same reasoning can be repeated in the \textit{magnetic} case.

Thus, only ``large'' BHs do exhibit a (classical) \textit{Attractor Mechanism%
}, implemented through \textit{non-trivial} (\textit{alias non-degenerate})
critical points of the effective potential itself \cite{FGK}.

\subsubsection{\label{New-Attr}\textit{``New Attractor''} Approach}

Through the so-called \textit{``new attractor''} approach \cite{K1}, an
equivalent form of the $n_{V}$ real criticality conditions (\textit{i.e.} of
the so-called \textit{Attractor Eqs.}) for the various classes of critical
points of $V$ can be obtained by plugging the criticality conditions
themselves into the $n_{V}+1$ real RSG identities\footnote{%
The extra real degree of freedom is only apparent, and removed by the
homogeneity of degree one of the RSG identities (\ref{RSG-electric-Ids})
under a \textit{real} overall shift of charges
\begin{equation*}
q_{i}\longrightarrow \eta q_{i},~\eta \in \mathbb{R}.
\end{equation*}
} (\ref{RSG-electric-Ids}). By so doing, one respectively obtains:

\begin{itemize}
\item  BPS Attractor Eqs.:
\begin{equation}
q_{i}=\widehat{\lambda }_{i}Z.  \label{1/2-BPS-new-attr}
\end{equation}
While Eqs. (\ref{1/2-BPS}) are $n_{V}$ real differential ones, the $n_{V}+1$
real Eqs. (\ref{1/2-BPS-new-attr}) are purely algebraic.

\item  Non-BPS Attractor Eqs.:
\begin{equation}
q_{i}=\widehat{\lambda }_{i}Z-\frac{1}{2}\left( \frac{3}{2}\right) ^{3/2}%
\frac{1}{Z}T^{xyz}Z_{y}Z_{z}\widehat{\lambda }_{i,x}.
\label{nBPS-Z<>0-new-attr}
\end{equation}
\end{itemize}

\subsection{\label{Higher-Order-T}Derivatives of $T_{xyz}$}

Now, in order to proceed further, it is convenient to compute the
Riemann-covariant derivative of the invariant tensor $T_{xyz}$, namely $%
T_{xyz;w}$, a quantity which will be relevant in the subsequent treatment.
By using the definition (\ref{T}), one obtains
\begin{equation}
T_{xyz;w}=T_{(xyz);w}=-\sqrt{6}\left[ -\frac{1}{2}%
g_{(yz}g_{xw)}+T_{r(yz}T_{xw)s}g^{rs}\right] =T_{(xyz;w)}.  \label{DT}
\end{equation}
Consequently, the condition (\ref{T-symm}) for the real special manifold $%
M_{5}$ to be a \textit{symmetric} coset can be equivalently recast as
follows (see \textit{e.g.} page 14 of \cite{FG2}, and Eq. (3.2.1.9) of \cite
{AFMT1}):
\begin{equation}
T_{r(yz}T_{xw)s}g^{rs}=\frac{1}{2}g_{(yz}g_{xw)}.  \label{DT-symm}
\end{equation}

One can then proceed further, and compute $T_{xyz;w;q}$. Starting from Eq. (%
\ref{DT}) one obtains (within the \textit{metric postulate})
\begin{equation}
T_{xyz;w;q}=T_{(xyz;w);q}=-2\sqrt{6}T_{r\left( yz\right| ;q}T_{\left|
xw\right) s}g^{rs}=-2\sqrt{6}T_{r(yz;q}T_{xw)s}g^{rs}=T_{(xyz;w;q)}.
\label{DDT-1}
\end{equation}
Through Eq. (\ref{DT}), this result can be further elaborated to give:
\begin{equation}
T_{xyz;w;q}=12\left[ -\frac{1}{2}g_{(yz}T_{xwq)}+T_{\left( q\right|
vr}T_{p\mid yz}T_{xw)s}g^{pv}g^{rs}\right] .  \label{DDT-2}
\end{equation}
One can now introduce the following rank-$5$ completely symmetric tensor $%
\widetilde{E}_{xyzwq}$, which is the ``RSG analogue'' of the so-called $E$%
-tensor\footnote{%
The $E$-tensor of SKG was firstly introduced in \cite{dWVVP}, and it has
been recently considered in the theory of extremal $d=4$ BH attractors in
\cite{ADFT-rev,DFT-Hom-non-Symm,Kallosh-rev,CFMZ1,Raju-1}.
\par
{}} of SKG:
\begin{equation}
\widetilde{E}_{xyzwq}\equiv \frac{1}{12}T_{xyz;w;q}=\frac{1}{12}%
T_{(xyz;w;q)}=\widetilde{E}_{(xyzwq)},  \label{E-tilde-1}
\end{equation}
satisfying by definition the relation
\begin{equation}
T_{\left( q\right| vr}T_{p\mid yz}T_{xw)s}g^{pv}g^{rs}=\frac{1}{2}%
g_{(yz}T_{xwq)}+\widetilde{E}_{xyzwq},  \label{E-tilde-2}
\end{equation}
holding globally in RSG.

By recalling the symmetricity condition (\ref{T-symm}), Eqs. (\ref{DDT-1})-(%
\ref{E-tilde-2}) yield
\begin{equation}
T_{xyz;w}=0\Rightarrow T_{xyz;w;q}=0\Leftrightarrow \widetilde{E}%
_{xyzwq}=0\Leftrightarrow T_{\left( q\right| vr}T_{p\mid
yz}T_{xw)s}g^{pv}g^{rs}=\frac{1}{2}g_{(yz}T_{xwq)}.
\end{equation}

\subsection{\label{Generic-RSG}Generic RSG}

Let us now consider the value of the effective potential $V$ at the various
classes of its critical points. By recalling its very definition (\ref{V}),
Eqs. (\ref{1/2-BPS}) and (\ref{nBPS-Z<>0-large}) yield the following results:

\subsubsection{BPS}

Recall Eq. (\ref{pppp-1}):
\begin{equation}
V=Z^{2}.  \label{V-1/2-BPS}
\end{equation}
Through Eq. (\ref{Bek-Haw}), this yields to
\begin{equation}
\frac{S_{BH,d=5}}{\pi }=\frac{A_{H}}{4\pi }\equiv R_{H}^{2}=V^{3/4}=\left|
Z\right| ^{3/2}.  \label{S-1/2-BPS}
\end{equation}

\subsubsection{Non-BPS and the \textit{``Dressed''} Charges' Sum Rule}

\begin{equation}
V=Z^{2}+\frac{3}{2}g^{xy}Z_{x}Z_{y}=Z^{2}+\frac{3}{8}\frac{1}{Z^{2}}%
g^{xy}T_{xzt}T_{wsy}Z^{z}Z^{t}Z^{w}Z^{s}.
\end{equation}
By recalling Eq. (\ref{DT}), the second term in the r.h.s. of Eq. (\ref{1})
can be further elaborated as follows:
\begin{equation}
Z_{x}Z^{x}=-\frac{1}{8}\sqrt{\frac{3}{2}}\frac{1}{Z^{2}}%
T_{(ztw;s)}Z^{z}Z^{t}Z^{w}Z^{s}+\frac{3}{16}\frac{1}{Z^{2}}\left(
Z_{x}Z^{x}\right) ^{2},
\end{equation}
yielding ($Z_{x}Z^{x}\neq 0$):
\begin{equation}
\frac{3}{2}Z_{x}Z^{x}=8Z^{2}+\sqrt{\frac{3}{2}}\frac{%
T_{(xyz;w)}Z^{x}Z^{y}Z^{z}Z^{w}}{Z_{u}Z^{u}}.  \label{rule of 8}
\end{equation}
Consequently at non-BPS $Z\neq 0$ critical points of $V$ it generally holds
that:
\begin{equation}
V=9Z^{2}+\widetilde{\Delta },  \label{V-nBPS-Z<>0-gen}
\end{equation}
where the real quantity
\begin{equation}
\widetilde{\Delta }\equiv \sqrt{\frac{3}{2}}\frac{%
T_{(xyz;w)}Z^{x}Z^{y}Z^{z}Z^{w}}{Z_{u}Z^{u}}  \label{Delta-tilde}
\end{equation}
has been introduced. This latter is the ``RSG analogue'' of the complex
quantity $\Delta $ introduced in SKG \cite{ADFT-rev} (see also \cite
{DFT-Hom-non-Symm,Kallosh-rev,CFMZ1,Raju-1}). As $\Delta $ enters the
\textit{``dressed''} charges' sum rule at non-BPS ($Z\neq 0$) critical
points of $V_{BH}$ in SKG (see \textit{e.g.} Eqs. (282)-(284) of \cite
{ADFT-rev}), so $\widetilde{\Delta }$ enters the \textit{``dressed''}
charges' sum rule (\ref{V-nBPS-Z<>0-gen}) at non-BPS critical points of $V$
in RSG, which further simplifies to (\ref{pppp-2}) \textit{at least} in
symmetric RSG (having $\widetilde{\Delta }=0$ globally). Notice that,
through Eq. (\ref{rule of 8}) and definition (\ref{Delta-tilde}), the
(assumed) strictly positive definiteness of $g_{xy}$ (throughout all $M_{5}$%
, and in particular at the considered class of critical points of $V$
itself) yields
\begin{equation}
Z^{2}+\frac{\widetilde{\Delta }}{8}>0.  \label{bound-1}
\end{equation}
Through Eq. (\ref{Bek-Haw}), Eq. (\ref{V-nBPS-Z<>0-gen}) yields
\begin{equation}
\frac{S_{BH,d=5}}{\pi }=\frac{A_{H}}{4\pi }\equiv R_{H}^{2}=V^{3/4}=\left(
9Z^{2}+\widetilde{\Delta }\right) ^{3/4}.  \label{S-nBPS-Z<>0-gen}
\end{equation}

\subsubsection{\label{I_3-hat-d}The Functional $\widehat{\mathcal{I}}_{3}$}

Within a generic RSG, let us now consider the function
\begin{equation}
\widehat{\mathcal{I}}_{3}\equiv \frac{1}{6}Z^{3}-\frac{3}{8}ZZ_{x}Z^{x}-%
\frac{1}{4}\sqrt{\frac{3}{2}}T_{xyz}Z^{x}Z^{y}Z^{z}.  \label{I_3-hat}
\end{equation}
In general, $\widehat{\mathcal{I}}_{3}$ is a diffeomorphism- and symplectic-
invariant function of the scalars $\phi ^{x}$ in $M_{5}$, or equivalently a
functional of the \textit{``dressed''} charges $\left\{ Z,Z_{x}\right\} $ in
$M_{5}$. Its derivative reads (recalling Eq. (\ref{DT}))
\begin{eqnarray}
\widehat{\mathcal{I}}_{3,w} &=&\widehat{\mathcal{I}}_{3;w}=-\sqrt{\frac{3}{2}%
}T_{xyz;w}Z^{x}Z^{y}Z^{z}  \notag \\
&=&-\frac{1}{2}Z_{x}Z^{x}Z_{w}+\frac{1}{3}g^{rs}\left(
T_{rzy}T_{xws}+T_{rzx}T_{yws}+T_{rzw}T_{xys}\right) Z^{x}Z^{y}Z^{z}.
\label{DI_3-hat}
\end{eqnarray}
From the definition (\ref{Delta-tilde}), it thus follows that
\begin{equation}
\widetilde{\Delta }=-\frac{\widehat{\mathcal{I}}_{3,x}Z^{x}}{Z_{y}Z^{y}}.
\end{equation}

The computation of $\widehat{\mathcal{I}}_{3}$ and $\widehat{\mathcal{I}}%
_{3,x}$ (respectively given by Eqs. (\ref{I_3-hat}) and (\ref{DI_3-hat})) at
the various classes of critical points of $V$ (specified by Eqs. (\ref
{1/2-BPS})-(\ref{nBPS-Z<>0-large})) respectively yield to the following
results.

\paragraph{BPS}

\begin{eqnarray}
\widehat{\mathcal{I}}_{3} &=&\frac{1}{6}Z^{3};  \label{I_3-hat-BPS} \\
\widehat{\mathcal{I}}_{3,x} &=&0.
\end{eqnarray}
Thus, by recalling Eqs. (\ref{V-1/2-BPS}) and (\ref{S-1/2-BPS}), it follows
that
\begin{equation}
\frac{S_{BH,d=5}}{\pi }=\frac{A_{H}}{4\pi }\equiv R_{H}^{2}=\left| Z\right|
^{3/2}=V^{3/4}=\sqrt{6}\left| \widehat{\mathcal{I}}_{3}\right| ^{1/2}.
\end{equation}

\paragraph{Non-BPS}

Eq. (\ref{rule of 8}) and definition (\ref{Delta-tilde}) yield
\begin{equation}
Z_{x}Z^{x}=\frac{16}{3}Z^{2}+\frac{2}{3}\widetilde{\Delta }.
\end{equation}
On the other hand, by recalling Eqs. (\ref{DT}) and (\ref{nBPS-Z<>0-large}),
the term $T_{xyz}Z^{x}Z^{y}Z^{z}$ can be further elaborated at non-BPS $%
Z\neq 0$ critical points of $V$ as follows:
\begin{equation}
T_{xyz}Z^{x}Z^{y}Z^{z}=-\frac{1}{2\sqrt{6}}\frac{\left( Z_{x}Z^{x}\right) }{Z%
}\left( \widetilde{\Delta }-\frac{3}{2}Z_{y}Z^{y}\right) .
\end{equation}
Thus, definition (\ref{I_3-hat}) yields the following expression of $%
\widehat{\mathcal{I}}_{3}$ at non-BPS $Z\neq 0$ critical points of $V$:
\begin{equation}
\widehat{\mathcal{I}}_{3}=-\frac{9}{2}Z^{3}\left( 1+\frac{7}{6}\frac{%
\widetilde{\Delta }}{3^{2}Z^{2}}\right) \Leftrightarrow \frac{\widetilde{%
\Delta }}{3^{2}Z^{2}}=-\frac{6}{7}\left( \frac{2}{9}\frac{\widehat{\mathcal{I%
}}_{3}}{Z^{3}}+1\right) .  \label{I_3-hat-nBPS-Z<>0}
\end{equation}
Thus, by recalling Eqs. (\ref{V-nBPS-Z<>0-gen}) and (\ref{S-nBPS-Z<>0-gen}),
it follows that
\begin{equation}
\frac{S_{BH,d=5}}{\pi }=\frac{A_{H}}{4\pi }\equiv R_{H}^{2}=\left( 9Z^{2}+%
\widetilde{\Delta }\right) ^{3/4}=V^{3/4}=\frac{3^{3/2}}{7^{3/4}}\left|
Z\right| ^{3/2}\left( 1-\frac{4}{3}\frac{\widehat{\mathcal{I}}_{3}}{Z^{3}}%
\right) ^{3/4},
\end{equation}
thus necessarily yielding
\begin{equation}
\frac{3}{4}>\frac{\widehat{\mathcal{I}}_{3}}{Z^{3}}.
\end{equation}

\subsection{\label{Symmetric-RSG}Symmetric RSG and ``Large'' Charge Orbits}

Let us now consider the case in which\footnote{%
``$mcs$'' is acronym for \textit{maximal compact subgroup} (with \textit{%
symmetric} embedding). Unless otherwise noted, all considered embeddings are
symmetric. Moreover, the subscript ``$\max $'' denotes the maximality of the
embedding throughout.}
\begin{equation}
M_{5}=\frac{G_{5}}{H_{5}}=\frac{G_{5}}{mcs\left( H_{5}\right) }
\end{equation}
is a \textit{symmetric} coset.

(\textit{At least}) in this case, $d_{ijk}$ is the unique $G_{5}$ -invariant
rank-$3$ completely symmetric tensor, whereas $T_{xyz}$ is the unique $H_{5}$
-invariant rank-$3$ completely symmetric tensor.

``Magic'' \textit{symmetric} $M_{5}$'s are reported in Table 1 (see \textit{%
e.g.} \cite{dWVVP}, and Refs. therein; see also \cite{LA08} for a brief
review and \ list of Refs.).

Besides these four isolated cases, there are two infinite sequences of other
\textit{symmetric} real special manifolds, namely the so-called \textit{%
Jordan symmetric} sequence
\begin{equation}
M_{J,5,n}\equiv SO\left( 1,1\right) \times \frac{SO\left( 1,n\right) }{%
SO\left( n\right) },~n=n_{V}-1\in \mathbb{N},  \label{N=2-d=5-Jordan-Symm}
\end{equation}
and the \textit{non-Jordan symmetric} sequence \cite{dWVP-3}
\begin{equation}
M_{nJ,5,n}\equiv \frac{SO\left( 1,n\right) }{SO\left( n\right) },~n=n_{V}\in
\mathbb{N},  \label{non-Jordan-symm}
\end{equation}
$n_{V}$ being the number of Abelian vector supermultiplets coupled to the $%
\mathcal{N}=2$, $d=5$ supergravity one.

The sequence (\ref{non-Jordan-symm}) is the only (sequence of) symmetric RSG
which is not related to Jordan algebras of degree three. It is usually
denoted by $L\left( -1,n-1\right) $ in the classification of homogeneous
Riemannian $d$-spaces (see \textit{e.g.} \cite{dWVVP}, and Refs. therein).
It will not be further considered here, because it does not correspond to
symmetric spaces in four dimensions.

$G_{5}$ and $H_{5}$ can respectively be interpreted as the reduced structure
group $Str_{0}$ and the automorphism group $Aut$ of the corresponding
Euclidean Jordan algebra of degree three (see \textit{e.g.} \cite
{Gunaydin-rec-rev} for a recent review, and Refs. therein):
\begin{equation}
M_{5}=\frac{G_{5}}{H_{5}}=\frac{Str_{0}\left( J_{3}\right) }{Aut\left(
J_{3}\right) }.
\end{equation}
\textit{\ }

\begin{table}[t]
\begin{center}
\begin{tabular}{|c||c|c|c|}
\hline
$
\begin{array}{c}
~ \\
J_{3}^{\mathbb{A}} \\
~
\end{array}
$ & $
\begin{array}{c}
~ \\
M_{5}=\frac{G_{5}}{H_{5}}=\frac{Str_{0}\left( J_{3}\right) }{Aut\left(
J_{3}\right) } \\
=~\mathcal{O}_{BPS,large}, \\
r=3 \\
H_{5}\equiv mcs\left( G_{5}\right) ~
\end{array}
$ & $
\begin{array}{c}
~ \\
M_{5}^{\ast }=\frac{G_{5}}{\widetilde{H}_{5}} \\
=~\mathcal{O}_{nBPS,large} \\
~r=3
\end{array}
$ & $
\begin{array}{c}
~ \\
\mathcal{M}_{nBPS,large}=\frac{\widetilde{H}_{5}}{\widetilde{h}_{5}}, \\
~ \\
\widetilde{h}_{5}\equiv mcs\left( \widetilde{H}_{5}\right) \\
~
\end{array}
$ \\ \hline
$
\begin{array}{c}
~ \\
J_{3}^{\mathbb{O}},~q=8 \\
~
\end{array}
$ & $
\begin{array}{c}
~ \\
\frac{E_{6(-26)}}{F_{4\left( -52\right) }} \\
~
\end{array}
$ & $
\begin{array}{c}
~ \\
\frac{E_{6(-26)}}{F_{4\left( -20\right) }} \\
~
\end{array}
$ & $
\begin{array}{c}
~ \\
\frac{F_{4(-20)}}{SO\left( 9\right) } \\
~
\end{array}
$ \\ \hline
$
\begin{array}{c}
~ \\
J_{3}^{\mathbb{H}},~q=4 \\
~
\end{array}
$ & $
\begin{array}{c}
~ \\
\frac{SU^{\ast }(6)}{USp\left( 6\right) } \\
~
\end{array}
$ & $
\begin{array}{c}
~ \\
\frac{SU^{\ast }(6)}{USp\left( 4,2\right) } \\
~
\end{array}
$ & $
\begin{array}{c}
~ \\
\frac{USp(4,2)}{USp\left( 4\right) \times USp\left( 2\right) } \\
~
\end{array}
$ \\ \hline
$
\begin{array}{c}
~ \\
J_{3}^{\mathbb{C}},~q=2 \\
~
\end{array}
$ & $
\begin{array}{c}
~ \\
\frac{SL(3,\mathbb{C})}{SU(3)} \\
~
\end{array}
$ & $
\begin{array}{c}
~ \\
\frac{SL(3,\mathbb{C})}{SU(2,1)} \\
~
\end{array}
$ & $
\begin{array}{c}
~ \\
\frac{SU(2,1)}{SU\left( 2\right) \times U\left( 1\right) } \\
~
\end{array}
$ \\ \hline
$
\begin{array}{c}
~ \\
J_{3}^{\mathbb{R}},~q=1 \\
~
\end{array}
$ & $
\begin{array}{c}
~ \\
\frac{SL(3,\mathbb{R})}{SO(3)} \\
~
\end{array}
$ & $
\begin{array}{c}
~ \\
\frac{SL(3,\mathbb{R})}{SO(2,1)} \\
~
\end{array}
$ & $
\begin{array}{c}
~ \\
\frac{SO(2,1)}{SO\left( 2\right) } \\
~
\end{array}
$ \\ \hline
\end{tabular}
\end{center}
\caption{\textbf{Homogeneous \textit{symmetric} real special vector
multiplets' scalar manifolds }$M_{5}$ \textbf{of }$\mathcal{N}$$=2$\textbf{,}
$d=5$\textbf{\ ``magic'' supergravity. }$M_{5}$\textbf{'s} \textbf{also are:
1) the non-BPS }$Z\neq 0$ \textbf{\textit{moduli spaces} of }$\mathcal{N}$%
\textbf{$=2$, $d=4$ special K\"{a}hler symmetric vector multiplets' scalar
manifolds \protect\cite{Ferrara-Marrani-2}; and 2) the \textit{``large''}}$%
\frac{1}{2}$\textbf{-BPS charge orbits }$\mathcal{O}_{BPS,large}$\textbf{'s
of }$\mathcal{N}=2$, $d=5$ \textbf{Maxwell-Einstein supergravity itself
\protect\cite{FG2}. The \textit{``large''}non-BPS }$Z\neq 0$\textbf{\ charge
orbits }$\mathcal{O}_{nBPS,large}\mathbf{=}M_{5}^{\ast }$ \textbf{(see
\textit{e.g.} Table 5 of \protect\cite{Trigiante-1} and Refs. therein)}
\textbf{and the related} \textbf{non-BPS }$Z\neq 0$\textbf{\ \textit{moduli
spaces}} $\mathcal{M}_{nBPS,large}$ \textbf{are reported, as well.} \textbf{%
The \textit{rank} }$r$ \textbf{of the orbit is} \textbf{defined as the
minimal number of charges defining a representative solution. As observed in
\protect\cite{Ferrara-Marrani-2}, for \textit{``magic''} supergravities }$%
n_{V}=\dim _{\mathbb{R}}M_{5}=3q+2$\textbf{, whereas} $dim_{\mathbb{R}}%
\mathcal{M}_{nBPS,large}=2q$\textbf{, and} $Spin\left( 1+q\right) \subset
\widetilde{h}_{5}$\textbf{. See text for more details}}
\end{table}

Furthermore, (\textit{at least}\footnote{%
Notice that, from Eq. (\ref{DI_3-hat}), it follows that
\begin{equation*}
\widehat{\mathcal{I}}_{3,w}=0\Leftrightarrow T_{xyz;w}Z^{x}Z^{y}Z^{z}=0,
\end{equation*}
whose (\ref{T-symm}) is a solution.}) in symmetric RSG, due to Eqs. (\ref
{T-symm}) and (\ref{DI_3-hat}), it holds that
\begin{equation}
\widehat{\mathcal{I}}_{3,x}=\widehat{\mathcal{I}}_{3;x}=0.
\end{equation}
In other words, $\widehat{\mathcal{I}}_{3}$ is \textit{independent} on all
scalars $\phi ^{x}$. Furthermore:
\begin{equation}
\widehat{\mathcal{I}}_{3}=\mathcal{I}_{3},  \label{I-I}
\end{equation}
where $\mathcal{I}_{3}$ is the unique cubic invariant of the relevant
\textit{electric} (ir)repr. $\mathbf{R}_{Q}$ of $d=5$ $U$-duality $G_{5}$,
defined by (\ref{I_3-el}). As mentioned above, $d_{ijk}$ is $G_{5}$%
-invariant in all RSG, whereas $d^{ijk}$ is $G_{5}$-invariant \textit{at
least} in \textit{symmetric} RSG.

In this framework, by virtue of the relations (\ref{2}) or (\ref{3}), the
Bekenstein-Hawking entropy-area formula (\ref{Bek-Haw}) can be completed as
follows (recall Eq. (\ref{I-I})):
\begin{equation}
\frac{S_{BH,d=5}}{\pi }=\frac{A_{H}}{4\pi }\equiv R_{H}^{2}=\left( \left.
V\right| _{\partial V=0}\right) ^{3/4}=\sqrt{6}\left| \mathcal{I}_{3}\right|
^{1/2}=\sqrt{6}\left| \widehat{\mathcal{I}}_{3}\right| ^{1/2}.
\label{Bek-Haw-2}
\end{equation}

Furthermore, in RSG based on \textit{symmetric} cosets $\frac{G_{5}}{H_{5}}$
the representation space of the irrepr. of $G_{5}$ in which the (electric
\textit{or} magnetic) charges sit admit a stratification in disjoint orbits
\cite{FG1,FG2}. Such orbits are homogeneous, in some case symmetric,
manifolds.

The charge orbits supporting \textit{non-degenerate} (in the sense specified
above; see the end of Subsect. \ref{Critical-Points-V}) critical points of $%
V $ are called ``large'' orbits, because they correspond to the previously
introduced class of ``large'' BHs with \textit{non-vanishing}
Bekenstein-Hawking entropy-area (see Eq. (\ref{Bek-Haw})). On the other
hand, charge orbits corresponding to ``small'' BHs (having \textit{vanishing}
Bekenstein-Hawking entropy-area) are correspondingly dubbed ``small'' orbits.

In the treatment of \textit{symmetric} RSG performed in present Subsection,
only ``large'' orbits, firstly found in \cite{FG2}, are considered.

In Sect. \ref{Small-Orbits-Symmetric-RSG}, through the properties of the
function $\widehat{\mathcal{I}}_{3}$ defined by Eq. (\ref{I_3-hat}), the
stabilizers of all ``small'' charge orbits of \textit{symmetric} RSG will be
derived, by suitably solving $G_{5}$-invariant (sets of) defining
differential constraints, as well as by performing suitable group
theoretical procedures.\bigskip \smallskip

We can now specialize the results obtained in Subsect. \ref{Generic-RSG} and
in Subsubsect. \ref{I_3-hat-d} to ``magic'' \textit{symmetric} RSG. The
detailed treatment of $\mathcal{N}=2$ Jordan symmetric sequence (\ref
{N=2-d=5-Jordan-Symm}) will be given in Sect. \ref{N=2,d=5-Jordan-Symm-Seq}.
Actually, the ``large'' charge orbits of (\ref{N=2-d=5-Jordan-Symm}) have
been already considered in \cite{FG2} (see also \cite{Ferrara-Marrani-2} for
the study of corresponding \textit{moduli spaces}), but in Sect. \ref
{N=2,d=5-Jordan-Symm-Seq} the treatment is further refined.

\subsubsection{\label{symm-BPS-large}BPS}

Eqs. (\ref{I_3-hat-BPS}) and (\ref{I-I}) yield to
\begin{equation}
\widehat{\mathcal{I}}_{3}=\frac{1}{6}Z^{3}=\mathcal{I}_{3},
\end{equation}
and thus:
\begin{equation}
\frac{S_{BH,d=5}}{\pi }=\frac{A_{H}}{4\pi }\equiv R_{H}^{2}=\left( \left.
V\right| _{\partial V=0}\right) ^{3/4}=\sqrt{6}\left| \mathcal{I}_{3}\right|
^{1/2}=\sqrt{6}\left| \widehat{\mathcal{I}}_{3}\right| ^{1/2}=\left|
Z\right| ^{3/2}.
\end{equation}
Such a ``large'' BH is supported by (electric) charges belonging to the
``large''\textit{\ }charge orbit (homogeneous symmetric manifold) \cite{FG2}
\begin{equation}
\mathcal{O}_{BPS,large}=\frac{G_{5}}{H_{5}}=M_{5}.  \label{O-1/2-BPS-large}
\end{equation}
The compactness of $H_{5}$ yields the absence of a \textit{moduli space}
related to $\frac{1}{2}$-BPS ``large'' attractor solutions, a fact that can
be seen also from the expression of the Hessian matrix of $V$ evaluated
along the BPS criticality constraints (\ref{1/2-BPS}) (see Eq. (\ref
{DDV-1/2-BPS}) below).

It is worth remarking that $M_{5}$'s also are the non-BPS $Z\neq 0$ \textit{%
moduli spaces} of $\mathcal{N}=2$, $d=4$ special K\"{a}hler symmetric vector
multiplets' scalar manifolds \cite{Ferrara-Marrani-2}.

Notice that in general
\begin{equation}
dim_{\mathbb{R}}M_{5}=n_{V}.
\end{equation}
As observed in \cite{Ferrara-Marrani-2}, for \textit{``magic''}
supergravities (based on Euclidean Jordan algebras of degree three $J_{3}^{%
\mathbb{A}}$ over the division algebras $\mathbb{A}$) it holds:
\begin{equation}
\begin{array}{l}
dim_{\mathbb{R}}M_{5}=3q+2, \\
\\
q\equiv dim_{\mathbb{R}}\left( \mathbb{A}=\mathbb{O},\mathbb{H},\mathbb{C},%
\mathbb{R}\right) =\left( 8,4,2,1\right) .
\end{array}
\label{q-def}
\end{equation}

\subsubsection{\label{symm-nBPS-large}Non-BPS}

Eqs. (\ref{I_3-hat-nBPS-Z<>0}) and (\ref{I-I}) yield to
\begin{equation}
\widehat{\mathcal{I}}_{3}=-\frac{9}{2}Z^{3}=\mathcal{I}_{3}.
\end{equation}
Indeed, from its very definition, in this framework it globally holds that
\begin{equation}
\widetilde{\Delta }=0,
\end{equation}
and thus (recall Eq. (\ref{pppp-2})):
\begin{equation}
\frac{3}{2}Z_{x}Z^{x}=8Z^{2}\Leftrightarrow V=9Z^{2}.
\end{equation}
Through Eq. (\ref{Bek-Haw-2}), it thus follows that
\begin{equation}
\frac{S_{BH,d=5}}{\pi }=\frac{A_{H}}{4\pi }\equiv R_{H}^{2}=\left( \left.
V\right| _{\partial V=0}\right) ^{3/4}=\sqrt{6}\left| \mathcal{I}_{3}\right|
^{1/2}=\sqrt{6}\left| \widehat{\mathcal{I}}_{3}\right| ^{1/2}=3^{3/2}\left|
Z\right| ^{3/2}.
\end{equation}
Such a ``large'' BH is supported by (electric) charges belonging to the
``large''\textit{\ }charge orbit (homogeneous symmetric manifold)
\begin{equation}
\mathcal{O}_{nBPS,large}=\frac{G_{5}}{\widetilde{H}_{5}}=M_{5}^{\ast },
\label{O-nBPS-Z<>0-large}
\end{equation}
where $\widetilde{H}_{5}$ is the unique non-compact, real form of $%
H_{5}=mcs\left( G_{5}\right) $ which admits a \textit{maximal symmetric}
embedding into $G_{5}$:
\begin{equation}
G_{5}\supsetneq _{\max }\widetilde{H}_{5}.
\end{equation}
The homogeneous symmetric pseudo-Riemannian manifold $M_{5}^{\ast }$ is the
\textit{``}$\ast $\textit{-version'' }of $M_{5}$, obtained through \textit{%
timelike} $d=6\rightarrow 5$ reduction from the corresponding \textit{%
anomaly-free} uplifted $\mathcal{N}=\left( 1,0\right) $, $d=6$ chiral theory
(see \textit{e.g.} Table 5 of \cite{Trigiante-1}, and Refs. therein). Notice
that Eq. (\ref{O-nBPS-Z<>0-large}) yields to
\begin{equation}
\mathcal{O}_{nBPS,large}=\mathcal{O}_{BPS,large}^{\ast },
\end{equation}
in the sense we have just specified.

The non-compactness of $\widetilde{H}_{5}$ implies the existence of a
non-BPS \textit{moduli space} \cite{Ferrara-Marrani-2}
\begin{equation}
\mathcal{M}_{nBPS,large}\equiv \frac{\widetilde{H}_{5}}{mcs\left( \widetilde{%
H}_{5}\right) }\equiv \frac{\widetilde{H}_{5}}{\widetilde{h}_{5}}.
\label{nBPS-Z<>0-large-moduli-space}
\end{equation}

As observed in \cite{Ferrara-Marrani-2}, for \textit{``magic''}
supergravities it holds (see \textit{e.g.} also Table 8 of \cite{LA08}, and
Refs. therein):
\begin{equation}
\begin{array}{l}
dim_{\mathbb{R}}\mathcal{M}_{nBPS,large}=2q; \\
Spin\left( 1+q\right) \subset \widetilde{h}_{5},
\end{array}
\label{Pal-2}
\end{equation}
where $Spin\left( 1+q\right) $ is the spin group in $1+q$ dimensions. Notice
that $2q$ is the number of $d=6$ (scalarless) vector multiplets needed for
an \textit{anomaly-free} uplift of the considered $\mathcal{N}=2$, $d=5$
\textit{``magic''} Maxwell-Einstein supergravity to the corresponding $%
\mathcal{N}=\left( 1,0\right) $ chiral quarter-minimal \textit{``magic''}
supergravity in $d=6$ (see \textit{e.g.} Sect. 5 of \cite{AFMT1}, and Refs.
therein).

Thus, by recalling (\ref{q-def}), the number $\sharp $ of ``non-flat''
scalar degrees of freedom along $\mathcal{O}_{nBPS,large}$ is
\begin{equation}
\sharp _{nBPS,large}\equiv dim_{\mathbb{R}}M_{5}-dim_{\mathbb{R}}\mathcal{M}%
_{nBPS,large}=q+2.  \label{n-nBPS-large}
\end{equation}

The ``large''\textit{\ }non-BPS $Z\neq 0$\ charge orbits $\mathcal{O}%
_{nBPS,large}=M_{5}^{\ast }$, and the related non-BPS $Z\neq 0$\ \textit{%
moduli spaces} $\mathcal{M}_{nBPS,large}$ for ``magic'' models are reported
in Table 1. Furthermore, it should be recalled that the \textit{Jordan
symmetric }sequence (\ref{N=2-d=5-Jordan-Symm}) is related to the reducible
rank-$3$ Euclidean Jordan algebra $\mathbb{R}\oplus \mathbf{\Gamma }_{1,n}$,
where $\mathbf{\Gamma }_{1,n}$ is the rank-$2$ Jordan algebra with a
quadratic form of Lorentzian signature $\left( 1,n\right) $, \textit{i.e.}
the Clifford algebra of $O\left( n,1\right) $ \cite{Jordan}.

\subsection{\label{Hessian-Matrix}Hessian Matrix of $V$}

From its very definition (\ref{V}), the first derivative of $V$ reads
(recall Eq. (\ref{RSG-AEs}))
\begin{equation}
V_{x}\equiv V_{,x}=V_{;x}=2\left( 2ZZ_{x}-\sqrt{\frac{3}{2}}%
T_{xyz}g^{ys}g^{zt}Z_{s}Z_{t}\right) .
\end{equation}
By further differentiating, the global expression of the real Hessian $%
n_{V}\times n_{V}$ matrix of $V$ in a generic RSG can be computed as
follows:
\begin{eqnarray}
V_{x;y} &=&V_{;x;y}  \notag \\
&=&\frac{8}{3}g_{xy}\left( Z^{2}-\frac{3}{8}Z_{w}Z^{w}\right) +2Z_{x}Z_{y}-8%
\sqrt{\frac{2}{3}}ZT_{xyz}Z^{z}+  \notag \\
&&+2\left( T_{xys}T_{rzw}+4T_{xzr}T_{yws}\right) g^{rs}Z^{z}Z^{w}  \notag \\
&=&V_{(x;y)},  \label{DDV}
\end{eqnarray}
where Eqs. (\ref{DDZ}) and (\ref{DT}) have been used.

On the other hand, by recalling definition (\ref{E-tilde-1}) and Eq. (\ref
{DI_3-hat}), it can be computed that
\begin{equation}
\widehat{\mathcal{I}}_{3;x;y}=-3\sqrt{\frac{3}{2}}\left( 4\widetilde{E}%
_{xyzwr}Z^{r}+\frac{2}{3}ZT_{xyz;w}-\sqrt{\frac{2}{3}}T_{zws;x}T_{ys^{\prime
}r}g^{ss^{\prime }}Z^{r}\right) Z^{z}Z^{w}.  \label{DDI_3-hat}
\end{equation}
Then, further elaboration of Eq. (\ref{DDV}) is possible for $Z\neq 0$.
Indeed, in such a case Eq. (\ref{DDI_3-hat}) implies that (recall Eq. (\ref
{DT}))
\begin{eqnarray}
&&T_{xzw;y}Z^{z}Z^{w}  \notag \\
&=&-\frac{1}{\sqrt{6}}\frac{1}{Z}\widehat{\mathcal{I}}_{3;x;y}-\frac{6}{Z}%
\widetilde{E}_{xyzwr}Z^{z}Z^{w}Z^{r}  \notag \\
&&+\frac{1}{2Z}\left( Z_{w}Z^{w}\right) T_{xyz}Z^{z}+\frac{1}{Z}%
Z_{y}T_{xzw}Z^{z}Z^{w}  \notag \\
&&-\frac{1}{Z}\left( T_{xrp}T_{yr^{\prime }s}T_{tzw}+2T_{xwp}T_{yr^{\prime
}s}T_{tzr}\right) g^{rr^{\prime }}g^{tp}Z^{s}Z^{z}Z^{w}.  \label{pp}
\end{eqnarray}
Notice that the symmetry properties $\widehat{\mathcal{I}}_{3;x;y}=\widehat{%
\mathcal{I}}_{3;\left( x;y\right) }$ and $T_{xzw;y}Z^{z}Z^{w}=T_{\left(
xzw;y\right) }Z^{z}Z^{w}$ are not manifest respectively from Eqs. (\ref
{DDI_3-hat}) and (\ref{pp}), due to the presence of $\widetilde{E}_{xyzwr}$,
$T_{xyz;w}$, and $\widehat{\mathcal{I}}_{3;x;y}$ itself. By plugging Eq. (%
\ref{pp}) back into Eq. (\ref{DDV}), the following result is achieved:
\begin{eqnarray}
V_{x;y} &=&V_{;x;y}  \notag \\
&=&4Z_{x}Z_{y}+\frac{8}{3}Z^{2}g_{xy}-8\sqrt{\frac{2}{3}}ZT_{xyz}Z^{z}
\notag \\
&&+\frac{1}{Z}\widehat{\mathcal{I}}_{3;x;y}+\frac{6\sqrt{6}}{Z}\widetilde{E}%
_{xyzwr}Z^{z}Z^{w}Z^{r}  \notag \\
&&-\sqrt{\frac{3}{2}}\frac{1}{Z}\left( Z_{w}Z^{w}\right) T_{xyz}Z^{z}-\frac{%
\sqrt{6}}{Z}Z_{y}T_{xzw}Z^{z}Z^{w}  \notag \\
&&+\frac{\sqrt{6}}{Z}\left( T_{xrp}T_{yr^{\prime
}s}T_{tzw}+2T_{xwp}T_{yr^{\prime }s}T_{tzr}\right) g^{rr^{\prime
}}g^{tp}Z^{s}Z^{z}Z^{w}  \notag \\
&&+4T_{xzw}T_{ysw^{\prime }}g^{ww^{\prime }}Z^{z}Z^{s},  \label{DDV-Z<>0}
\end{eqnarray}
holding true for $Z\neq 0$. Once again, notice that the symmetry property $%
V_{x;y}=V_{(x;y)}$ is not manifest from Eq. (\ref{DDV-Z<>0}), due to the
presence of $\widetilde{E}_{xyzwr}$ and $\widehat{\mathcal{I}}_{3;x;y}$.

By inserting the global condition (\ref{T-symm}) into Eq. (\ref{DDV}), one
obtains that
\begin{equation}
V_{x;y}=V_{;x;y}=4Z_{x}Z_{y}+\frac{8}{3}Z^{2}g_{xy}-8\sqrt{\frac{2}{3}}%
ZT_{xyz}Z^{z}+4T_{xzw}T_{ysw^{\prime }}g^{ww^{\prime }}Z^{z}Z^{s}\equiv
V_{;x;y,symm.}.  \label{DDV-symm}
\end{equation}
This is the global expression of the real Hessian $n_{V}\times n_{V}$ matrix
of $V$ (\textit{at least}) in symmetric RSG, and indeed it matches the
result given by Eq. (5-1) of \cite{FG2} (see also \cite{AFMT1}). Thus, Eqs. (%
\ref{DDV}) and (\ref{DDV-symm}) yield to the following result:
\begin{equation}
V_{;x;y}=V_{;x;y,symm.}-g_{xy}\left( Z_{w}Z^{w}\right) -2Z_{x}Z_{y}+2\left(
2T_{xwz}T_{ysz^{\prime }}+T_{xyz}T_{swz^{\prime }}\right) g^{zz^{\prime
}}Z^{w}Z^{s}.
\end{equation}

\subsubsection{\label{Hessian-Matrix-Critical}Evaluation at Critical Points
of $V$}

We will now proceed to evaluate the Hessian matrix of $V$ given by Eq. (\ref
{DDV}) at the various classes of critical points of $V$ itself, as given by
Eqs. (\ref{1/2-BPS})-(\ref{nBPS-Z<>0-large}).

\paragraph{BPS}

The necessary and sufficient BPS criticality constraints (\ref{1/2-BPS})
plugged into Eq. (\ref{DDV}) yield
\begin{equation}
V_{;x;y}=\frac{8}{3}g_{xy}Z^{2}.  \label{DDV-1/2-BPS}
\end{equation}
Eq. (\ref{DDV-1/2-BPS}) holds for a \textit{generic} RSG, and it matches the
result given by Eq. (5-2) of \cite{FG2}. For a strictly positive definite $%
g_{xy}$ (as it is usually assumed), it implies that the Hessian matrix of $V$
at its BPS critical points has \textit{all} strictly positive eigenvalues.

As mentioned above, the lack of \textit{Hessian massless modes} at $\frac{1}{%
2}$-BPS critical points of $V$ determines the absence of a \textit{moduli
space} in BPS attractor solutions, which thus have \textit{all} scalar
fields $\phi ^{x}$ stabilized at the (unique) event horizon of the
considered (\textit{electric}) $d=5$ extremal BH.

\paragraph{Non-BPS}

It is here worth noticing that Eq. (\ref{nBPS-Z<>0-large}) yields to
\begin{equation}
Z_{x}Z^{x}=\sqrt{\frac{3}{2}}\frac{1}{2Z}T_{xyz}Z^{x}Z^{y}Z^{z}.
\label{ML-1}
\end{equation}
By recalling the \textit{``dressed''} charges' sum rule given by Eq. (\ref
{rule of 8}) and definition (\ref{Delta-tilde}), Eq. (\ref{ML-1}) implies
\begin{equation}
\frac{32}{3}Z^{2}+\widetilde{\Delta }=\sqrt{\frac{3}{2}}\frac{1}{Z}%
T_{xyz}Z^{x}Z^{y}Z^{z}.  \label{ML-2}
\end{equation}
On the other hand, by using Eq. (\ref{DT}), one can compute also that
\begin{equation}
Z_{x}Z^{x}=-\frac{1}{8}\sqrt{\frac{3}{2}}\frac{1}{Z^{2}}%
T_{xyz;w}Z^{x}Z^{y}Z^{z}Z^{w}+\frac{3}{16}\frac{1}{Z^{2}}\left(
Z_{x}Z^{x}\right) ^{2}.  \label{ML-3}
\end{equation}
By dividing by $Z_{x}Z^{x}\neq 0$, one then obtains the \textit{``dressed''}
charges' sum rule given by Eq. (\ref{rule of 8}). However, one can also
interpret Eq. (\ref{ML-3}) as a quadratic Eq. in the unknown $Z_{x}Z^{x}$,
obtaining the result
\begin{equation}
0<Z_{x}Z^{x}=\frac{8}{3}Z^{2}\pm \sqrt{\frac{64}{9}Z^{4}-\frac{2}{3}\widehat{%
\mathcal{I}}_{3;x}Z^{x}}.  \label{ML-4}
\end{equation}
When $\widehat{\mathcal{I}}_{3;x}=0$ (\textit{i.e.} \ - \textit{at least} -
for symmetric RSG) Eq. (\ref{ML-4}) consistently yields\textit{\ }\cite{FG2}%
:
\begin{equation}
\frac{3}{2}Z_{x}Z^{x}=8Z^{2}.
\end{equation}

\section{\label{Small-Orbits-Symmetric-RSG}``Small'' Charge Orbits and
\textit{Moduli Spaces} in Symmetric ``Magic'' RSG}

In the treatment of \textit{symmetric} RSG performed in Subsect. \ref
{Symmetric-RSG}, only ``large'' charge orbits, supporting solutions to the
corresponding Attractor Eqs. (and firstly found in \cite{FG2}; see also \cite
{AFMT1}), have been considered.

In the present Section, by exploiting the properties of the functional $%
\widehat{\mathcal{I}}_{3}$ introduced in Subsubsect. \ref{I_3-hat-d}, all
``small'' charge orbits of ``magic'' \textit{symmetric} RSG will be
explicitly determined through the resolution of $G_{5}$-invariant defining
(differential) constraints both in ``bare'' and ``dressed'' charges bases,
as well as through group theoretical techniques.\medskip

By definition, $\widehat{\mathcal{I}}_{3}$($=\mathcal{I}_{3}$ in symmetric
RSG, as discussed in Subsect. \ref{Symmetric-RSG}, see Eq. (\ref{I-I}))
vanishes for all ``small'' charge orbits. Consequently, such orbits do not
support solutions to the Attractor Eqs. (\textit{alias} criticality
conditions of the effective potential $V$; see Eqs. (\ref{1/2-BPS})-(\ref
{nBPS-Z<>0-large}), or Eqs. (\ref{1/2-BPS-new-attr})-(\ref
{nBPS-Z<>0-new-attr}) in the so-called \textit{``new attractor''} approach).
In other words, the (classical) \textit{Attractor Mechanism} does not hold
for ``small'' charge orbits, which indeed do support BH states which are
\textit{intrinsically quantum}, in the sense that the effective description
through Einstein supergravity fail for them.

Besides the condition of vanishing $\widehat{\mathcal{I}}_{3}$, further
conditions, formulated in terms of derivatives of $\widehat{\mathcal{I}}_{3}$
in some charge basis, may be needed to fully characterize the class of
``small'' orbits under consideration. It is here worth pointing out that the
(sets of) $G_{5}$-invariant constraints which define ``small'' charge orbits
in homogeneous symmetric real special manifolds $\frac{G_{5}}{H_{5}}$ are
characterizing \textit{equations} for charges (in both \textit{``bare''} and
\textit{``dressed''} bases), but they actually are \textit{identities} in
\textit{all} scalar fields $\phi ^{x}$, and thus they hold globally in $%
\frac{G_{5}}{H_{5}}$. This is to be contrasted with ``large'' charge orbits,
which are defined through the Attractor Eqs. themselves, which are at the
same time characterizing \textit{equations} for charges (in both \textit{%
``bare''} and \textit{``dressed''} bases) and \textit{stabilization equations%
} for the scalars $\phi ^{x}$ at the event horizon of the extremal BH.

As it is well known \cite{Ferrara-Marrani-2}, at non-BPS $Z\neq 0$ critical
points of $V$ some scalars are actually unstabilized at the (unique) event
horizon of the corresponding ``large'' extremal BH solutions. Such
unstabilized $\phi ^{x}$'s span the \textit{moduli space }$\mathcal{M}%
_{nBPS,large}$ (given by Eq. (\ref{nBPS-Z<>0-large-moduli-space})),
associated with an hidden compact symmetry of the non-BPS $Z\neq 0$
Attractor Eqs. themselves, which can be traced back to the non-compactness
of the stabilizer of the non-BPS $Z\neq 0$ ``large'' charge orbit $\mathcal{O%
}_{nBPS,large}$ (see Eq. (\ref{O-nBPS-Z<>0-large}), to be contrasted with
Eq. (\ref{O-1/2-BPS-large})).

The ``small'' charge orbits are homogeneous manifolds of the form:
\begin{equation}
\mathcal{O}_{small}=\frac{G_{5}}{\mathcal{S}_{max}\rtimes \mathcal{T}},
\label{O-small-gen-struct}
\end{equation}
where $\rtimes $ denotes semi-direct group product throughout, and $\mathcal{%
T}$ is the non-semi-simple part of the stabilizer of $\mathcal{O}_{small}$,
which in all symmetric RSG (with some extra features characterizing the
symmetric Jordan sequence, see Sect. \ref{N=2,d=5-Jordan-Symm-Seq}) can be
identified with an Abelian translational subgroup of $G_{5}$ itself.

One can associate a \textit{moduli space} also to ``small'' charge orbits,
by observing that the non-compactness of $\mathcal{S}_{max}\rtimes \mathcal{T%
}$ yields the existence of a corresponding \textit{moduli space} defined as%
\footnote{%
We thank M. Trigiante for a discussion on the ``flat'' directions of
``small'' charge orbits.}
\begin{equation}
\mathcal{M}_{small}\equiv \frac{\mathcal{S}_{max}}{mcs\left( \mathcal{S}%
_{max}\right) }\rtimes \mathcal{T}.  \label{M-small-gen-struct}
\end{equation}
Note that, differently from ``large'' orbits, for ``small'' orbits there
exists a moduli space $\mathcal{M}_{small}=\mathcal{T}$ also when $\mathcal{S%
}_{max}$ is compact. As found in \cite{GLS-1,stu-unveiled} for ``large''
charge orbits of $\mathcal{N}=2$, $d=4$ $stu$ model, and recently proved in
a model-independent way in \cite{ADFT-FO-1}, the \textit{moduli spaces} of
charge orbits are defined all along the scalar flows, and thus they can be
interpreted as \textit{moduli spaces} of unstabilized scalars at the event
horizon (\textit{if any}) of the extremal BH, as well as \textit{moduli
spaces} of the ADM mass of the extremal BH at spatial infinity. In the
``small'' case, the interpretation at the event horizon breaks down, simply
because such an horizon does not exist at all (\textit{at least} in
Einsteinian supergravity approximation).

In general, the number $\sharp $ of ``non-flat'' scalar degrees of freedom
supported by a (``large'' or ``small'') charge orbit $\mathcal{O}$ with
associated \textit{moduli space} $\mathcal{M}$ is defined as follows:
\begin{equation}
\sharp \equiv dim_{\mathbb{R}}M_{d=5}-dim_{\mathbb{R}}\mathcal{M}.
\label{jjjjj-1}
\end{equation}

As an example, let us briefly consider the maximal $\mathcal{N}=8$, $d=5$
supergravity, whose ``large'' and ``small'' charge orbits have been
classified in \cite{FG1}. The scalar manifold of the theory is
\begin{equation}
M_{\mathcal{N}=8,d=5}=\frac{E_{6\left( 6\right) }}{USp\left( 8\right) }%
,~dim_{\mathbb{R}}=42.
\end{equation}

\begin{enumerate}
\item  The unique ``large'' charge orbit is $\frac{1}{8}$-BPS:
\begin{equation}
\mathcal{O}_{\frac{1}{8}-BPS}=\frac{E_{6\left( 6\right) }}{F_{4\left(
4\right) }},~dim_{\mathbb{R}}=26,
\end{equation}
with corresponding \textit{moduli space} \cite{Ferrara-Marrani-2}
\begin{equation}
\mathcal{M}_{\frac{1}{8}-BPS}=\frac{F_{4\left( 4\right) }}{USp\left(
6\right) \times USp\left( 2\right) },~dim_{\mathbb{R}}=28.
\end{equation}
Thus, the number of ``non-flat'' directions along $\mathcal{O}_{\frac{1}{8}%
-BPS}$ reads
\begin{equation}
\sharp _{\frac{1}{8}-BPS}\equiv dim_{\mathbb{R}}M_{\mathcal{N}=8,d=5}-dim_{%
\mathbb{R}}\mathcal{M}_{\frac{1}{8}-BPS}=14.
\end{equation}
Since the charge orbit is ``large'', $\sharp _{\frac{1}{8}-BPS}$ also
expresses the actual number of scalar degrees of freedom which are
stabilized in terms of the electric (magnetic) charges in the near-horizon
geometry of the extremal black hole (black string) under consideration.

\item  The ``small'' $\frac{1}{4}$-BPS orbit is
\begin{equation}
\mathcal{O}_{\frac{1}{4}-BPS}=\frac{E_{6\left( 6\right) }}{SO\left(
5,4\right) \rtimes \mathbb{R}^{16}},~dim_{\mathbb{R}}=26,
\end{equation}
with corresponding \textit{moduli space}
\begin{equation}
\mathcal{M}_{\frac{1}{4}-BPS}=\frac{SO\left( 5,4\right) }{SO\left( 5\right)
\times SO\left( 4\right) }\rtimes \mathbb{R}^{16},~dim_{\mathbb{R}}=36.
\end{equation}
Thus, the number of ``non-flat'' directions along $\mathcal{O}_{\frac{1}{4}%
-BPS}$ reads
\begin{equation}
\sharp _{\frac{1}{4}-BPS}\equiv dim_{\mathbb{R}}M_{\mathcal{N}=8,d=5}-%
\mathcal{M}_{\frac{1}{4}-BPS}=6.
\end{equation}

\item  The ``small'' $\frac{1}{2}$-BPS orbit is
\begin{equation}
\mathcal{O}_{\frac{1}{2}-BPS}=\frac{E_{6\left( 6\right) }}{SO\left(
5,5\right) \rtimes \mathbb{R}^{16}},~dim_{\mathbb{R}}=17,
\end{equation}
with corresponding \textit{moduli space}
\begin{equation}
\mathcal{M}_{\frac{1}{2}-BPS}=\frac{SO\left( 5,5\right) }{SO\left( 5\right)
\times SO\left( 5\right) }\rtimes \mathbb{R}^{16}=M_{\left( 2,2\right)
,d=6}\rtimes \mathbb{R}^{16},~dim_{\mathbb{R}}=41,
\end{equation}
where $M_{\left( 2,2\right) ,d=6}$ is the scalar manifold of maximal
(non-chiral) supergravity in $d=6$. Thus, the number of ``non-flat''
directions along $\mathcal{O}_{\frac{1}{2}-BPS}$ reads
\begin{equation}
\sharp _{\frac{1}{2}-BPS}\equiv dim_{\mathbb{R}}M_{\mathcal{N}=8,d=5}-%
\mathcal{M}_{\frac{1}{2}-BPS}=1.  \label{ress-1}
\end{equation}
As we will point out more than once in the treatment below, result (\ref
{ress-1}) expresses the pretty general fact that the unique ``non-flat''
direction along maximally supersymmetric (namely, $\frac{1}{2}$-BPS) charge
orbits is the Kaluza-Klein radius in the dimensional reduction $%
d=5\rightarrow d=4$.
\end{enumerate}

In the treatment of Subsect. \ref{Constraints-Small-Orbits}, the $G_{5}$%
-invariant constraints defining all classes of ``small'' charge orbits in
\textit{all} symmetric RSG will be derived. Then they will be solved both in
\textit{``bare''} and \textit{``dressed''} charge bases in Subsect. \ref
{Solutions-Constraints-Small-Orbits}. Furthermore, in App. \ref
{Group-Theory-Small-Orbits} the origin of ``small'' charge orbits (and in
particular of $\mathcal{T}$) will be elucidated through group theoretical
procedures (namely, \.{I}n\"{o}n\"{u}-Wigner contractions \cite{IW-1,IW-2}
and $SO\left( 1,1\right) $-three grading).

While the treatment of Subsect. \ref{Constraints-Small-Orbits} holds for
\textit{all} symmetric RSG, the treatments given in Apps. \ref
{Solutions-Constraints-Small-Orbits} and \ref{Group-Theory-Small-Orbits}
strictly fit only the isolated cases of symmetric RSG provided by the
so-called \textit{``magic''} symmetric RSG's \cite{GST1,GST2,GST3,GST4}. The
main results of Apps. \ref{Solutions-Constraints-Small-Orbits} and \ref
{Group-Theory-Small-Orbits} are reported in Tables 3 and 4 (the symmetric
Jordan sequence (\ref{N=2-d=5-Jordan-Symm}) is considered in Sect. \ref
{N=2,d=5-Jordan-Symm-Seq}). In the \textit{``magic''} octonionic case $%
J_{3}^{\mathbb{O}}$ ($q=8$), the results of \cite{FG1} are matched.\medskip

Below we summarize the main results of Apps. \ref
{Solutions-Constraints-Small-Orbits} and \ref{Group-Theory-Small-Orbits}.

\begin{itemize}
\item  The ``small''\textit{\ lightlike} BPS charge orbit ($dim_{\mathbb{R}%
}=3q+2$)
\begin{equation}
\mathcal{O}_{lightlike,BPS}=\frac{G_{5}}{\left( SO\left( q+1\right) \times
\mathcal{A}_{q}\right) \rtimes \mathbb{R}^{\left( spin\left( q+1\right)
,spin\left( Q_{q}\right) \right) }},  \label{magic-orbit-lightlike-Z<>0}
\end{equation}
thus with
\begin{eqnarray}
\mathcal{S}_{max,lightlike,BPS} &=&SO\left( q+1\right) \times \mathcal{A}%
_{q}; \\
&&  \notag \\
\mathcal{T}_{lightlike,BPS} &=&\mathbb{R}^{\left( spin\left( q+1\right)
,spin\left( Q_{q}\right) \right) }.  \label{rrr-1}
\end{eqnarray}

\begin{table}[t]
\begin{center}
\begin{tabular}{|c||c|c|}
\hline
$
\begin{array}{c}
~q
\end{array}
$ & $Q_{q}$ & $\mathcal{A}_{q}$ \\ \hline\hline
$8~$ & $-$ & \multicolumn{1}{|l|}{$-$} \\ \hline
$4$ & $2$ & \multicolumn{1}{|l|}{$SO\left( 3\right) $} \\ \hline
$2$ & $2$ & $SO\left( 2\right) $ \\ \hline
$1$ & $-$ & $-$ \\ \hline
\end{tabular}
\end{center}
\caption{$Q_{q}$ \textbf{and }$\mathcal{A}_{q}$ \textbf{for the various }$%
\mathcal{N}=2$, $d=5$ \textbf{``magic'' supergravities (based on }$J_{3}^{%
\mathbb{A}}$, $\mathbb{A}=\mathbb{O},\mathbb{H},\mathbb{C},\mathbb{R}$%
\textbf{), classified by }$q\mathbf{\equiv }$dim$_{\mathbb{R}}\mathbb{A}%
=8,4,2,1$}
\end{table}
$Q_{q}$ and $\mathcal{A}_{q}$, a further factor group in $\mathcal{S}_{max}$%
, are given by Table 2. Furthermore, we define
\begin{eqnarray}
spin\left( q+1\right) &\equiv &\dim _{\mathbb{R}}\left( \mathbf{Spin}\left(
q+1\right) \right) ;  \label{def-1} \\
spin\left( Q_{q}\right) &\equiv &\dim _{\mathbb{R}}\left( \mathbf{Spin}%
\left( Q_{q}\right) \right) ,  \label{def-2}
\end{eqnarray}

with $\mathbf{Spin}\left( q+1\right) $ and $\mathbf{Spin}\left( Q_{q}\right)
$ respectively denoting the spinor irreprs. in $q+1$ and $Q_{q}$ dimensions.
It is worth remarking that $\mathcal{A}_{q}$ is independent on the
space-time dimension ($d=3,4,5,6$) in which the quarter-minimal symmetric
\textit{``magic''} (Maxwell-Einstein) supergravity (classified by $q=8,4,2,1$%
) is considered. It also holds that
\begin{eqnarray}
d &=&5,6:\widehat{G}_{cent}=SO\left( 1,1\right) \times SO\left( q-1\right)
\times \mathcal{A}_{q}; \\
d &=&3,4:\widehat{G}_{cent}=G_{paint}=SO\left( q\right) \times \mathcal{A}%
_{q},
\end{eqnarray}
where the groups $\widehat{G}_{cent}$ and $G_{paint}$ are usually introduced
in the treatment of supergravity billiards and timelike reductions (for \
recent treatment and set of related Refs., see \textit{e.g.} \cite
{Trigiante-1}; see also Table 5 therein, also for subtleties concerning the
case $q=8$ in $d=5,6$). The \textit{moduli space} corresponding to (\ref
{magic-orbit-lightlike-Z<>0}) is purely translational:
\begin{equation}
\mathcal{M}_{lightlike,BPS}=\mathbb{R}^{\left( spin\left( q+1\right)
,spin\left( Q_{q}\right) \right) },  \label{magic-mod-space-lightlike-Z<>0}
\end{equation}
with real dimension
\begin{equation}
spin\left( q+1\right) \cdot spin\left( Q_{q}\right) =2q.
\end{equation}
Thus, by recalling (\ref{q-def}), the number $\sharp $ of scalar degrees of
freedom on which the ADM mass depends along $\mathcal{O}_{lightlike,BPS}$ is
(recall Eq. (\ref{n-nBPS-large}))
\begin{eqnarray}
\sharp _{light,BPS} &\equiv &dim_{\mathbb{R}}M_{5}-dim_{\mathbb{R}}\mathcal{M%
}_{lightlike,BPS}  \notag \\
&=&3q+2-\left( spin\left( q+1\right) \cdot spin\left( Q_{q}\right) \right)
=q+2.
\end{eqnarray}
By recalling Eq. (\ref{Pal-2}), it is worth noting that $\mathcal{M}%
_{nBPS,large}$ and $\mathcal{M}_{lightlike,BPS}$ have the same real
dimension, but they are completely different, as yielded by Eqs. (\ref
{nBPS-Z<>0-large-moduli-space}) and (\ref{magic-mod-space-lightlike-Z<>0}).

\item  The ``small''\textit{\ lightlike} non-BPS charge orbit ($dim_{\mathbb{%
R}}=3q+2$)
\begin{equation}
\mathcal{O}_{lightlike,nBPS}=\frac{G_{5}}{\left( SO\left( q,1\right) \times
\mathcal{A}_{q}\right) \rtimes \mathbb{R}^{\left( spin\left( q+1\right)
,spin\left( Q_{q}\right) \right) }},  \label{magic-orbit-lightlike-Z=0}
\end{equation}
thus with
\begin{eqnarray}
\mathcal{S}_{max,lightlike,nBPS} &=&SO\left( q,1\right) \times \mathcal{A}%
_{q}; \\
&&  \notag \\
\mathcal{T}_{lightlike,nBPS} &=&\mathbb{R}^{\left( spin\left( q+1\right)
,spin\left( Q_{q}\right) \right) }=\mathcal{T}_{lightlike,BPS}.
\label{rrr-2}
\end{eqnarray}
The related \textit{moduli space} reads ($dim_{\mathbb{R}}=3q$)
\begin{eqnarray}
\mathcal{M}_{lightlike,nBPS} &=&\frac{SO\left( q,1\right) }{SO\left(
q\right) }\rtimes \mathbb{R}^{\left( spin\left( q+1\right) ,spin\left(
Q_{q}\right) \right) }  \notag \\
&=&M_{nJ,5,q}\rtimes \mathbb{R}^{\left( spin\left( q+1\right) ,spin\left(
Q_{q}\right) \right) },  \label{magic-mod-space-lightlike-Z=0}
\end{eqnarray}
where $M_{nJ,5,q}$ is the $q$-th element of the \textit{generic non-Jordan
symmetric sequence} (\ref{non-Jordan-symm}). Thus, by recalling (\ref{q-def}%
), the number $\sharp $ of scalar degrees of freedom on which the ADM mass
depends along $\mathcal{O}_{lightlike,nBPS}$ is
\begin{eqnarray}
\sharp _{light,nBPS} &\equiv &dim_{\mathbb{R}}M_{5}-dim_{\mathbb{R}}\mathcal{%
M}_{lightlike,nBPS}  \notag \\
&=&2q+2-\left( spin\left( q+1\right) \cdot spin\left( Q_{q}\right) \right)
=2.
\end{eqnarray}

\item  The ``small''\textit{\ critical} BPS charge orbit ($dim_{\mathbb{R}%
}=2q+1$)
\begin{equation}
\mathcal{O}_{critical,BPS}=\frac{G_{5}}{\left( G_{6}\times \mathcal{A}%
_{q}\right) \rtimes \mathbb{R}^{\left( spin\left( q+1\right) ,spin\left(
Q_{q}\right) \right) }},  \label{magic-orbit-critical-Z<>0}
\end{equation}
where
\begin{equation}
G_{6}=SO\left( 1,q+1\right)  \label{G_6}
\end{equation}
is the $U$-duality group of the corresponding $\left( 1,0\right) $, $d=6$
chiral supergravity theory. Thus:
\begin{equation}
\begin{array}{l}
\mathcal{S}_{max,critical,BPS}=G_{6}\times \mathcal{A}_{q}; \\
\mathcal{T}_{critical,BPS}=\mathcal{T}_{lightlike,nBPS}=\mathcal{T}%
_{lightlike,BPS}.
\end{array}
\end{equation}
The related \textit{moduli space }reads ($dim_{\mathbb{R}}=3q+1$)
\begin{eqnarray}
\mathcal{M}_{critical,BPS} &=&\frac{SO\left( q+1,1\right) }{SO\left(
q+1\right) }\rtimes \mathbb{R}^{\left( spin\left( q+1\right) ,spin\left(
Q_{q}\right) \right) }  \notag \\
&=&M_{nJ,5,q+1}\rtimes \mathbb{R}^{\left( spin\left( q+1\right) ,spin\left(
Q_{q}\right) \right) }.  \label{magic-mod-space-critical-Z<>0}
\end{eqnarray}
Thus, by recalling (\ref{q-def}), the number $\sharp $ of scalar degrees of
freedom on which the ADM mass depends along $\mathcal{O}_{critical,BPS}$ is
\begin{eqnarray}
\sharp _{crit,BPS} &\equiv &dim_{\mathbb{R}}M_{5}-dim_{\mathbb{R}}\mathcal{M}%
_{critical,BPS}  \notag \\
&=&2q+1-\left( spin\left( q+1\right) \cdot spin\left( Q_{q}\right) \right)
=1.  \label{jjjj-1}
\end{eqnarray}
The unique scalar degree of freedom on which the ADM mass depends can be
interpreted as the Kaluza-Klein radius in the $d=5\rightarrow d=4$
reduction. Furthermore, it is worth observing that:
\begin{equation}
\mathcal{M}_{critical,BPS}=M_{\left( 1,0\right) ,d=6,J_{3}^{\mathbb{A}%
}}\rtimes \mathbb{R}^{\left( spin\left( q+1\right) ,spin\left( Q_{q}\right)
\right) },  \label{MM-1}
\end{equation}
where $M_{\left( 1,0\right) ,d=6,J_{3}^{\mathbb{A}}}$ is the manifold of
tensor multiplets' scalars in the corresponding $\left( 1,0\right) $, $d=6$
theory (see \textit{e.g.} Sect. 5 of \cite{AFMT1} for a recent treatment).
\end{itemize}

It should also be noticed that $\mathcal{O}_{nBPS,large}$ (given by Eq. (\ref
{O-nBPS-Z<>0-large})) and $\mathcal{O}_{critical,BPS}$ (given by Eq. (\ref
{magic-orbit-critical-Z<>0})) share the same compact symmetry, or
equivalently that $\mathcal{M}_{nBPS,large}$ (given by Eq. (\ref
{nBPS-Z<>0-large-moduli-space})) and $\mathcal{M}_{critical,BPS}$ (given by
Eq. (\ref{magic-mod-space-critical-Z<>0})) share the same stabilizer group
(apart from an $\mathcal{A}_{q}$ commuting factor), but they do not
coincide. This is due to the fact that $\widetilde{H}_{5}$ and $G_{6}\times
\mathcal{A}_{q}$ share the same $mcs$, namely
\begin{equation}
\widetilde{h}_{5}\equiv mcs\left( \widetilde{H}_{5}\right) =mcs\left(
G_{6}\times \mathcal{A}_{q}\right) =SO\left( q+1\right) \times \mathcal{A}%
_{q}.  \label{rr-1}
\end{equation}

In the case $\mathbb{A}=\mathbb{R}$ ($q=1$), the following further results
holds (see also Tables 3 and 4):
\begin{equation}
\mathcal{M}_{nBPS,large,J_{3}^{\mathbb{R}}}\rtimes \left\{
\begin{array}{l}
\mathbb{R}^{2} \\
\\
\mathbb{R}^{\left( 2,2\right) }
\end{array}
\right. =M_{\mathcal{N}=\left( 1,0\right) ,d=6,J_{3}^{\mathbb{R}}}\rtimes
\left\{
\begin{array}{l}
\mathbb{R}^{2} \\
\\
\mathbb{R}^{\left( 2,2\right) }
\end{array}
\right. =\left\{
\begin{array}{l}
\mathcal{M}_{critical,BPS,J_{3}^{\mathbb{R}}}; \\
\\
\mathcal{M}_{lightlike,nBPS,J_{3}^{\mathbb{C}}}.
\end{array}
\right.
\end{equation}
Notice that $J_{3}^{\mathbb{R}}$ is the unique case, among $J_{3}^{\mathbb{A}%
}$ in $d=5$, in which $\mathcal{M}_{nBPS,large}$ and $\mathcal{M}%
_{critical,BPS}$ not only share the same stabilizer, but they actually do
coincide (up to $\rtimes \mathbb{R}^{2}$). Moreover, $\mathcal{M}%
_{nBPS,large,J_{3}^{\mathbb{R}}}$ also coincides with $\mathcal{M}%
_{lightlike,nBPS,J_{3}^{\mathbb{C}}}$ (up to $\rtimes \mathbb{R}^{\left(
2,2\right) }$), because the respective charge orbits \linebreak $\mathcal{O}%
_{nBPS,large,J_{3}^{\mathbb{R}}}$ and $\mathcal{O}_{lightlike,nBPS,J_{3}^{%
\mathbb{C}}}$ share the same semi-simple, namely non-translational, part of
the stabilizer (apart from a commuting $\mathcal{A}_{2}=SO\left( 2\right) $
factor), \textit{i.e.} $SO\left( 2,1\right) $.

\begin{table}[p]
\begin{center}
\begin{tabular}{|c||c|c|c|c|}
\hline
$
\begin{array}{c}
~ \\
J_{3}^{\mathbb{A}}~ \\
\left( +~rel.~data\right) \\
~
\end{array}
$ & $
\begin{array}{c}
~ \\
\mathcal{O}_{lightlike,BPS}, \\
~r=2
\end{array}
$ & $
\begin{array}{c}
~ \\
\mathcal{M}_{lightlike,BPS} \\
~
\end{array}
$ & $
\begin{array}{c}
~ \\
\mathcal{O}_{lightlike,nBPS}, \\
~r=2
\end{array}
$ & $
\begin{array}{c}
~ \\
\mathcal{M}_{lightlike,nBPS} \\
~
\end{array}
$ \\ \hline\hline
\multicolumn{1}{|l||}{$
\begin{array}{l}
\\
\mathbb{A}=\mathbb{O},q=8 \\
\\
\mathbf{Spin}\left( 9\right) =\mathbf{16} \\
~\sharp _{light,BPS}=10 \\
~\sharp _{light,nBPS}=2
\end{array}
$} & $\frac{E_{6(-26)}}{SO\left( 9\right) \rtimes \mathbb{R}^{16}}$ & $%
\mathbb{R}^{16}$ & $~\frac{E_{6(-26)}}{SO\left( 8,1\right) \rtimes \mathbb{R}%
^{16}}~$ & $~\frac{SO\left( 8,1\right) }{SO\left( 8\right) }\rtimes \mathbb{R%
}^{16}~$ \\ \hline
$
\begin{array}{l}
\\
\mathbb{A}=\mathbb{H},q=4 \\
\\
\mathcal{A}_{4}=SO\left( 3\right) \\
Q_{4}=2 \\
\mathbf{Spin}\left( 5\right) =\mathbf{4} \\
\mathbf{Spin}\left( Q_{4}\right) =\mathbf{2} \\
\sharp _{light,BPS}=6 \\
\sharp _{light,nBPS}=2
\end{array}
$ & $\frac{SU^{\ast }\left( 6\right) }{\left( SO\left( 5\right) \times
SO\left( 3\right) \right) \rtimes \mathbb{R}^{\left( 4,2\right) }}$ & $%
\mathbb{R}^{\left( 4,2\right) }$ & $\frac{SU^{\ast }\left( 6\right) }{\left(
SO\left( 4,1\right) \times SO\left( 3\right) \right) \rtimes \mathbb{R}%
^{\left( 4,2\right) }}$ & $\frac{SO\left( 4,1\right) }{SO\left( 4\right) }%
\rtimes \mathbb{R}^{\left( 4,2\right) }$ \\ \hline
$
\begin{array}{l}
\\
\mathbb{A}=\mathbb{C},q=2 \\
\\
\mathcal{A}_{2}=SO\left( 2\right) \\
Q_{2}=2, \\
\mathbf{Spin}\left( 3\right) =\mathbf{2}, \\
\mathbf{Spin}\left( Q_{2}\right) =\mathbf{2} \\
\sharp _{light,BPS}=4 \\
\sharp _{light,nBPS}=2
\end{array}
$ & $\frac{SL\left( 3,\mathbb{C}\right) }{\left( SO\left( 3\right) \times
SO\left( 2\right) \right) \rtimes \mathbb{R}^{\left( 2,2\right) }}$ & $%
\mathbb{R}^{\left( 2,2\right) }$ & $\frac{SL\left( 3,\mathbb{C}\right) }{%
\left( SO\left( 2,1\right) \times SO\left( 2\right) \right) \rtimes \mathbb{R%
}^{\left( 2,2\right) }}$ & $\frac{SO\left( 2,1\right) }{SO\left( 2\right) }%
\rtimes \mathbb{R}^{\left( 2,2\right) }$ \\ \hline
$
\begin{array}{l}
\\
\mathbb{A}=\mathbb{R},q=1 \\
\\
\mathbf{Spin}\left( 2\right) =\mathbf{2} \\
\sharp _{light,BPS}=3 \\
\sharp _{light,nBPS}=2
\end{array}
$ & $\frac{SL\left( 3,\mathbb{R}\right) }{SO\left( 2\right) \rtimes \mathbb{R%
}^{2}}$ & $\mathbb{R}^{2}$ & $\frac{SL\left( 3,\mathbb{R}\right) }{SO\left(
1,1\right) \rtimes \mathbb{R}^{2}}$ & $SO\left( 1,1\right) \rtimes \mathbb{R}%
^{2}$ \\ \hline
\end{tabular}
\end{center}
\caption{\textbf{``Small'' lightlike charge orbits} $\mathcal{O}%
_{lightlike,BPS}$ \textbf{and} $\mathcal{O}_{lightlike,nBPS}$ \textbf{(with
associated \textit{moduli spaces}) in symmetric \textit{``magic'' }RSG}}
\end{table}

\begin{table}[p]
\begin{center}
\begin{tabular}{|c||c|c|}
\hline
$
\begin{array}{c}
~ \\
J_{3}^{\mathbb{A}} \\
~ \\
\left( +~rel.~data\right) \\
~
\end{array}
$ & $
\begin{array}{c}
~ \\
\mathcal{O}_{critical,BPS}, \\
~r=1
\end{array}
$ & $
\begin{array}{c}
~ \\
\mathcal{M}_{critical,BPS} \\
~
\end{array}
$ \\ \hline\hline
\multicolumn{1}{|l||}{$
\begin{array}{l}
~ \\
\mathbb{A}=\mathbb{O},q=8 \\
~ \\
\mathbf{Spin}\left( 9\right) =\mathbf{16} \\
~\sharp _{crit,BPS}=1
\end{array}
$} & $
\begin{array}{c}
~ \\
\frac{E_{6\left( -26\right) }}{SO\left( 9,1\right) \rtimes \mathbb{R}^{16}}
\\
~
\end{array}
$ & $
\begin{array}{c}
~ \\
\frac{SO\left( 9,1\right) }{SO\left( 9\right) }\rtimes \mathbb{R}^{16} \\
~
\end{array}
$ \\ \hline
$
\begin{array}{l}
\\
\mathbb{A}=\mathbb{H},q=4 \\
\\
\mathcal{A}_{4}=SO\left( 3\right) , \\
Q_{4}=2, \\
\mathbf{Spin}\left( 5\right) =\mathbf{4}, \\
\mathbf{Spin}\left( Q_{4}\right) =\mathbf{2} \\
\sharp _{crit,BPS}=1
\end{array}
$ & $
\begin{array}{c}
~ \\
\frac{SU^{\ast }\left( 6\right) }{\left( SO\left( 5,1\right) \times SO\left(
3\right) \right) \rtimes \mathbb{R}^{\left( 4,2\right) }} \\
~
\end{array}
$ & $
\begin{array}{c}
~ \\
\frac{SO\left( 5,1\right) }{SO\left( 5\right) }\rtimes \mathbb{R}^{\left(
4,2\right) } \\
~
\end{array}
$ \\ \hline
$
\begin{array}{l}
\\
\mathbb{A}=\mathbb{C},q=2 \\
\\
\mathcal{A}_{2}=SO\left( 2\right) , \\
Q_{2}=2, \\
\mathbf{Spin}\left( 3\right) =\mathbf{2}, \\
\mathbf{Spin}\left( Q_{2}\right) =\mathbf{2} \\
\sharp _{crit,BPS}=1
\end{array}
$ & $
\begin{array}{c}
~ \\
\frac{SL\left( 3,\mathbb{C}\right) }{\left( SO\left( 3,1\right) \times
SO\left( 2\right) \right) \rtimes \mathbb{R}^{\left( 2,2\right) }} \\
~
\end{array}
$ & $
\begin{array}{c}
~ \\
\frac{SO\left( 3,1\right) }{SO\left( 3\right) }\rtimes \mathbb{R}^{\left(
2,2\right) } \\
~
\end{array}
$ \\ \hline
$
\begin{array}{l}
~ \\
\mathbb{A}=\mathbb{R},q=1 \\
~ \\
\mathbf{Spin}\left( 2\right) =\mathbf{2} \\
~\sharp _{crit,BPS}=1
\end{array}
$ & $
\begin{array}{c}
~ \\
\frac{SL\left( 3,\mathbb{R}\right) }{SO\left( 2,1\right) \rtimes \mathbb{R}%
^{2}} \\
~
\end{array}
$ & $
\begin{array}{c}
~ \\
\frac{SO\left( 2,1\right) }{SO\left( 2\right) }\rtimes \mathbb{R}^{2} \\
~
\end{array}
$ \\ \hline
\end{tabular}
\end{center}
\caption{\textbf{``Small'' critical charge orbit} $\mathcal{O}%
_{critical,BPS} $ \textbf{(with associated moduli space}\textit{\ }$\mathcal{%
M}_{critical,BPS}$\textbf{) in symmetric \textit{``magic'' }RSG}}
\end{table}
\medskip

The \textit{Jordan symmetric} infinite sequence \cite
{GST1,GST2,GST3,GST4,dWVVP,dWVP-1,dWVP-2} given by Eq. (\ref
{N=2-d=5-Jordan-Symm}) needs some extra care (also at the level of ``large''
charge orbits), because of the factorization of $G_{5}$. The ``large''%
\textit{\ and }``small'' charge orbits for such a sequence will be treated
in Sect. \ref{N=2,d=5-Jordan-Symm-Seq}. This treatment refines and complete
the ones given \textit{e.g.} in \cite{FG2,Ferrara-Marrani-2,AFMT1}).

\subsection{\label{Constraints-Small-Orbits}$G_{5}$-invariant Defining
Constraints}

As mentioned above, ``small'' charge orbits in \textit{all} symmetric RSG
are all characterized by the constraint (recall Eq. (\ref{I-I})):
\begin{equation}
\widehat{\mathcal{I}}_{3}=\mathcal{I}_{3}=0,  \label{Small-constraint}
\end{equation}
where $\widehat{\mathcal{I}}_{3}=\mathcal{I}_{3}$ is the unique cubic scalar
invariant of the relevant electric representation $\mathbf{R}_{q}$ of the $%
d=5$ $U$-duality group $G_{5}$ (in which the electric charges $q_{i}$'s
sit). By recalling definitions (\ref{I_3-hat}) and (\ref{I_3-el}), the
\textit{``smallness''} condition (\ref{Small-constraint}) can be recast as
follows:
\begin{eqnarray}
\widehat{\mathcal{I}}_{3} &=&0\Leftrightarrow Z^{3}-\left( \frac{3}{2}%
\right) ^{2}ZZ_{x}Z^{x}-\left( \frac{3}{2}\right) ^{\frac{3}{2}%
}T_{xyz}Z^{x}Z^{y}Z^{z}=0;  \label{Small-constraint-dressed} \\
\mathcal{I}_{3} &=&0\Leftrightarrow d^{ijk}q_{i}q_{j}q_{k}=0,
\label{Small-constraint-bare}
\end{eqnarray}
in the \textit{``dressed''} and \textit{``bare''} charges basis,
respectively.

It is here worth noticing that Eq. (\ref{Small-constraint-dressed}) can be
recast as a cubic algebraic equation:
\begin{equation}
Z^{3}+\mathbf{p}Z-\mathbf{q}=0;~\left\{
\begin{array}{c}
\mathbf{p}\equiv -\left( \frac{3}{2}\right) ^{2}Z_{x}Z^{x}<0; \\
\\
\mathbf{q}\equiv \left( \frac{3}{2}\right) ^{\frac{3}{2}%
}T_{xyz}Z^{x}Z^{y}Z^{z},
\end{array}
\right.  \label{cubic-1}
\end{equation}
with polynomial discriminant
\begin{equation}
D\equiv \frac{\mathbf{p}^{3}}{9}+\frac{\mathbf{q}^{2}}{4}=\frac{3^{3}}{2^{6}}%
\left[ 2\left( T_{xyz}Z^{x}Z^{y}Z^{z}\right) ^{2}-\left( Z_{x}Z^{x}\right)
^{3}\right] .
\end{equation}
Thus, for $D>0$ one gets one real and two complex conjugate (unacceptable)
roots, whereas for $D<0$ all roots are real and unequal. In the particular
case
\begin{equation}
D=0\Leftrightarrow 2\left( T_{xyz}Z^{x}Z^{y}Z^{z}\right) ^{2}=\left(
Z_{x}Z^{x}\right) ^{3},  \label{discr}
\end{equation}
all roots are real, and at least two equal.

Let us proceed further, by differentiating the functional $\widehat{\mathcal{%
I}}_{3}$ with respect to the \textit{``dressed''} charges
\begin{equation}
\mathcal{Z}\equiv \left\{ Z,Z_{x}\right\} ,  \label{Z-call-def}
\end{equation}
as well as function $\mathcal{I}_{3}$ with respect to the \textit{``bare''}
charges $\left\{ q_{i}\right\} $. One respectively obtains:
\begin{eqnarray}
\frac{\partial \widehat{\mathcal{I}}_{3}}{\partial \mathcal{Z}} &=&\left\{
\begin{array}{l}
\frac{\partial \widehat{\mathcal{I}}_{3}}{\partial Z}=\frac{1}{2}Z^{2}-\frac{%
3}{8}Z_{x}Z^{x}; \\
\\
\frac{\partial \widehat{\mathcal{I}}_{3}}{\partial Z_{x}}=-\frac{3}{4}ZZ^{x}-%
\frac{1}{2}\left( \frac{3}{2}\right) ^{3/2}T_{~yz}^{x}Z^{y}Z^{z};
\end{array}
\right. \\
&&  \notag \\
\frac{\partial \mathcal{I}_{3}}{\partial q_{i}} &=&\frac{1}{2}%
d^{ijk}q_{j}q_{k},  \label{crit-bare}
\end{eqnarray}
where it should be recalled once again that here we are considering
symmetric real special manifolds $\frac{G_{5}}{H_{5}}$, where Eqs. (\ref
{d-symm}) and (\ref{T-symm}) all hold true.

A further differentiation with respect to $\mathcal{Z}$ or $\left\{
q_{i}\right\} $ respectively yields
\begin{eqnarray}
\frac{\partial ^{2}\widehat{\mathcal{I}}_{3}}{\left( \partial \mathcal{Z}%
\right) ^{2}} &=&\left\{
\begin{array}{l}
\frac{\partial ^{2}\widehat{\mathcal{I}}_{3}}{\left( \partial Z\right) ^{2}}%
=Z; \\
\\
\frac{\partial ^{2}\widehat{\mathcal{I}}_{3}}{\partial Z\partial Z_{x}}=-%
\frac{3}{4}Z^{x}; \\
\\
\frac{\partial ^{2}\widehat{\mathcal{I}}_{3}}{\partial Z_{x}\partial Z_{y}}=-%
\frac{3}{4}Zg^{xy}-\left( \frac{3}{2}\right) ^{3/2}T_{~~z}^{xy}Z^{z}=\frac{%
\partial ^{2}\widehat{\mathcal{I}}_{3}}{\partial Z_{(x}\partial Z_{y)}};
\end{array}
\right. \\
&&  \notag \\
\frac{\partial ^{2}\mathcal{I}_{3}}{\partial q_{i}\partial q_{j}}
&=&d^{ijk}q_{k}=\frac{\partial ^{2}\mathcal{I}_{3}}{\partial q_{(i}\partial
q_{j)}}.  \label{d-crit-bare}
\end{eqnarray}

By further differentiating, one then obtains:
\begin{eqnarray}
\frac{\partial ^{3}\widehat{\mathcal{I}}_{3}}{\left( \partial \mathcal{Z}%
\right) ^{3}} &=&\left\{
\begin{array}{l}
\frac{\partial ^{3}\widehat{\mathcal{I}}_{3}}{\left( \partial Z\right) ^{3}}%
=1; \\
\\
\frac{\partial ^{3}\widehat{\mathcal{I}}_{3}}{\left( \partial Z\right)
^{2}\partial Z_{x}}=0; \\
\\
\frac{\partial ^{3}\widehat{\mathcal{I}}_{3}}{\partial Z\partial
Z_{x}\partial Z_{y}}=-\frac{3}{4}g^{xy}=\frac{\partial ^{3}\widehat{\mathcal{%
I}}_{3}}{\partial Z\partial Z_{(x}\partial Z_{y)}}; \\
\\
\frac{\partial ^{3}\widehat{\mathcal{I}}_{3}}{\partial Z_{x}\partial
Z_{y}\partial Z_{z}}=-\left( \frac{3}{2}\right) ^{3/2}T^{xyz}=\frac{\partial
^{3}\widehat{\mathcal{I}}_{3}}{\partial Z_{(x}\partial Z_{y}\partial Z_{z)}};
\end{array}
\right.  \label{t-crit-1} \\
&&  \notag \\
\frac{\partial ^{3}\mathcal{I}_{3}}{\partial q_{i}\partial q_{j}\partial
q_{k}} &=&d^{ijk}=\frac{\partial ^{3}\mathcal{I}_{3}}{\partial
q_{(i}\partial q_{j}\partial q_{k)}}.  \label{t-crit-bare}
\end{eqnarray}
Starting from the fourth order of differentiation, all derivatives vanish.
This is no surprise, because $\widehat{\mathcal{I}}_{3}$ is a functional
polynomial homogeneous of degree three in \textit{``dressed''} charges $%
\mathcal{Z}$, as well as (equivalently) $\mathcal{I}_{3}$ is a polynomial
homogeneous of degree three in \textit{``bare''} charges $q_{i}$'s.\smallskip

At this point, it is possible to classify the various ``small'' charge
orbits through $G_{5}$-invariant conditions involving $\widehat{\mathcal{I}}%
_{3}$ and its non-vanishing functional derivatives with respect to $\mathcal{%
Z}$, or equivalently through $G_{5}$-invariant conditions involving $%
\mathcal{I}_{3}$ and its non-vanishing derivatives with respect to $q_{i}$'s.

\subsubsection{``Small'' \textit{Lightlike} Orbits}

The ``small'' \textit{lightlike} charge orbits are defined by the
constraints (recall Eqs. (\ref{Small-constraint-dressed}) and (\ref
{Small-constraint-bare})):
\begin{equation}
\left\{
\begin{array}{l}
\widehat{\mathcal{I}}_{3}=0\Leftrightarrow Z^{3}-\left( \frac{3}{2}\right)
^{2}ZZ_{x}Z^{x}-\left( \frac{3}{2}\right) ^{\frac{3}{2}%
}T_{xyz}Z^{x}Z^{y}Z^{z}=0; \\
\\
\frac{\partial \widehat{\mathcal{I}}_{3}}{\partial \mathcal{Z}}\neq
0\Leftrightarrow \left\{
\begin{array}{l}
Z^{2}-\frac{3}{4}Z_{x}Z^{x}\neq 0; \\
\text{\textit{and}}~/~\text{\textit{or}} \\
ZZ^{x}+\sqrt{\frac{3}{2}}T_{~yz}^{x}Z^{y}Z^{z}\neq 0~\text{(\textit{at~least}%
~for~some~}x\text{)},
\end{array}
\right.
\end{array}
\right.  \label{lightlike-dressed}
\end{equation}
or equivalently:
\begin{equation}
\left\{
\begin{array}{l}
\mathcal{I}_{3}=0\Leftrightarrow d^{ijk}q_{i}q_{j}q_{k}=0; \\
\\
\frac{\partial \mathcal{I}_{3}}{\partial q_{i}}\neq 0\Leftrightarrow
d^{ijk}q_{j}q_{k}\neq 0~\text{(\textit{at~least}~for~some~}i\text{)};
\end{array}
\right.  \label{lightlike-bare}
\end{equation}
The sets of constraints (\ref{lightlike-dressed}) and (\ref{lightlike-bare})
are both $G_{5}$-invariant, but their \textit{manifest} invariance is
different. Indeed, the \textit{``dressed''} charge basis $\mathcal{Z}$ is
covariant with respect to $H_{5}$, and thus the set of constraints (\ref
{lightlike-dressed}) exhibits a manifest $H_{5}$-invariance. Instead, the
\textit{``bare''} charge basis $\left\{ q_{i}\right\} $ is $G_{5}$%
-covariant, and thus the set of constraints (\ref{lightlike-dressed}) is
manifestly $G_{5}$-invariant.

In the \textit{``dressed''} charge basis, it is immediate to realize that
two classes of ``small''\textit{\ lightlike} charge orbits exist, namely:

\begin{itemize}
\item  ``small''\textit{\ lightlike} charge orbit for which the constraints (%
\ref{lightlike-dressed}) are solved with $Z=0$:
\begin{equation}
\left\{
\begin{array}{l}
\left. \widehat{\mathcal{I}}_{3}\right| _{Z=0}=0\Leftrightarrow
T_{xyz}Z^{x}Z^{y}Z^{z}=0; \\
\\
\left. \frac{\partial \widehat{\mathcal{I}}_{3}}{\partial \mathcal{Z}}%
\right| _{Z=0}\neq 0\Leftrightarrow \left\{
\begin{array}{l}
Z_{x}Z^{x}\neq 0; \\
\text{\textit{and}}~/~\text{\textit{or}} \\
T_{~yz}^{x}Z^{y}Z^{z}\neq 0~\text{(\textit{at~least}~for~some~}x\text{)}.
\end{array}
\right.
\end{array}
\right.  \label{lightlike-Z=0}
\end{equation}
Notice that the constraint $Z_{x}Z^{x}\neq 0$ is automatically satisfied,
because: 1) $g_{xy}$ is assumed to be strictly positive definite, and 2) $%
Z_{x}\neq 0$ \textit{at least} for some $x$ (otherwise, since $Z=0$, one
would obtain the trivial limit in which all charges vanish).

\item  ``small''\textit{\ lightlike} charge orbit for which the constraints (%
\ref{lightlike-dressed}) are solved with $Z\neq 0$ (also recall Eqs. (\ref
{cubic-1})-(\ref{discr})):
\begin{equation}
\left\{
\begin{array}{l}
\left. \widehat{\mathcal{I}}_{3}\right| _{Z\neq 0}=0\Leftrightarrow
Z^{3}-\left( \frac{3}{2}\right) ^{2}ZZ_{x}Z^{x}-\left( \frac{3}{2}\right) ^{%
\frac{3}{2}}T_{xyz}Z^{x}Z^{y}Z^{z}=0; \\
\\
\left. \frac{\partial \widehat{\mathcal{I}}_{3}}{\partial \mathcal{Z}}%
\right| _{Z\neq 0}\neq 0\Leftrightarrow \left\{
\begin{array}{l}
Z^{2}-\frac{3}{4}Z_{x}Z^{x}\neq 0; \\
\text{\textit{and}}~/~\text{\textit{or}} \\
ZZ^{x}+\sqrt{\frac{3}{2}}T_{~yz}^{x}Z^{y}Z^{z}\neq 0~\text{(\textit{at~least}%
~for~some~}x\text{)}.
\end{array}
\right.
\end{array}
\right.  \label{lightlike-Z<>0}
\end{equation}
\end{itemize}

\subsubsection{``Small''\textit{\ Critical} Orbit}

The ``small'' \textit{critical} charge orbit is defined by the constraints
(recall Eqs. (\ref{Small-constraint-dressed}) and (\ref
{Small-constraint-bare})):
\begin{eqnarray}
&&\left\{
\begin{array}{l}
\widehat{\mathcal{I}}_{3}=0\Leftrightarrow Z^{3}-\left( \frac{3}{2}\right)
^{2}ZZ_{x}Z^{x}-\left( \frac{3}{2}\right) ^{\frac{3}{2}%
}T_{xyz}Z^{x}Z^{y}Z^{z}=0; \\
\\
\frac{\partial \widehat{\mathcal{I}}_{3}}{\partial \mathcal{Z}}%
=0\Leftrightarrow \left\{
\begin{array}{l}
Z^{2}-\frac{3}{4}Z_{x}Z^{x}=0; \\
\\
ZZ^{x}+\sqrt{\frac{3}{2}}T_{~yz}^{x}Z^{y}Z^{z}=0,
\end{array}
\right.
\end{array}
\right.  \notag \\
&&  \label{critical-dressed}
\end{eqnarray}
or equivalently:
\begin{equation}
\left\{
\begin{array}{l}
\mathcal{I}_{3}=0\Leftrightarrow d^{ijk}q_{i}q_{j}q_{k}=0; \\
\\
\frac{\partial \mathcal{I}_{3}}{\partial q_{i}}=0\Leftrightarrow
d^{ijk}q_{j}q_{k}=0.
\end{array}
\right.  \label{critical-bare}
\end{equation}
As noticed above for the sets of constraints (\ref{lightlike-dressed}) and (%
\ref{lightlike-bare}), the sets of constraints (\ref{critical-dressed}) and (%
\ref{critical-bare}) are both $G_{5}$-invariant: while (\ref
{critical-dressed}) is manifestly invariant only under $H_{5}=mcs\left(
G_{5}\right) $, (\ref{critical-bare}) is actually manifestly $G_{5}$%
-invariant.

Once again, in the \textit{``dressed''} charges basis it is immediate to
realize that only one class of ``small''\textit{\ critical} charge orbits
exists, namely:

\begin{itemize}
\item  ``small''\textit{\ critical} charge orbit for which the constraints (%
\ref{critical-dressed}) are solved with $Z\neq 0$:
\begin{eqnarray}
&&\left\{
\begin{array}{l}
\left. \widehat{\mathcal{I}}_{3}\right| _{Z\neq 0}=0\Leftrightarrow
Z^{3}-\left( \frac{3}{2}\right) ^{2}ZZ_{x}Z^{x}-\left( \frac{3}{2}\right) ^{%
\frac{3}{2}}T_{xyz}Z^{x}Z^{y}Z^{z}=0; \\
\\
\left. \frac{\partial \widehat{\mathcal{I}}_{3}}{\partial \mathcal{Z}}%
\right| _{Z\neq 0}=0\Leftrightarrow \left\{
\begin{array}{l}
Z^{2}-\frac{3}{4}Z_{x}Z^{x}=0; \\
\\
ZZ^{x}+\sqrt{\frac{3}{2}}T_{~yz}^{x}Z^{y}Z^{z}=0.
\end{array}
\right.
\end{array}
\right.  \notag \\
&&  \label{critical-Z<>0}
\end{eqnarray}
\end{itemize}

Notice that, for the same reason the constraint $\left. \frac{\partial
\widehat{\mathcal{I}}_{3}}{\partial Z}\right| _{Z=0}\neq 0$ is automatically
satisfied for the ``small''\textit{\ lightlike} charge orbit whose a
representative in the \textit{``dressed''} charges basis is given by Eq. (%
\ref{lightlike-Z=0}), a ``small''\textit{\ critical} charge orbit with
representative having $Z=0$ cannot exist. Indeed, such an orbit should have $%
Z=0$ and $Z_{x}Z^{x}=0$. Due to the assumed strictly positive definiteness
of $g_{xy}$, this would be possible only in the trivial limit of the theory
in which \textit{all} charges do vanish. This can be formally stated as
follows:
\begin{equation}
\left. \frac{\partial \widehat{\mathcal{I}}_{3}}{\partial \mathcal{Z}}%
\right| _{Z=0}=0\Leftrightarrow \mathcal{Z}=0.
\end{equation}

\section{\label{N=2-magic-H-d=5---N=6-d=5}$J_{3}^{\mathbb{H}}:$ $\mathcal{N}%
=2$ \textit{versus} $\mathcal{N}=6$}

The rank-$3$ Euclidean Jordan algebra $J_{3}^{\mathbb{H}}$ ($q=4$) is
related to two different theories, namely an $\mathcal{N}=2$ theory coupled
to $14$ Abelian vector multiplets and the $\mathcal{N}=6$ \textit{``pure''}
theory. These two theories share the same bosonic sector \cite
{FG1,FG2,Samtleben-twin}, but their fermionic sectors, exploiting the
supersymmetric completion of the bosonic one, are different.

Thus, it also follows that the supersymmetry-preserving features of the
``large'' and ``small'' charge orbits of the relevant irrepr. $\mathbf{15}$
of $G_{5}=SU^{\ast }\left( 6\right) $ are different. The $\mathcal{N}$%
-dependent supersymetry properties of the various orbits are given in Table
5 (notice they are consistent with the results of \cite{BMP-1}). In the
``large'' (attractor) cases, these match the results of \cite{AFMT1}.

\begin{table}[t]
\begin{center}
\begin{tabular}{|c||c|c|}
\hline
$
\begin{array}{c}
\\
~J_{3}^{\mathbb{H}} \\
~
\end{array}
$ & $\mathcal{N}=2$ & $\mathcal{N}=6$ \\ \hline\hline
$
\begin{array}{c}
\\
\frac{SU^{\ast }\left( 6\right) }{USp\left( 6\right) }~ \\
~_{``large",~\mathcal{I}_{3}\neq 0}~
\end{array}
$ & $\frac{1}{2}-BPS$ & \multicolumn{1}{|l|}{$
\begin{array}{c}
nBPS, \\
Z_{AB,H}=0, \\
~X_{H}\neq 0
\end{array}
$} \\ \hline
$
\begin{array}{c}
\\
\frac{SU^{\ast }\left( 6\right) }{USp\left( 4,2\right) }~ \\
~~_{``large",~\mathcal{I}_{3}\neq 0}
\end{array}
$ & $nBPS,Z_{H}\neq 0$ & \multicolumn{1}{|l|}{$
\begin{array}{c}
\frac{1}{6}-BPS, \\
Z_{AB,H}\neq 0,~ \\
~X_{H}\neq 0
\end{array}
$} \\ \hline
$
\begin{array}{c}
\\
\frac{SU^{\ast }\left( 6\right) }{\left( SO\left( 5\right) \times SO\left(
3\right) \right) \rtimes \mathbb{R}^{\left( 4,2\right) }}~ \\
~_{``small",~\mathcal{I}_{3}=0}
\end{array}
$ & $\frac{1}{2}-BPS$ & $\frac{1}{6}-BPS~$ \\ \hline
$
\begin{array}{c}
\\
\frac{SU^{\ast }\left( 6\right) }{\left( SO\left( 4,1\right) \times SO\left(
3\right) \right) \rtimes \mathbb{R}^{\left( 4,2\right) }}~ \\
~_{``small",~\mathcal{I}_{3}=0}
\end{array}
$ & $nBPS$ & $\frac{1}{3}-BPS$ \\ \hline
$
\begin{array}{c}
\\
\frac{SU^{\ast }\left( 6\right) }{\left( SO\left( 5,1\right) \times SO\left(
3\right) \right) \rtimes \mathbb{R}^{\left( 4,2\right) }}~ \\
~_{``small",~\partial \mathcal{I}_{3}=0}
\end{array}
$ & $\frac{1}{2}-BPS$ & $\frac{1}{2}-BPS$ \\ \hline
\end{tabular}
\end{center}
\caption{$\mathcal{N}$\textbf{-dependent supersymmetry-preserving features
of \textit{``large''} and \textit{``small''} charge orbits of the irrepr. }$%
\mathbf{15}$ \textbf{of the }$d=5$\textbf{\ }$U$\textbf{-duality group }$%
SU^{\ast }\left( 6\right) $\textbf{, related to }$J_{3}^{\mathbb{H}}$\textbf{%
. This corresponds to two \textit{``twin''} theories, sharing the same
bosonic sector: an }$\mathcal{N}=2$\textbf{\ Maxwell-Einstein theory and the
}$\mathcal{N}=6$\textbf{\ ``pure'' theory. The subscript ``}$H$\textbf{''
stands for ``(evaluated at the) horizon''}}
\end{table}

\section{\label{N=2,d=5-Jordan-Symm-Seq}$\mathcal{N}=2$, $d=5$ Jordan
Symmetric Sequence}

The Jordan symmetric sequence of $\mathcal{N}=2$, $d=5$ supergravity coupled
to $n_{V}=n+1$ vector multiplets reads ($dim_{\mathbb{R}}=n+1,$ rank$=2$, $%
n\in \mathbb{N}\cup \left\{ 0\right\} $)
\begin{equation}
M_{\mathcal{N}=2,d=5,Jordan,symm}=SO\left( 1,1\right) \times \frac{SO\left(
1,n\right) }{SO\left( n\right) }.
\label{N=2-d=5-Jordan-symm-scalar-manifold}
\end{equation}
This sequence is associated to the rank-$3$ Euclidean \textit{reducible}
Jordan algebra $\mathbb{R}\oplus \mathbf{\Gamma }_{1,n}$. In the following
treatment, we will determine the ``large'' and ``small'' orbits of the
irrepr. $\left( \mathbf{1},\mathbf{1+n}\right) $ of the $U$-duality group $%
SO\left( 1,1\right) \times SO\left( 1,n\right) $.

For brevity's sake, we will do this only through an analysis in the ``bare''
charges' basis.

Without any loss in generality, one can choose to treat only $d=5$ extremal (%
\textit{electric}) BHs. Indeed, due to the symmetricity of the reducible
coset (\ref{N=2-d=5-Jordan-symm-scalar-manifold}), the treatment of $d=5$
extremal (\textit{magnetic}) black strings is essentially analogous.

Two disconnected geometric structures emerge in the treatment, namely:

\begin{itemize}
\item  Timelike two-sheet hyperboloid $T_{n}$, with the two disconnected
sheets $T_{n}^{\pm }$ respectively related to $q_{0}\gtrless 0$:
\begin{equation}
T_{n}\equiv \left. \frac{SO\left( 1,n\right) }{SO\left( n\right) }\right|
_{q_{I}^{2}>0}=\underset{q_{0}>0}{T_{n}^{+}}\cup \underset{q_{0}<0}{T_{n}^{-}%
};~T_{n}^{+}\cap T_{n}^{-}=\emptyset .  \label{hyperb-def}
\end{equation}

\item  Forward/backward light-cone $\Lambda _{n}$ of $\left( n+1\right) $%
-dimensional Minkowski space with metric $\eta _{IJ}$ defined by (\ref
{eta-IJ}), with two (forward $\Lambda _{n}^{+}$ and backward $\Lambda
_{n}^{-}$) cone branches, respectively related to $q_{0}\gtrless 0$:
\begin{equation}
\Lambda _{n}\equiv \frac{SO\left( 1,n\right) }{SO\left( n-1\right) \rtimes
\mathbb{R}^{n-1}}=\Lambda _{n}^{+}\cup \Lambda _{n}^{-};~\Lambda
_{n}^{+}\cap \Lambda _{n}^{-}=0,  \label{lc-def}
\end{equation}
$0$ here denoting the origin of $\Lambda _{n}$ itself.
\end{itemize}

Due to such structures, as well as to the lower ($\mathcal{N}=2$)
supersymmetry, the case study of ``large'' and ``small'' charge orbits in $%
\mathcal{N}=2$, $d=5$ Jordan symmetric sequence exhibits some subtleties
absent in the $\mathcal{N}=4$, $d=5$ theory analyzed in Sect. \ref{N=4,d=5}.

In the ``bare'' charges' basis, the electric cubic invariant of the $\left(
\mathbf{1},\mathbf{1+n}\right) $ of $SO\left( 1,1\right) \times SO\left(
1,n\right) $ reads as follows ($I=0,i$, where $i=1,...,n$, throughout; ``$0$%
'' pertains to the $d=5$ graviphoton field, which through the dimensional
reduction $d=5\rightarrow d=4$ becomes the Maxwell vector field of the
axio-dilatonic vector multiplet):
\begin{equation}
\mathcal{I}_{3,el}\equiv q_{H}q_{I}q_{J}\eta ^{IJ}\equiv
q_{H}q_{I}^{2}=q_{H}\left( q_{0}^{2}-\sum_{i=1}^{n}q_{i}^{2}\right)
\label{I_3-el-Jordan-symm-seq}
\end{equation}
where $q_{H}$ is the electric charge of the dilatonic vector multiplet: it
is an $SO\left( 1,n\right) $-singlet, with $SO\left( 1,1\right) $-weight $+2$%
. On the other hand, the $SO\left( 1,n\right) $-vector $q_{I}$ has $SO\left(
1,1\right) $-weight $-1$, such that $\mathcal{I}_{3,el}$ defined by (\ref
{I_3-el-Jordan-symm-seq}) is $SO\left( 1,1\right) \times SO\left( 1,n\right)
$-invariant. Notice that the action of the $U$-duality group does not mix $%
q_{H}$ and $q_{I}$, and this originates more charge orbits with respect to
the \textit{irreducible} cases. Moreover, $\eta _{IJ}=\eta ^{IJ}$ is the
Lorentzian metric of $SO\left( 1,n\right) $:
\begin{equation}
\eta _{IJ}=\eta ^{IJ}\equiv diag\left( +1,\overset{n}{\overbrace{-1,...,-1}}%
\right) .  \label{eta-IJ}
\end{equation}

In $\mathcal{N}=2$, $d=5$ Jordan symmetric sequence, as well as in $\mathcal{%
N}=4$, $d=5$ theory, the \textit{reducibility} of the associated rank-$3$
Jordan algebra gives rise to many subtleties and differences with respect to
the theories associated to \textit{irreducible} Euclidean rank-$3$ Jordan
algebras. In the $\mathcal{N}=2$ case under consideration, the major
difference consists in a higher number of ``large'' and ``small'' orbits
with respect to the ``magic'' supergravities.

\subsection{\label{N=2-d=5-Jordan-Symm-Seq-Large}``Large'' Orbits}

\begin{itemize}
\item  BPS ($3$-charge) orbit, defined as follows:
\begin{equation}
\left\{
\begin{array}{l}
q_{H}>0; \\
\\
q_{0}^{2}-\sum_{i=1}^{n}q_{i}^{2}>0; \\
\\
q_{0}>0;
\end{array}
\right. ~~or~~\left\{
\begin{array}{l}
q_{H}<0; \\
\\
q_{0}^{2}-\sum_{i=1}^{n}q_{i}^{2}>0; \\
\\
q_{0}<0.
\end{array}
\right.  \label{cconstr-1}
\end{equation}
By recalling definition (\ref{hyperb-def}), the orbit reads ($n\geqslant 0$%
):
\begin{equation}
\mathcal{O}_{BPS,large}=\left[ SO\left( 1,1\right) ^{+}\times T_{n}^{+}%
\right] \cup \left[ SO\left( 1,1\right) ^{-}\times T_{n}^{-}\right] ,
\label{ppp-1}
\end{equation}
with no related \textit{moduli space}. In particular, for $n=0$, namely in
the so-called $\mathcal{N}=2$, $d=5$ $SO\left( 1,1\right) $ model ($d=5$
uplift of the $d=4$ $st^{2}$ model), in which only the dilatonic vector
multiplet is coupled to the gravity multiplet, this orbit is actually $2$%
-charge, and it is given by
\begin{equation}
\mathcal{O}_{BPS,large,SO(1,1)}=\left\{ \left( q_{H},q_{0}\right) =\left(
+,+\right) ,\left( -,-\right) \right\} .
\end{equation}
On the other hand, for $n=1$, \textit{i.e.} in the so-called $\mathcal{N}=2$%
, $d=5$ $\left[ SO\left( 1,1\right) \right] ^{2}$ model ($d=5$ uplift of $%
stu $ model), the cubic invariant (\ref{I_3-el-Jordan-symm-seq}) can be
rewritten as follows:
\begin{equation}
\begin{array}{l}
\mathcal{I}_{3,el}\equiv q_{H}q_{I}q_{J}\eta ^{IJ}\equiv q_{H}\left(
q_{0}^{2}-q_{1}^{2}\right) =q_{H}q_{+}q_{-}; \\
q_{\pm }\equiv q_{0}\pm q_{1},
\end{array}
\end{equation}
and thus the hyperboloid (\ref{hyperb-def}) and light-cone (\ref{lc-def})
structures gets respectively factorized as follows (``$+$'', ``$-$'' and ``$%
0 $'' respectively denote strictly positive, strictly negative and vanishing
values):
\begin{eqnarray}
&&
\begin{array}{l}
T_{1}=\left. SO\left( 1,1\right) \right| _{q_{+}q_{-}>0}=\underset{q_{0}>0}{%
T_{1}^{+}}\cup \underset{q_{0}<0}{T_{1}^{-}};~T_{1}^{+}\cap
T_{1}^{-}=\emptyset , \\
~ \\
\left\{
\begin{array}{l}
T_{1}^{+}=\left\{ \left( q_{+},q_{-}\right) =\left( +,+\right) \right\} ; \\
\\
T_{1}^{-}=\left\{ \left( q_{+},q_{-}\right) =\left( -,-\right) \right\} .
\end{array}
\right.
\end{array}
\label{hyperb-n=1} \\
&&  \notag \\
&&
\begin{array}{l}
\Lambda _{1}=SO\left( 1,1\right) =\Lambda _{1}^{+}\cup \Lambda
_{1}^{-};~\Lambda _{1}^{+}\cap \Lambda _{1}^{-}=0, \\
~ \\
\left\{
\begin{array}{l}
\Lambda _{1}^{+}=\left\{ \left( q_{+},q_{-}\right) =\left( +,0\right)
,\left( 0,+\right) \right\} ; \\
\\
\Lambda _{1}^{-}=\left\{ \left( q_{+},q_{-}\right) =\left( -,0\right)
,\left( 0,-\right) \right\} .
\end{array}
\right.
\end{array}
\label{lc-def-n=1}
\end{eqnarray}
For $n=1$, orbit (\ref{ppp-1}) reads
\begin{equation}
\mathcal{O}_{BPS,3\text{-charge},\left[ SO(1,1)\right] ^{2}}=\left\{ \left(
q_{H},q_{+},q_{-}\right) =\left( +,+,+\right) ,\left( -,-,-\right) \right\} .
\label{ppp-1-n=1}
\end{equation}
This is invariant under triality permutation symmetry of $q_{H}$, $q_{+}$
and $q_{-}$, and it is consistent with the analysis of \cite{CFM1}.

\item  non-BPS ($3$-charge) orbit, with $Z\neq 0$ at the horizon, defined as
follows:
\begin{equation}
\left\{
\begin{array}{l}
q_{H}>0; \\
\\
q_{0}^{2}-\sum_{i=1}^{n}q_{i}^{2}>0; \\
\\
q_{0}<0;
\end{array}
\right. ~~or~~\left\{
\begin{array}{l}
q_{H}<0; \\
\\
q_{0}^{2}-\sum_{i=1}^{n}q_{i}^{2}>0; \\
\\
q_{0}>0.
\end{array}
\right.  \label{cconstr-2}
\end{equation}
By recalling definition (\ref{hyperb-def}), the orbit reads ($n\geqslant 0$%
):
\begin{equation}
\mathcal{O}_{nBPS,large,I}=\left[ SO\left( 1,1\right) ^{+}\times T_{n}^{-}%
\right] \cup \left[ SO\left( 1,1\right) ^{-}\times T_{n}^{+}\right] ,
\label{ppp-2}
\end{equation}
with no related \textit{moduli space}. In particular, for $n=0$, this orbit
is actually $2$-charge, and it is given by
\begin{equation}
\mathcal{O}_{nBPS,large,SO(1,1)}=\left\{ \left( q_{H},q_{0}\right) =\left(
+,-\right) ,\left( -,+\right) \right\} .
\end{equation}
On the other hand, for $n=1$, orbit (\ref{ppp-2}) reads
\begin{equation}
\mathcal{O}_{nBPS,large,I,\left[ SO(1,1)\right] ^{2}}=\left\{ \left(
q_{H},q_{+},q_{-}\right) =\left( +,-,-\right) ,\left( -,+,+\right) \right\} .
\label{ppp-2-n=1}
\end{equation}
\end{itemize}

The supersymmetry properties of $\mathcal{O}_{BPS,large}$ and $\mathcal{O}%
_{nBPS,large,I}$ can be understood by noticing that the flip of the sign of $%
q_{H}$ amounts, in the dressed charges' basis, to the exchange $%
Z\longleftrightarrow \partial _{s}Z$, where $s$ is the real dilaton scalar
field, parametrizing $SO\left( 1,1\right) $ of (\ref
{N=2-d=5-Jordan-symm-scalar-manifold}).

It is worth pointing out both the $\mathcal{N}=2$ orbits $\mathcal{O}%
_{BPS,large}$ and $\mathcal{O}_{nBPS,large,I}$ (respectively given by (\ref
{ppp-1}) and (\ref{ppp-2})) uplift to the same $\mathcal{N}=4$ orbit $%
\mathcal{O}_{\frac{1}{4}-BPS,large,\mathcal{N}=4,d=5}$ given by Eq. (\ref
{PAA-1}). As mentioned, this is due to the fact that in $\mathcal{N}=4$, $%
d=5 $ $q_{H}>0\longleftrightarrow q_{H}<0$ amounts to exchanging the two
gravitinos in the gravity multiplet, \textit{i.e.} the two (opposite)
skew-eigenvalues of the skew-traceless central charge matrix \r{Z}$_{AB}$ ($%
A,B=1,...,4$).

\begin{itemize}
\item  Another non-BPS ($3$-charge) orbit, with $Z\neq 0$ at the horizon, is
defined as follows \cite{FG2}:
\begin{equation}
\left\{
\begin{array}{l}
q_{H}\gtrless 0; \\
\\
q_{0}^{2}-\sum_{i=1}^{n}q_{i}^{2}<0.
\end{array}
\right.  \label{cconstr-3}
\end{equation}
%Notice that both signs of $q_{H}$ are allowed, due to the fact that the
%non-BPS \r{Z}$_{AB}=0$ Attractor Eqs. are quadratic in $q_{H}$ (see \textit{%
%e.g.} \cite{AFMT1}).
Thus, the resulting orbit reads (existing only for $n\geqslant 1$)
\begin{equation}
\mathcal{O}_{nBPS,large,II}=SO\left( 1,1\right) \times \frac{SO\left(
1,n\right) }{SO\left( 1,n-1\right) },  \label{PAA-2}
\end{equation}
with related \textit{moduli space} (recall (\ref{N=2-d=5-Jordan-Symm}) and (%
\ref{non-Jordan-symm})):
\begin{eqnarray}
\mathcal{M}_{nBPS,large,II} &=&\frac{SO\left( 1,n-1\right) }{SO\left(
n-1\right) }  \notag \\
&=&\frac{M_{J,5,n-1}}{SO(1,1)}=M_{nJ,5,n-1}=\left. M_{\left( 1,0\right)
,d=6}\right| _{n-1},  \label{nBPS-large-III-mod-space}
\end{eqnarray}
where $M_{nJ,5,n-1}$ denotes the $\mathcal{N}=2$, $d=5$ \textit{non-Jordan}
symmetric sequence with $n-1$ vector multiplets \cite{dWVP-3}, and $\left.
M_{\left( 1,0\right) ,d=6}\right| _{n-1}$ is the scalar manifold of $\left(
1,0\right) $, $d=6$ supergravity with $n_{T}=n-1$ tensor multiplets. Thus,
by recalling (\ref{N=2-d=5-Jordan-symm-scalar-manifold}), the number $\sharp
$ of ``non-flat'' scalar degrees of freedom along $\mathcal{O}%
_{nBPS,large,II}$ is independent on $n>1$:
\begin{equation}
\sharp _{nBPS,large,II}\equiv dim_{\mathbb{R}}M_{\mathcal{N}%
=2,d=5,Jordan,symm}-dim_{\mathbb{R}}\mathcal{M}_{nBPS,large,II}=2.
\end{equation}
For $n=1$, orbit (\ref{PAA-2}) reads
\begin{equation}
\mathcal{O}_{nBPS,large,II,\left[ SO(1,1)\right] ^{2}}=\left\{ \left(
q_{H},q_{+},q_{-}\right) =\left( +,+,-\right) ,\left( +,-,+\right) ,\left(
-,+,-\right) ,\left( -,-,+\right) \right\} ,  \label{PAA-2-n=1}
\end{equation}
with no corresponding \textit{moduli space}. (\ref{PAA-2-n=1}) is equivalent
to (\ref{ppp-2-n=1}) through triality permutation symmetry of $q_{H}$, $%
q_{+} $ and $q_{-}$. Thus, consistent with the analysis of \cite{CFM1}, the
non-BPS ``large'' orbit of $\left[ SO\left( 1,1\right) \right] ^{2}$ model
is given, up to permutations of the triplet $\left( q_{H},q_{+},q_{-}\right)
$, by
\begin{equation}
\mathcal{O}_{nBPS,3\text{-charge},\left[ SO(1,1)\right] ^{2}}=\left\{ \left(
q_{H},q_{+},q_{-}\right) =\left( +,+,-\right) ,\left( +,-,-\right) \right\} .
\end{equation}
\smallskip
\end{itemize}

\subsection{\label{N=2-d=5-Jordan-Symm-Seq-Small}``Small'' Orbits}

Let us now consider the ``small'' orbits, and compute the criticality and
double-criticality conditions on $\mathcal{I}_{3,el}$ defined by (\ref
{I_3-el-Jordan-symm-seq}):
\begin{equation}
\frac{\partial \mathcal{I}_{3,el}}{\partial Q}=\left\{
\begin{array}{l}
\frac{\partial \mathcal{I}_{3,el}}{\partial q_{H}}=q_{I}^{2}; \\
\\
\frac{\partial \mathcal{I}_{3,el}}{\partial q_{I}}=2q_{H}q_{J}\eta ^{IJ};
\end{array}
\right.  \label{dI_3-el}
\end{equation}
\begin{equation}
\frac{\partial ^{2}\mathcal{I}_{3,el}}{\partial Q^{2}}=\left\{
\begin{array}{l}
\frac{\partial \mathcal{I}_{3,el}}{\left( \partial q_{H}\right) ^{2}}=0; \\
\\
\frac{\partial \mathcal{I}_{3,el}}{\partial q_{H}\partial q_{I}}=\frac{%
\partial \mathcal{I}_{3,el}}{\partial q_{I}\partial q_{H}}=2q_{J}\eta ^{IJ};
\\
\\
\frac{\partial \mathcal{I}_{3,el}}{\partial q_{I}\partial q_{J}}=2q_{H},
\end{array}
\right.  \label{ddI_3-el}
\end{equation}
where
\begin{equation}
Q\equiv \left( q_{H},q_{I}\right)
\end{equation}
is shorthand for the vector of electric charges. As expected from the fact
that $\mathcal{I}_{3,el}$ is homogeneous of degree three, (\ref{ddI_3-el})
implies that the unique doubly-critical orbit is the trivial one with all
charges vanishing, because
\begin{equation}
\frac{\partial ^{2}\mathcal{I}_{3,el}}{\partial Q^{2}}=0\Leftrightarrow Q=0.
\label{PAA-3}
\end{equation}

The ``small'' orbits of the $\left( \mathbf{1},\mathbf{1+n}\right) $ of the $%
U$-duality group $SO\left( 1,1\right) \times SO\left( 1,n\right) $ list as
follows:

\begin{enumerate}
\item  BPS lightlike ($\mathcal{I}_{3,el}=0$, $\frac{\partial \mathcal{I}%
_{3,el}}{\partial Q}\neq 0$: $2$-charge) orbit with vanishing $q_{H}$ and
timelike $q_{I}$:
\begin{equation}
\left\{
\begin{array}{l}
q_{H}=0; \\
\\
q_{0}^{2}-\sum_{i=1}^{n}q_{i}^{2}>0.
\end{array}
\right.  \label{cconstr-4}
\end{equation}
By recalling definition (\ref{hyperb-def}), the orbit reads ($n\geqslant 0$%
):
\begin{equation}
\mathcal{O}_{BPS,small,I}=SO\left( 1,1\right) \times T_{n},
\label{PPA-5-bis}
\end{equation}
with no corresponding \textit{moduli space.} In particular, for $n=0$ this
orbit is actually $1$-charge, and it is given by
\begin{equation}
\mathcal{O}_{BPS,small,I,SO(1,1)}=\left\{ \left( q_{H},q_{0}\right) =\left(
0,+\right) ,\left( 0,-\right) \right\} .  \label{ja-1}
\end{equation}
On the other hand, for $n=1$, the orbit (\ref{PPA-5-bis}) reads
\begin{equation}
\mathcal{O}_{BPS,small,I,\left[ SO(1,1)\right] ^{2}}=\left\{ \left(
q_{H},q_{+},q_{-}\right) =\left( 0,+,+\right) ,\left( 0,-,-\right) \right\} ,
\label{PPA-5-bis-n=1}
\end{equation}
with no corresponding \textit{moduli space},\textit{\ }and thus
\begin{equation}
\sharp _{BPS,small,I,\left[ SO(1,1)\right] ^{2}}=2.
\end{equation}

\item  Non-BPS lightlike ($\mathcal{I}_{3,el}=0$, $\frac{\partial \mathcal{I}%
_{3,el}}{\partial Q}\neq 0$: $2$-charge) orbit with vanishing $q_{H}$ and
spacelike $q_{I}$:
\begin{equation}
\left\{
\begin{array}{l}
q_{H}=0; \\
\\
q_{0}^{2}-\sum_{i=1}^{n}q_{i}^{2}<0.
\end{array}
\right.  \label{cconstr-5}
\end{equation}
It reads (existing only for $n\geqslant 1$)
\begin{equation}
\mathcal{O}_{nBPS,small,I}=SO\left( 1,1\right) \times \frac{SO\left(
1,n\right) }{SO\left( 1,n-1\right) },  \label{PPA-5}
\end{equation}
with corresponding \textit{moduli space} (recall Eq. (\ref
{nBPS-large-III-mod-space}))
\begin{equation}
\mathcal{M}_{nBPS,small,I}=\mathcal{M}_{nBPS,large,II}.  \label{PPA-6}
\end{equation}
Thus, by recalling (\ref{N=2-d=5-Jordan-symm-scalar-manifold}), the number $%
\sharp $ of ``non-flat'' scalar degrees of freedom along $\mathcal{O}%
_{nBPS,small,I}$ is independent on $n\geqslant 1$:
\begin{equation}
\sharp _{nBPS,small,I}\equiv dim_{\mathbb{R}}M_{\mathcal{N}%
=2,d=5,Jordan,symm}-dim_{\mathbb{R}}\mathcal{M}_{nBPS,small,I}=2.
\end{equation}
For $n=1$, orbit (\ref{PPA-5}) reads
\begin{equation}
\mathcal{O}_{nBPS,small,I,\left[ SO(1,1)\right] ^{2}}=\left\{ \left(
q_{H},q_{+},q_{-}\right) =\left( 0,+,-\right) ,\left( 0,-,+\right) \right\} ,
\label{PPA-5-n=1}
\end{equation}
with no corresponding \textit{moduli space}.

\item  BPS critical ($\mathcal{I}_{3,el}=0$, $\frac{\partial \mathcal{I}%
_{3,el}}{\partial Q}=0$: $1$-charge) orbit with vanishing $q_{H}$ and
lightlike $q_{I}$:
\begin{equation}
\left\{
\begin{array}{l}
q_{H}=0; \\
\\
q_{0}^{2}-\sum_{i=1}^{n}q_{i}^{2}=0.
\end{array}
\right.  \label{cconstr-6}
\end{equation}
By recalling definition (\ref{lc-def}), the orbit reads (existing only for $%
n\geqslant 1$)
\begin{equation}
\mathcal{O}_{BPS,small,II}=\Lambda _{n},  \label{PPA-1}
\end{equation}
and the corresponding \textit{moduli space} is ($n\geqslant 1$)
\begin{equation}
\mathcal{M}_{BPS,small,II}=SO\left( 1,1\right) \times \mathbb{R}^{n-1}.
\label{Pal-3}
\end{equation}
Thus, by recalling (\ref{N=2-d=5-Jordan-symm-scalar-manifold}), the number $%
\sharp $ of ``non-flat'' scalar degrees of freedom along $\mathcal{O}%
_{BPS,small,II}$ is independent on $n\geqslant 1$:
\begin{equation}
\sharp _{BPS,small,II}\equiv dim_{\mathbb{R}}M_{\mathcal{N}%
=2,d=5,Jordan,symm}-dim_{\mathbb{R}}\mathcal{M}_{BPS,small,II}=1.
\end{equation}
Analogously to what holds for symmetric ``magic'' RSG (noted below Eq. (\ref
{jjjj-1})), the unique scalar degree of freedom on which the ADM mass
depends can be interpreted as the Kaluza-Klein radius in the $d=5\rightarrow
d=4$ reduction. For $n=1$, orbit (\ref{PPA-1}) reads
\begin{equation}
\mathcal{O}_{BPS,small,II,\left[ SO\left( 1,1\right) \right] ^{2}}=\left\{
\left( q_{H},q_{+},q_{-}\right) =\left( 0,0,+\right) ,\left( 0,+,0\right)
,\left( 0,0,-\right) ,\left( 0,-,0\right) \right\} .  \label{PPA-1-n=1}
\end{equation}

\item  BPS lightlike ($\mathcal{I}_{3,el}=0$, $\frac{\partial \mathcal{I}%
_{3,el}}{\partial Q}\neq 0$: $2$-charge) orbit, defined as follows:
\begin{equation}
\left\{
\begin{array}{l}
q_{H}>0; \\
\\
q_{0}^{2}-\sum_{i=1}^{n}q_{i}^{2}=0; \\
\\
q_{0}>0;
\end{array}
\right. ~~or~~\left\{
\begin{array}{l}
q_{H}<0; \\
\\
q_{0}^{2}-\sum_{i=1}^{n}q_{i}^{2}=0; \\
\\
q_{0}<0.
\end{array}
\right.  \label{cconstr-7}
\end{equation}
By recalling definition (\ref{lc-def}), the orbit reads ($n\geqslant 2$)
\begin{equation}
\mathcal{O}_{BPS,small,III}=\left[ SO\left( 1,1\right) ^{+}\times \Lambda
_{n}^{+}\right] \cup \left[ SO\left( 1,1\right) ^{-}\times \Lambda _{n}^{-}%
\right] ,  \label{ppp-3}
\end{equation}
and the corresponding \textit{moduli space} is purely translational ($%
n\geqslant 2$):
\begin{equation}
\mathcal{M}_{BPS,small,III}=\mathbb{R}^{n-1}=\mathcal{M}_{BPS,small,II}.
\label{Pal-4}
\end{equation}
Thus, by recalling (\ref{N=2-d=5-Jordan-symm-scalar-manifold}), the number $%
\sharp $ of ``non-flat'' scalar degrees of freedom along $\mathcal{O}%
_{BPS,small,III}$ is independent on $n\geqslant 2$:
\begin{equation}
\sharp _{BPS,small,III}\equiv dim_{\mathbb{R}}M_{\mathcal{N}%
=2,d=5,Jordan,symm}-dim_{\mathbb{R}}\mathcal{M}_{BPS,small,III}=2.
\end{equation}
This orbit exists also for $n=1$, and it reads
\begin{equation}
\mathcal{O}_{BPS,small,III,\left[ SO(1,1)\right] ^{2}}=\left\{ \left(
q_{H},q_{+},q_{-}\right) =\left( +,0,+\right) ,\left( +,+,0\right) ,\left(
-,0,-\right) ,\left( -,-,0\right) \right\} ,  \label{ppp-3-n=1}
\end{equation}
with no corresponding \textit{moduli space}. (\ref{ppp-3-n=1}) is equivalent
to (\ref{PPA-5-bis-n=1}) through triality permutation symmetry of $q_{H}$, $%
q_{+}$ and $q_{-}$. Thus, the BPS $2$-charge orbit of $\left[ SO\left(
1,1\right) \right] ^{2}$ model is given, up to permutations of the triplet $%
\left( q_{H},q_{+},q_{-}\right) $, by
\begin{equation}
\mathcal{O}_{BPS,2\text{-charge},\left[ SO(1,1)\right] ^{2}}=\left\{ \left(
q_{H},q_{+},q_{-}\right) =\left( +,+,0\right) ,\left( -,-,0\right) \right\} .
\end{equation}

\item  Non-BPS lightlike ($\mathcal{I}_{3,el}=0$, $\frac{\partial \mathcal{I}%
_{3,el}}{\partial Q}\neq 0$: $2$-charge) orbit, defined as follows:
\begin{equation}
\left\{
\begin{array}{l}
q_{H}<0; \\
\\
q_{0}^{2}-\sum_{i=1}^{n}q_{i}^{2}=0; \\
\\
q_{0}>0;
\end{array}
\right. ~~or~~\left\{
\begin{array}{l}
q_{H}>0; \\
\\
q_{0}^{2}-\sum_{i=1}^{n}q_{i}^{2}=0; \\
\\
q_{0}<0.
\end{array}
\right.  \label{cconstr-8}
\end{equation}
By recalling definition (\ref{lc-def}), the orbit reads ($n\geqslant 2$)
\begin{equation}
\mathcal{O}_{nBPS,small,II}=\left[ SO\left( 1,1\right) ^{+}\times \Lambda
_{n}^{-}\right] \cup \left[ SO\left( 1,1\right) ^{-}\times \Lambda _{n}^{+}%
\right] ,  \label{ppp-4}
\end{equation}
with corresponding \textit{moduli space }($n\geqslant 2$)
\begin{equation}
\mathcal{M}_{nBPS,small,II}=\mathbb{R}^{n-1}=\mathcal{M}_{BPS,small,II}=%
\mathcal{M}_{BPS,small,III}  \label{Pal-5}
\end{equation}
Thus, by recalling (\ref{N=2-d=5-Jordan-symm-scalar-manifold}), the number $%
\sharp $ of ``non-flat'' scalar degrees of freedom along $\mathcal{O}%
_{nBPS,small,II}$ is independent on $n\geqslant 2$:
\begin{equation}
\sharp _{nBPS,small,II}\equiv dim_{\mathbb{R}}M_{\mathcal{N}%
=2,d=5,Jordan,symm}-dim_{\mathbb{R}}\mathcal{M}_{nBPS,small,II}=2.
\end{equation}
This orbit exists also for $n=1$, and it reads
\begin{equation}
\mathcal{O}_{nBPS,small,II,\left[ SO\left( 1,1\right) \right] ^{2}}=\left\{
\left( q_{H},q_{+},q_{-}\right) =\left( +,0,-\right) ,\left( +,-,0\right)
,\left( -,0,+\right) ,\left( -,+,0\right) \right\} ,  \label{Pal-1}
\end{equation}
with no corresponding \textit{moduli space}. (\ref{Pal-1}) is equivalent to (%
\ref{PPA-5-n=1}) through triality permutation symmetry of $q_{H}$, $q_{+}$
and $q_{-}$. Thus, the non-BPS $2$-charge orbit of $\left[ SO\left(
1,1\right) \right] ^{2}$ model is given, up to permutations of the triplet $%
\left( q_{H},q_{+},q_{-}\right) $, by
\begin{equation}
\mathcal{O}_{nBPS,2\text{-charge},\left[ SO\left( 1,1\right) \right]
^{2}}=\left\{ \left( q_{H},q_{+},q_{-}\right) =\left( +,-,0\right) \right\} .
\end{equation}

\item  BPS critical ($\mathcal{I}_{3,el}=0$, $\frac{\partial \mathcal{I}%
_{3,el}}{\partial Q}=0$: $1$-charge) orbit with vanishing $q_{I}$ and
non-vanishing $q_{H}$:
\begin{equation}
\left\{
\begin{array}{l}
q_{H}\in \mathbb{R}_{0}; \\
\\
q_{I}=0.
\end{array}
\right.  \label{PPA-3}
\end{equation}
It exists for every $n\geqslant 0$, and it reads
\begin{equation}
\mathcal{O}_{BPS,small,IV}=SO\left( 1,1\right) ,  \label{PPPA-2}
\end{equation}
with \textit{moduli space }($n\geqslant 1$; recall (\ref{non-Jordan-symm}))
\begin{equation}
\mathcal{M}_{BPS,small,IV}=\frac{SO\left( 1,n\right) }{SO\left( n\right) }%
=M_{nJ,5,n}.
\end{equation}
Thus, by recalling (\ref{N=2-d=5-Jordan-symm-scalar-manifold}), the number $%
\sharp $ of ``non-flat'' scalar degrees of freedom along $\mathcal{O}%
_{BPS,small,IV}$ is independent on $n\geqslant 1$:
\begin{equation}
\sharp _{BPS,small,IV}\equiv dim_{\mathbb{R}}M_{\mathcal{N}%
=2,d=5,Jordan,symm}-\mathcal{M}_{BPS,small,IV}=1.
\end{equation}
Analogously to what holds for symmetric ``magic'' RSG (noted below Eq. (\ref
{jjjj-1})), the unique scalar degree of freedom on which the ADM mass
depends can be interpreted as the Kaluza-Klein radius in the $d=5\rightarrow
d=4$ reduction. Furthermore, as in the corresponding $\mathcal{N}=4$, $d=5$
``small'' orbit (given by Eq. (\ref{PA-2})), the sign of $q_{H}$\ does not
matter here. Orbit (\ref{PPPA-2}) is originated by the $d=6\rightarrow d=5$
reduction of $\left( 1,0\right) $ theory with \textit{all} charges switched
off. Indeed, $q_{H}$ is the electric charge of the Kaluza-Klein vector in
the reduction $d=6\rightarrow d=5$.\medskip\ In particular, for $n=0$, this
orbit reads
\begin{equation}
\mathcal{O}_{BPS,small,IV,SO(1,1)}=\left\{ \left( q_{H},q_{0}\right) =\left(
+,0\right) ,\left( -,0\right) \right\} ,  \label{ja-2}
\end{equation}
with no corresponding \textit{moduli space}. On the other hand, for $n=1$
the orbit (\ref{PPPA-2}) reads
\begin{equation}
\mathcal{O}_{BPS,small,IV,\left[ SO(1,1)\right] ^{2}}=\left\{ \left(
q_{H},q_{+},q_{-}\right) =\left( +,0,0\right) ,\left( -,0,0\right) \right\} ,
\label{PPPA-2-n=1}
\end{equation}
which is equivalent to (\ref{PPA-1-n=1}) through triality permutation
symmetry of $q_{H}$, $q_{+}$ and $q_{-}$. Thus, the BPS $1$-charge orbit of $%
\left[ SO\left( 1,1\right) \right] ^{2}$ model is given, up to permutations
of the triplet $\left( q_{H},q_{+},q_{-}\right) $, by
\begin{equation}
\mathcal{O}_{BPS,1\text{-charge},\left[ SO(1,1)\right] ^{2}}=\left\{ \left(
q_{H},q_{+},q_{-}\right) =\left( +,0,0\right) ,\left( -,0,0\right) \right\} .
\end{equation}
\medskip \smallskip
\end{enumerate}

Thus, the stratification structure of the $\left( \mathbf{1},\mathbf{1+n}%
\right) $-repr. space of the $d=5$ $U$-duality group $SO\left( 1,1\right)
\times SO\left( 1,n\right) $ can be given through the following two chains
of relations, proceeding (left to right) from $1$-charge orbits to $2$%
-charge and then $3$-charge orbits:
\begin{eqnarray}
\mathcal{O}_{BPS,small,II} &\rightarrow &\left\{
\begin{array}{l}
\mathcal{O}_{BPS,small,I}\rightarrow \left\{
\begin{array}{l}
\mathcal{O}_{BPS,large} \\
\mathcal{O}_{nBPS,large,I}
\end{array}
\right. \\
\\
\mathcal{O}_{nBPS,small,I}\rightarrow \mathcal{O}_{nBPS,large,II} \\
\\
\mathcal{O}_{BPS,small,III}\rightarrow \left\{
\begin{array}{l}
\mathcal{O}_{BPS,large} \\
\mathcal{O}_{nBPS,large,II}
\end{array}
\right. \\
\\
\mathcal{O}_{nBPS,small,II}\rightarrow \left\{
\begin{array}{l}
\mathcal{O}_{nBPS,large,I} \\
\mathcal{O}_{nBPS,large,II};
\end{array}
\right.
\end{array}
\right.  \label{strat-1} \\
&&  \notag \\
&&  \notag \\
\mathcal{O}_{BPS,small,IV} &\rightarrow &\left\{
\begin{array}{l}
\mathcal{O}_{BPS,small,III}\rightarrow \left\{
\begin{array}{l}
\mathcal{O}_{BPS,large} \\
\mathcal{O}_{nBPS,large,II}
\end{array}
\right. \\
\\
\mathcal{O}_{nBPS,small,II}\rightarrow \left\{
\begin{array}{l}
\mathcal{O}_{nBPS,large,I} \\
\mathcal{O}_{nBPS,large,II}.
\end{array}
\right.
\end{array}
\right.  \label{strat-2}
\end{eqnarray}
For the $SO\left( 1,1\right) $ model ($n=0$), such a stratification
structure simplifies as follows:
\begin{equation}
SO\left( 1,1\right) :\left. \overset{1\text{-charge}}{
\begin{array}{l}
\mathcal{O}_{BPS,small,I} \\
\\
\mathcal{O}_{BPS,small,IV}
\end{array}
}\right\} \rightarrow \left\{ \overset{2\text{-charge}}{
\begin{array}{l}
\mathcal{O}_{BPS,large} \\
\\
\mathcal{O}_{nBPS,large}.
\end{array}
}\right.
\end{equation}
On the other hand, for the $\left[ SO\left( 1,1\right) \right] ^{2}$ model ($%
n=1$), stratification structure (\ref{strat-1})-(\ref{strat-2}) reads:
\begin{equation}
\left[ SO\left( 1,1\right) \right] ^{2}:\mathcal{O}_{BPS,1\text{-charge}%
}\rightarrow \left\{
\begin{array}{l}
\mathcal{O}_{BPS,2\text{-charge}}\rightarrow \left\{
\begin{array}{l}
\mathcal{O}_{BPS,3\text{-charge}} \\
\mathcal{O}_{nBPS,3\text{-charge}}
\end{array}
\right. \\
\\
\mathcal{O}_{nBPS,2\text{-charge}}\rightarrow \mathcal{O}_{nBPS,3\text{%
-charge}}.
\end{array}
\right.
\end{equation}
\smallskip

Thus, summarizing, $\mathcal{N}=2$, $d=5$ Jordan symmetric sequence admits
six ``small'' charge orbits describing the flux configurations supporting
static, spherically symmetric, asymptotically flat ``small'' BHs: four $%
\frac{1}{2}$-BPS and two non-BPS. Furthermore, the ``large'' orbits are
three, namely one $\frac{1}{2}$-BPS and two non-BPS (with $Z\neq 0$ at the
horizon).

\section{\label{N=4,d=5}$\mathcal{N}=4$, $d=5$ Supergravity}

The scalar manifold of $\mathcal{N}=4$, $d=5$ supergravity coupled to $%
n_{V}=n\in \mathbb{N}\cup \left\{ 0\right\} $ matter (vector) multiplets
reads ($dim_{\mathbb{R}}=1+5n$, rank$=1+min\left( 5,n\right) $)
\begin{equation}
M_{\mathcal{N}=4,d=5}=SO\left( 1,1\right) \times \frac{SO\left( 5,n\right) }{%
SO\left( 5\right) \times SO\left( n\right) }.  \label{N=4-scalar-manifold}
\end{equation}
This theory is associated to the rank-$3$ Euclidean \textit{reducible}
Jordan algebra $\mathbb{R}\oplus \mathbf{\Gamma }_{5,n}$. In the following
treatment, we will determine the ``large'' and ``small'' orbits of the
irrepr. $\left( \mathbf{1},\mathbf{5+n}\right) $ of the $U$-duality group $%
SO\left( 1,1\right) \times SO\left( 5,n\right) $.

For brevity's sake, we will do this only through an analysis in the ``bare''
charges' basis.

Without any loss in generality, one can choose to treat only $d=5$ extremal (%
\textit{electric}) BHs. Indeed, due to the symmetricity of the reducible
coset (\ref{N=4-scalar-manifold}), the treatment of $d=5$ extremal (\textit{%
magnetic}) black strings is essentially analogous.

In the ``bare'' charges' basis, the electric cubic invariant of the $\left(
\mathbf{1},\mathbf{5+n}\right) $ of $SO\left( 1,1\right) \times SO\left(
5,n\right) $ reads as follows ($I=1,...,5+n$ throughout; the indices $%
1,...,5 $, with positive signature, pertain to the five $\mathcal{N}=4$, $%
d=5 $ graviphotons):
\begin{equation}
\mathcal{I}_{3,el}\equiv q_{H}q_{I}q_{J}\eta ^{IJ}\equiv q_{H}q_{I}^{2},
\label{I_3-el}
\end{equation}
where $q_{H}$ is the electric charge of the $3$-form field strength of the $%
2 $-form $B_{\mu \nu }$ ($\mu $, $\nu =0,1,...,4$) in the gravity multiplet
(see \textit{e.g.} \cite{ADF-central-diverse,ADF-U-duality-revisited}). $%
q_{H}$ is an $SO\left( 5,n\right) $-singlet, with $SO\left( 1,1\right) $%
-weight $+2$. On the other hand, the $SO\left( 5,n\right) $-vector $q_{I}$
has $SO\left( 1,1\right) $-weight $-1$, such that $\mathcal{I}_{3,el}$
defined by (\ref{I_3-el}) is $SO\left( 1,1\right) \times SO\left( 5,n\right)
$-invariant. Notice that the action of the $U$-duality group does not mix $%
q_{H}$ and $q_{I}$, and this originates more charge orbits with respect to
the \textit{irreducible} cases. Moreover, $\eta _{IJ}=\eta ^{IJ}$ is the
pseudo-Euclidean metric of $SO\left( 5,n\right) $, with signature $\left(
\overset{5}{\overbrace{+,...,+}},\overset{n}{\overbrace{-,...,-}}\right) $.

\subsection{\label{N=4-d=5-Large}``Large'' Orbits}

\begin{itemize}
\item  $\frac{1}{4}$-BPS ($3$-charge) orbit, defined by a timelike $q_{I}$
vector, with $q_{H}$ of any sign:
\begin{equation}
q_{H}\in \mathbb{R}_{0},~q_{I}q_{J}\eta ^{IJ}>0.
\end{equation}
The resulting form of the orbit reads \cite{AFMT1} ($n\geqslant 0$)
\begin{equation}
\mathcal{O}_{\frac{1}{4}-BPS,large}=SO\left( 1,1\right) \times \frac{%
SO\left( 5,n\right) }{SO\left( 4,n\right) },  \label{PAA-1}
\end{equation}
with related \textit{moduli space}:
\begin{eqnarray}
\mathcal{M}_{\frac{1}{4}-BPS,large} &=&\frac{SO\left( 4,n\right) }{SO\left(
4\right) \times SO\left( n\right) }  \notag \\
&=&\frac{M_{\left( 1,1\right) ,d=6}}{SO\left( 1,1\right) },
\label{1/4-BPS-large-mod-space}
\end{eqnarray}
where $M_{\left( 1,1\right) ,d=6}$ is the scalar manifold of non-chiral
half-maximal supergravity in $d=6$ with $n$ matter (vector) multiplets. The
exchange between $q_{H}>0$ and $q_{H}<0$ amounts to exchanging the two
gravitinos in the gravity multiplet, \textit{i.e.} the two (opposite)
skew-eigenvalues of the skew-traceless central charge matrix \r{Z}$_{AB}$ ($%
A,B=1,...,4$). Thus, the number $\sharp $ of ``non-flat'' scalar degrees of
freedom along $\mathcal{O}_{\frac{1}{4}-BPS,large}$ is (for $n\geqslant 1$)
\begin{equation}
\sharp _{\frac{1}{4}-BPS,large}\equiv dim_{\mathbb{R}}M_{\mathcal{N}%
=4,d=5}-dim_{\mathbb{R}}\mathcal{M}_{\frac{1}{4}-BPS,large}=n+1.
\end{equation}
In $\mathcal{N}>2$-extended supergravity theories, in general $\frac{1}{%
\mathcal{N}}$-BPS attractors have a related \textit{moduli space} \cite
{Ferrara-Marrani-2}. It corresponds to the hypermultiplets' scalar manifold
in the supersymmetry reduction $\mathcal{N}>2\longrightarrow \mathcal{N}=2$
of the theory under consideration. In this case, it is amusing to observe
that $\mathcal{M}_{\frac{1}{4}-BPS,large}$ given by (\ref
{1/4-BPS-large-mod-space}) is the $c$-map of the vector multiplets' scalar
manifold of the $\mathcal{N}=2$, $d=4$ Jordan symmetric sequence:
\begin{equation}
\mathcal{M}_{\frac{1}{4}-BPS,large}=c\left( \frac{SU\left( 1,1\right) }{%
U\left( 1\right) }\times \frac{SO\left( 2,n-2\right) }{SO\left( 2\right)
\times SO\left( n-2\right) }\right) .
\end{equation}
Thus, $\mathcal{M}_{\frac{1}{4}-BPS,large}$ admits an interpretation either
as \textbf{1}) scalar manifold of $\mathcal{N}=4$, $d=3$ Jordan symmetric
sequence in $d=3$, or as \textbf{2}) the hypermultiplets' scalar manifold of
Jordan symmetric sequence in $d=4$, $5$ ($\mathcal{N}=2$) and $6$ ($\left(
1,0\right) $). In particular, $\mathcal{M}_{\frac{1}{4}-BPS,large}$
parametrizes the $\mathcal{N}=2$ hyperscalar degrees of freedom in the
supersymmetry/Jordan algebra reduction:
\begin{equation}
d=5:
\begin{array}{c}
\mathcal{N}=4 \\
\mathbb{R}\oplus \mathbf{\Gamma }_{5,n}
\end{array}
\longrightarrow
\begin{array}{c}
\mathcal{N}=2 \\
\mathbb{R}\oplus \mathbf{\Gamma }_{1,n-3}
\end{array}
.  \label{N=6-reduction-1}
\end{equation}
The \textit{pure} theory (\textit{i.e.} $n=0$) limit of orbit (\ref{PAA-1})
is actually $2$-charge (indeed, $SO\left( 5\right) $ symmetry can be used to
make only one component of the Euclidean vector $q_{I}$ non-vanishing), and
it reads
\begin{equation}
\mathcal{O}_{\frac{1}{4}-BPS,large,n=0}=SO\left( 1,1\right) \times \frac{%
SO\left( 5\right) }{SO\left( 4\right) }\equiv SO\left( 1,1\right) \times
S^{4},  \label{PAA-1-n=0}
\end{equation}
with no corresponding \textit{moduli space}, and thus trivially
\begin{equation}
\sharp _{\frac{1}{4}-BPS,large,n=0}=1.  \label{jjj-1}
\end{equation}

\item  non-BPS ($3$-charge) orbit with \r{Z}$_{AB}=0$ (at the horizon),
defined by a spacelike $q_{I}$ vector, and $q_{H}$ of any sign:
\begin{equation}
q_{H}\in \mathbb{R}_{0},~q_{I}q_{J}\eta ^{IJ}<0.
\end{equation}
Notice that both signs of $q_{H}$ are allowed, due to the fact that the
non-BPS \r{Z}$_{AB}=0$ Attractor Eqs. are quadratic in $q_{H}$ (see \textit{%
e.g.} \cite{AFMT1}). The resulting orbit reads ($n\geqslant 1$, not existing
in \textit{pure} theory) \cite{AFMT1}
\begin{equation}
\mathcal{O}_{nBPS,large}=SO\left( 1,1\right) \times \frac{SO\left(
5,n\right) }{SO\left( 5,n-1\right) },
\end{equation}
with related \textit{moduli space}:
\begin{eqnarray}
\mathcal{M}_{nBPS,large} &=&\frac{SO\left( 5,n-1\right) }{SO\left( 5\right)
\times SO\left( n-1\right) }  \notag \\
&=&\left. M_{\left( 2,0\right) ,d=6}\right| _{n-1},
\label{nBPS-large-mod-space}
\end{eqnarray}
where $\left. M_{\left( 2,0\right) ,d=6}\right| _{n-1}$ is the scalar
manifold of $\left( 2,0\right) $, $d=6$ supergravity with $n_{T}=n-1$ tensor
multiplets. Note that $\mathcal{N}=4$, $d=5$ and $\left( 2,0\right) $, $d=6$
supergravities share the same $\mathcal{R}$-symmetry $SO\left( 5\right) \sim
USp\left( 4\right) $. Thus, the number $\sharp $ of ``non-flat'' scalar
degrees of freedom along $\mathcal{O}_{nBPS,large}$ is independent on $%
n\geqslant 2$:
\begin{equation}
\sharp _{nBPS,large}\equiv dim_{\mathbb{R}}M_{\mathcal{N}=4,d=5}-dim_{%
\mathbb{R}}\mathcal{M}_{nBPS,large}=6.
\end{equation}
\texttt{\ }
\end{itemize}

\subsection{\label{N=4-d=5-Small}``Small'' Orbits}

The conditions on $\mathcal{I}_{3,el}$ defined by (\ref{I_3-el}) are
formally the same as the ones holding in $\mathcal{N}=2$, $d=5$ Jordan
symmetric sequence, and given by Eqs. (\ref{dI_3-el}) and (\ref{ddI_3-el}).
Thus, analogously to the case of $\mathcal{N}=2$, $d=5$ Jordan symmetric
sequence, and as expected from the fact that $\mathcal{I}_{3,el}$ is
homogeneous of degree three, (\ref{ddI_3-el}) implies that the unique
doubly-critical orbit is the trivial one with all charges vanishing (namely,
$0$-charge orbit; recall Eq. (\ref{PAA-3})).

The ``small'' orbits of the $\left( \mathbf{1},\mathbf{5+n}\right) $ of the $%
U$-duality group $SO\left( 1,1\right) \times SO\left( 5,n\right) $ list as
follows:

\begin{enumerate}
\item  Lightlike ($\mathcal{I}_{3,el}=0$, $\frac{\partial \mathcal{I}_{3,el}%
}{\partial Q}\neq 0$: $2$-charge) orbit with vanishing $q_{H}$ and timelike $%
q_{I}$:
\begin{equation}
\left\{
\begin{array}{l}
q_{H}=0; \\
\\
q_{I}^{2}>0.
\end{array}
\right.
\end{equation}
This orbit is $\frac{1}{2}$-BPS \cite{Ferrara-Maldacena}. It reads ($%
n\geqslant 0$)
\begin{equation}
\mathcal{O}_{\frac{1}{2}-BPS,small,I}=SO\left( 1,1\right) \times \frac{%
SO\left( 5,n\right) }{SO\left( 4,n\right) },  \label{PA-4}
\end{equation}
with corresponding \textit{moduli space} (recall Eq. (\ref
{1/4-BPS-large-mod-space}))
\begin{equation}
\mathcal{M}_{\frac{1}{2}-BPS,small,I}=\mathcal{M}_{\frac{1}{4}-BPS,large}.
\label{1}
\end{equation}
Thus, the number $\sharp $ of ``non-flat'' scalar degrees of freedom along $%
\mathcal{O}_{\frac{1}{2}-BPS,small,I}$ is (for $n\geqslant 1$):
\begin{equation}
\sharp _{\frac{1}{2}-BPS,small,I}\equiv dim_{\mathbb{R}}M_{\mathcal{N}%
=4,d=5}-dim_{\mathbb{R}}\mathcal{M}_{\frac{1}{2}-BPS,small,I}=n+1.
\end{equation}
The \textit{pure} theory (\textit{i.e.} $n=0$) limit of orbit (\ref{PA-4})
is actually $1$-charge, and it reads
\begin{equation}
\mathcal{O}_{\frac{1}{2}-BPS,small,I,n=0}=SO\left( 1,1\right) \times S^{4},
\label{PA-4-n=0}
\end{equation}
with no related \textit{moduli space}, and thus
\begin{equation}
\sharp _{\frac{1}{2}-BPS,small,I,n=0}=1.
\end{equation}

\item  Lightlike ($\mathcal{I}_{3,el}=0$, $\frac{\partial \mathcal{I}_{3,el}%
}{\partial Q}\neq 0$: $2$-charge) orbit with vanishing $q_{H}$ and spacelike
$q_{I}$:
\begin{equation}
\left\{
\begin{array}{l}
q_{H}=0; \\
\\
q_{I}^{2}<0.
\end{array}
\right.
\end{equation}
This orbit is non-BPS. It reads ($n\geqslant 1$, not existing in \textit{pure%
} theory)
\begin{equation}
\mathcal{O}_{nBPS,small}=SO\left( 1,1\right) \times \frac{SO\left(
5,n\right) }{SO\left( 5,n-1\right) },  \label{PA-5}
\end{equation}
with corresponding \textit{moduli space} (recall Eq. (\ref
{nBPS-large-mod-space}))
\begin{equation}
\mathcal{M}_{nBPS,small}=\mathcal{M}_{nBPS,large}.  \label{PA-6}
\end{equation}
Thus, the number $\sharp $ of ``non-flat'' scalar degrees of freedom along $%
\mathcal{O}_{nBPS,small}$ is independent on $n\geqslant 1$:
\begin{equation}
\sharp _{nBPS,small}\equiv dim_{\mathbb{R}}M_{\mathcal{N}=4,d=5}-dim_{%
\mathbb{R}}\mathcal{M}_{nBPS,small}=6.
\end{equation}
\texttt{\ }

\item  Critical ($\mathcal{I}_{3,el}=0$, $\frac{\partial \mathcal{I}_{3,el}}{%
\partial Q}=0$: $1$-charge) orbit with vanishing $q_{H}$ and lightlike $%
q_{I} $:
\begin{equation}
\left\{
\begin{array}{l}
q_{H}=0; \\
\\
q_{I}^{2}=0.
\end{array}
\right.
\end{equation}
This orbit is $\frac{1}{2}$-BPS \cite{Ferrara-Maldacena}. It reads ($%
n\geqslant 1$, not existing in \textit{pure} theory)
\begin{equation}
\mathcal{O}_{\frac{1}{2}-BPS,small,II}=\frac{SO\left( 5,n\right) }{SO\left(
4,n-1\right) \rtimes \mathbb{R}^{4,n-1}},  \label{PA-1}
\end{equation}
with corresponding \textit{moduli space} (recall Eq. (\ref{1}))
\begin{eqnarray}
\mathcal{M}_{\frac{1}{2}-BPS,small,II} &=&SO\left( 1,1\right) \times \left.
\mathcal{M}_{\frac{1}{2}-BPS,small,I}\right| _{n\rightarrow n-1}\rtimes
\mathbb{R}^{4,n-1}  \notag \\
&=&SO\left( 1,1\right) \times \left. \mathcal{M}_{\frac{1}{4}%
-BPS,large}\right| _{n\rightarrow n-1}\rtimes \mathbb{R}^{4,n-1}.  \label{2}
\end{eqnarray}
Thus, the number $\sharp $ of ``non-flat'' scalar degrees of freedom along $%
\mathcal{O}_{\frac{1}{2}-BPS,small,II}$ is independent on $n\geqslant 1$:
\begin{equation}
\sharp _{\frac{1}{2}-BPS,small,II}\equiv dim_{\mathbb{R}}M_{\mathcal{N}%
=4,d=5}-dim_{\mathbb{R}}\mathcal{M}_{\frac{1}{2}-BPS,small,II}=1.
\end{equation}
Analogously to what holds for symmetric ``magic'' RSG (noted below Eq. (\ref
{jjjj-1})) and for $N=2$, $d=5$ Jordan symmetric sequence treated in Sect.
\ref{N=2,d=5-Jordan-Symm-Seq}, the unique scalar degree of freedom on which
the ADM mass depends can be interpreted as the Kaluza-Klein radius in the $%
d=5\rightarrow d=4$ reduction.

\item  Lightlike ($\mathcal{I}_{3,el}=0$, $\frac{\partial \mathcal{I}_{3,el}%
}{\partial Q}\neq 0$: $2$-charge) orbit with non-vanishing $q_{H}$ and
lightlike $q_{I}$:
\begin{equation}
\left\{
\begin{array}{l}
q_{H}\in \mathbb{R}_{0}; \\
\\
q_{I}^{2}=0.
\end{array}
\right.  \label{jazz-1}
\end{equation}
This orbit is $\frac{1}{4}$-BPS. It reads ($n\geqslant 1$)
\begin{equation}
\mathcal{O}_{\frac{1}{4}-BPS,small}=SO\left( 1,1\right) \times \frac{%
SO\left( 5,n\right) }{SO\left( 4,n-1\right) \rtimes \mathbb{R}^{4,n-1}},
\label{2c-split-1}
\end{equation}
with corresponding \textit{moduli space} (recall Eq. (\ref{2}))
\begin{eqnarray}
\mathcal{M}_{\frac{1}{4}-BPS,small} &=&\left. \mathcal{M}_{\frac{1}{2}%
-BPS,small,I}\right| _{n\rightarrow n-1}\rtimes \mathbb{R}^{4,n-1}  \notag \\
&=&\left. \mathcal{M}_{\frac{1}{4}-BPS,large}\right| _{n\rightarrow
n-1}\rtimes \mathbb{R}^{4,n-1}.  \label{3}
\end{eqnarray}
Thus, the number $\sharp $ of ``non-flat'' scalar degrees of freedom along $%
\mathcal{O}_{\frac{1}{2}-BPS,small,II}$ is independent on $n\geqslant 1$:
\begin{equation}
\sharp _{\frac{1}{4}-BPS,small}\equiv dim_{\mathbb{R}}M_{\mathcal{N}%
=4,d=5}-dim_{\mathbb{R}}\mathcal{M}_{\frac{1}{4}-BPS,small}=2.
\end{equation}

\item  Critical ($\mathcal{I}_{3,el}=0$, $\frac{\partial \mathcal{I}_{3,el}}{%
\partial Q}=0$: $1$-charge) orbit with vanishing $q_{I}$ and non-vanishing $%
q_{H}$:
\begin{equation}
\left\{
\begin{array}{l}
q_{H}\in \mathbb{R}_{0}; \\
\\
q_{I}=0.
\end{array}
\right.  \label{PA-3}
\end{equation}
This orbit is $\frac{1}{2}$-BPS \cite{Ferrara-Maldacena}. It reads
(independent on $n\geqslant 0$)
\begin{equation}
\mathcal{O}_{\frac{1}{2}-BPS,small,III}=SO\left( 1,1\right) ,  \label{PA-2}
\end{equation}
with \textit{moduli space}
\begin{equation}
\mathcal{M}_{\frac{1}{2}-BPS,small,III}=\frac{SO\left( 5,n\right) }{SO\left(
5\right) \times SO\left( n\right) }.
\end{equation}
Thus, the number $\sharp $ of ``non-flat'' scalar degrees of freedom along $%
\mathcal{O}_{\frac{1}{2}-BPS,small,III}$ is independent on $n\geqslant 0$:
\begin{equation}
\sharp _{\frac{1}{2}-BPS,small,III}\equiv dim_{\mathbb{R}}M_{\mathcal{N}%
=4,d=5}-\mathcal{M}_{\frac{1}{2}-BPS,small,III}=1.
\end{equation}
Notice that $\mathcal{O}_{\frac{1}{2}-BPS,small,III}$ can also be seen as
the ``$n=0$ formal limit'' of $\mathcal{O}_{\frac{1}{4}-BPS,small}$ given by
Eq. (\ref{2c-split-1}). Indeed, the $n=0$ limit of (\ref{jazz-1}) is given
by (\ref{PA-3}) itself. Furthermore, analogously to what holds for symmetric
``magic'' RSG (noted below Eq. (\ref{jjjj-1})) and for $N=2$, $d=5$ Jordan
symmetric sequence treated in Sect. \ref{N=2,d=5-Jordan-Symm-Seq}, the
unique scalar degree of freedom on which the ADM mass depends can be
interpreted as the Kaluza-Klein radius in the $d=5\rightarrow d=4$
reduction. Orbit (\ref{PA-2}) is originated by the $d=6\rightarrow d=5$
reduction of $\left( 2,0\right) $ theory with \textit{all} charges switched
off. Indeed, $q_{H}$ is the electric charge of the Kaluza-Klein vector in
the reduction $d=6\rightarrow d=5$. Notice that in the \textit{pure} theory (%
\textit{i.e.} $n=0$) $\mathcal{M}_{\frac{1}{2}-BPS,small,III}$ vanishes, and
thus:
\begin{equation}
\sharp _{\frac{1}{2}-BPS,small,III,n=0}=1.  \label{jjj-2}
\end{equation}
\smallskip
\end{enumerate}

Thus, the stratification structure of the $\left( \mathbf{1},\mathbf{5+n}%
\right) $-repr. space of the $d=5$ $U$-duality group $SO\left( 1,1\right)
\times SO\left( 5,n\right) $ can be given through the two chains of
relations, proceeding (left to right) from $1$-charge orbits to $2$-charge
and then $3$-charge orbits:
\begin{eqnarray}
\mathcal{O}_{\frac{1}{2}-BPS,small,II} &\rightarrow &\left\{
\begin{array}{l}
\mathcal{O}_{\frac{1}{2}-BPS,small,I}\rightarrow \mathcal{O}_{\frac{1}{4}%
-BPS,large} \\
\\
\mathcal{O}_{nBPS,small}\rightarrow \mathcal{O}_{nBPS,large} \\
\\
\mathcal{O}_{\frac{1}{4}-BPS,small}\rightarrow \left\{
\begin{array}{l}
\mathcal{O}_{\frac{1}{4}-BPS,large} \\
\mathcal{O}_{nBPS,large};
\end{array}
\right.
\end{array}
\right. \\
&&  \notag \\
&&  \notag \\
\mathcal{O}_{\frac{1}{2}-BPS,small,III} &\rightarrow &\mathcal{O}_{\frac{1}{4%
}-BPS,small}\rightarrow \left\{
\begin{array}{l}
\mathcal{O}_{\frac{1}{4}-BPS,large} \\
\mathcal{O}_{nBPS,large}.
\end{array}
\right.
\end{eqnarray}
For \textit{pure} $\mathcal{N}=4$, $d=5$ supergravity, such a stratification
structure simplifies as follows:
\begin{equation}
\left. \overset{1\text{-charge}}{
\begin{array}{l}
\mathcal{O}_{\frac{1}{2}-BPS,small,I,n=0} \\
\\
\mathcal{O}_{\frac{1}{2}-BPS,small,III}
\end{array}
}\right\} \rightarrow \overset{2\text{-charge}}{\mathcal{O}_{\frac{1}{4}%
-BPS,large,n=0}}.\medskip
\end{equation}

\begin{table}[p]
\begin{center}
\begin{tabular}{|c||c|c|}
\hline
$\mathit{r~~}$ & $\mathcal{N}=4:\mathbb{R}\oplus \mathbf{\Gamma }_{5,n}$ & $%
\mathcal{N}=2:\mathbb{R}\oplus \mathbf{\Gamma }_{1,n}$ \\ \hline\hline
$3$ & $
\begin{array}{c}
\mathcal{O}_{\frac{1}{4}-BPS,large} \\
SO\left( 1,1\right) \times \frac{SO\left( 5,n\right) }{SO\left( 4,n\right) }
\\
\sharp =n+1
\end{array}
$ & $
\begin{array}{c}
\mathcal{O}_{BPS,large} \\
\left[ SO\left( 1,1\right) ^{+}\times T_{n}^{+}\right] \cup \left[ SO\left(
1,1\right) ^{-}\times T_{n}^{-}\right] \\
\sharp =n+1 \\
\updownarrow \ast \\
\mathcal{O}_{nBPS,large,I} \\
\left[ SO\left( 1,1\right) ^{+}\times T_{n}^{-}\right] \cup \left[ SO\left(
1,1\right) ^{-}\times T_{n}^{+}\right] \\
\sharp =n+1
\end{array}
$ \\ \hline
$3$ & $
\begin{array}{c}
\mathcal{O}_{nBPS,large} \\
SO\left( 1,1\right) \times \frac{SO\left( 5,n\right) }{SO\left( 5,n-1\right)
} \\
\sharp =6
\end{array}
$ & $
\begin{array}{c}
\mathcal{O}_{nBPS,large,II} \\
SO\left( 1,1\right) \times \frac{SO\left( 1,n\right) }{SO\left( 1,n-1\right)
} \\
\sharp =2
\end{array}
$ \\ \hline
$2$ & $
\begin{array}{c}
\mathcal{O}_{\frac{1}{2}-BPS,small,I} \\
SO\left( 1,1\right) \times \frac{SO\left( 5,n\right) }{SO\left( 4,n\right) }
\\
\sharp =n+1
\end{array}
$ & $
\begin{array}{c}
\mathcal{O}_{BPS,small,I} \\
SO\left( 1,1\right) \times T_{n} \\
\sharp =n+1
\end{array}
$ \\ \hline
$2$ & $
\begin{array}{c}
\mathcal{O}_{\frac{1}{4}-BPS,small} \\
SO\left( 1,1\right) \times \frac{SO\left( 5,n\right) }{SO\left( 4,n-1\right)
\rtimes \mathbb{R}^{4,n-1}} \\
\sharp =2
\end{array}
$ & $
\begin{array}{c}
\mathcal{O}_{BPS,small,III} \\
\left[ SO\left( 1,1\right) ^{+}\times \Lambda _{n}^{+}\right] \cup \left[
SO\left( 1,1\right) ^{-}\times \Lambda _{n}^{-}\right] \\
\sharp =2 \\
\updownarrow \ast \\
\mathcal{O}_{nBPS,small,II} \\
\left[ SO\left( 1,1\right) ^{+}\times \Lambda _{n}^{-}\right] \cup \left[
SO\left( 1,1\right) ^{-}\times \Lambda _{n}^{+}\right] \\
\sharp =2
\end{array}
$ \\ \hline
$2$ & $
\begin{array}{c}
\mathcal{O}_{nBPS,small} \\
SO\left( 1,1\right) \times \frac{SO\left( 5,n\right) }{SO\left( 5,n-1\right)
} \\
\sharp =6
\end{array}
$ & $
\begin{array}{c}
\mathcal{O}_{nBPS,small,I} \\
SO\left( 1,1\right) \times \frac{SO\left( 1,n\right) }{SO\left( 1,n-1\right)
} \\
\sharp =2
\end{array}
$ \\ \hline
$1$ & $
\begin{array}{c}
\mathcal{O}_{\frac{1}{2}-BPS,small,II} \\
\frac{SO\left( 5,n\right) }{SO\left( 4,n-1\right) \rtimes \mathbb{R}^{4,n-1}}
\\
\sharp =1
\end{array}
$ & $
\begin{array}{c}
\mathcal{O}_{BPS,small,II} \\
SO\left( 1,1\right) \times \mathbb{R}^{n-1} \\
\sharp =1
\end{array}
$ \\ \hline
$1$ & $
\begin{array}{c}
\mathcal{O}_{\frac{1}{2}-BPS,small,III} \\
SO\left( 1,1\right) \\
\sharp =1
\end{array}
$ & $
\begin{array}{c}
\mathcal{O}_{BPS,small,IV} \\
SO\left( 1,1\right) \\
\sharp =1
\end{array}
$ \\ \hline
\end{tabular}
\end{center}
\caption{\textbf{``\textit{Large}'' (rank}$=3$\textbf{) and ``\textit{small}%
'' (rank}$=1$\textbf{\ and }$2$\textbf{) charge orbits of the repr. }$\left(
\mathbf{1},\mathbf{5+n}\right) $ \textbf{and} $\left( \mathbf{1},\mathbf{1+n}%
\right) $ \textbf{of the }$d=5$\textbf{\ }$U$\textbf{-duality groups }$%
SO\left( 1,1\right) \times SO\left( 5,n\right) $\textbf{\ and }$SO\left(
1,1\right) \times SO\left( 1,n\right) $\textbf{\ of }$\mathcal{N}=4$ \textbf{%
supergravity (based on} $\mathbb{R}\oplus \mathbf{\Gamma }_{5,n}$\textbf{)
and} $\mathcal{N}=2$ \textbf{Jordan symmetric sequence (based on }$\mathbb{R}%
\oplus \mathbf{\Gamma }_{1,n}$\textbf{), respectively. The \textit{rank} }$r$
\textbf{of the orbit is} \textbf{defined as the minimal number of charges
defining a representative solution. ``}$\updownarrow \ast $\textbf{''
denotes the fact the orbits are related through a flip of the sign of }$%
q_{H} $\textbf{. The disconnected timelike hyperboloid }$T_{n}$\textbf{\ and
lightcone }$\Lambda _{n}$\textbf{\ structures are defined by (\ref
{hyperb-def}) and (\ref{lc-def}), respectively. }$\sharp $\textbf{, defined
in (\ref{jjjjj-1}),\ denotes the number of ``non-flat'' scalar degrees of
freedom supported by the charge orbit}}
\end{table}

Thus, summarizing, $\mathcal{N}=4$, $d=5$ supergravity theory admits five
``small'' charge orbits describing the flux configurations supporting
static, spherically symmetric, asymptotically flat ``small'' BHs: one $\frac{%
1}{4}$-BPS, three $\frac{1}{2}$-BPS and one non-BPS. The ``large'' orbits
are two, namely one $\frac{1}{4}$-BPS and one non-BPS (with \r{Z}$_{AB}=0$
at the horizon).\medskip

The relations among the charge orbits of $\mathcal{N}=4$, $d=5$ supergravity
and the charge orbits of $\mathcal{N}=2$, $d=5$ Jordan symmetric sequence
can be determined through the supersymmetry reduction
\begin{equation}
d=5:
\begin{array}{c}
\mathcal{N}=4 \\
\mathbb{R}\oplus \mathbf{\Gamma }_{5,n}
\end{array}
\longrightarrow
\begin{array}{c}
\mathcal{N}=2 \\
\mathbb{R}\oplus \mathbf{\Gamma }_{1,n}
\end{array}
,
\end{equation}
yielding to the results summarized in Table 6.

Finally, it is worth summarizing the results obtained about the number $%
\sharp $ of ``non-flat'' scalar degrees of freedom, within the symmetric RSG
studied in previous Sections. For the ``magic'' supergravities, it holds
\begin{equation}
J_{3}^{\mathbb{A}}:\left\{
\begin{array}{l}
\text{\textit{``large''~(rank}}=3\text{)}:\left\{
\begin{array}{l}
BPS:\sharp =3q+2; \\
nBPS:\sharp =q+2;
\end{array}
\right. \\
\\
\text{\textit{``small'':}}\left\{
\begin{array}{l}
\text{\textit{rank}}=3:\left\{
\begin{array}{l}
BPS:\sharp =q+2; \\
nBPS:\sharp =2;
\end{array}
\right. \\
\\
\text{\textit{rank}}=3:BPS:\sharp =1,
\end{array}
\right.
\end{array}
\right.  \label{magic-n's}
\end{equation}
whereas for $\mathcal{N}=4$ supergravity and $\mathcal{N}=2$ Jordan
symmetric sequence the results are reported in Table 6. As pointed out
above, in the symmetric RSG's under consideration the unique scalar degree
of freedom on which the ADM mass depends along the $1$-charge $\frac{1}{2}$%
-BPS (maximally symmetric) charge orbits can be interpreted as the
Kaluza-Klein radius in the $d=5\rightarrow d=4$ reduction.

\section*{Acknowledgments}

S. F. and A. M. would like to thank L. Borsten and M. Trigiante for
enlightening discussions.

B. L. C. and A. M. would like to thank the CTP of the University of
California, Berkeley, CA USA, where part of this work was done, for kind
hospitality and stimulating environment. A. M. would also like to
acknowledge the warm hospitality and inspiring environment of the Department
of Physics, Theory Unit Group at CERN, Geneva CH.

The work of B. L. C. has been supported in part by the European Commission
under the FP7-PEOPLE-IRG-2008 Grant n PIRG04-GA-2008-239412 \textit{``String
Theory and Noncommutative Geometry''} (\textit{STRING}).

The work of S. F. has been supported in part by the ERC Advanced Grant no.
226455, \textit{``Supersymmetry, Quantum Gravity and Gauge Fields''} (%
\textit{SUPERFIELDS}), and also in part by INFN - Frascati National
Laboratories, and by D.O.E.~grant DE-FG03-91ER40662, Task C.

The work of A. M. has been supported by an INFN visiting Theoretical
Fellowship at SITP, Stanford University, Stanford, CA, USA.

The work of B. Z. ~has been supported in part by the Director, Office of
Science, Office of High Energy and Nuclear Physics, Division of High Energy
Physics of the U.S. Department of Energy under Contract No.
DE-AC02-05CH11231, and in part by NSF grant 10996-13607-44 PHHXM.

\begin{appendix}

\section{\label{Solutions-Constraints-Small-Orbits}Resolution of $G_{5}$%
-invariant Constraints}

In this Appendix, we explicitly solve the $G_{5}$-invariant defining
constraints of ``small'' charge orbits in \textit{``magic''} symmetric RSG,
both in the \textit{``bare''} (Sub-App. \ref{Sol-Constrs-Bare}) and \textit{%
``dressed''} (Sub-App. \ref{Sol-Constrs-Dressed}) charges bases.

\subsection{\label{Sol-Constrs-Bare}\textit{``Bare''} Charges Basis}

Let us start by noticing that for each of the four \textit{``magic''}
symmetric RSG's a unique maximal symmetric embedding into $G_{5}$ exists
containing a factor $SO\left( 1,1\right) $. It reads (recall Eq. (\ref{G_6}%
)) \cite{Gilmore}
\begin{equation}
G_{5}\supsetneq _{\max }G_{6}\times \mathcal{A}_{q}\times SO\left(
1,1\right) ,  \label{bare-starting}
\end{equation}
where the group $\mathcal{A}_{q}$ has been defined in Table 2. Notice that,
in the cases $q=4$ and $2$, $G_{6}\times SO\left( 1,1\right) $ is not
embedded maximally (also considering non-symmetric embeddings \cite{Slansky}%
) into $G_{5}$ itself.

When removing $\mathcal{A}_{q}$ in the cases $q=4$ and $2$ (and thus losing
the maximality), the embedding (\ref{bare-starting}) has a nice
interpretation in terms of truncation of the ``magic'' supergravity to
theories belonging to the Jordan symmetric sequence (\ref
{N=2-d=5-Jordan-Symm}) \cite{FG2}:
\begin{equation}
\begin{array}{ll}
J_{3}^{\mathbb{O}}\supsetneq _{\max }\mathbb{R}\oplus J_{2}^{\mathbb{O}}: &
E_{6\left( -26\right) }\supsetneq _{\max }SO\left( 1,1\right) \times
SO\left( 1,9\right) ; \\
J_{3}^{\mathbb{H}}\supsetneq \mathbb{R}\oplus J_{2}^{\mathbb{H}}: & SU^{\ast
}\left( 6\right) \supsetneq SO\left( 1,1\right) \times SO\left( 1,5\right) ;
\\
J_{3}^{\mathbb{C}}\supsetneq \mathbb{R}\oplus J_{2}^{\mathbb{C}}: & SL\left(
3,\mathbb{C}\right) \supsetneq SO\left( 1,1\right) \times SO\left(
1,3\right) ; \\
J_{3}^{\mathbb{R}}\supsetneq _{\max }\mathbb{R}\oplus J_{2}^{\mathbb{R}}: &
SL\left( 3,\mathbb{R}\right) \supsetneq _{\max }SO\left( 1,1\right) \times
SO\left( 1,2\right) ,
\end{array}
\end{equation}
where it should be recalled that ($q=8,4,2,1$; see \textit{e.g.} \cite
{Gunaydin-rec-rev})
\begin{equation}
J_{2}^{\mathbb{A}}\sim \mathbf{\Gamma }_{1,q+1}.
\end{equation}

\subsubsection{$\mathcal{O}_{lightlike,BPS}$}

In order to solve the ``small''\textit{\ lightlike} $G_{5}$-invariant
defining constraints (\ref{lightlike-bare}) in \textit{``bare''} charges in
a way consistent with an orbit representative having $Z\neq 0$, let us
further embed the $mcs$ of the group in the right-hand side of Eq. (\ref
{bare-starting}), thus obtaining
\begin{equation}
G_{5}\supsetneq _{\max }G_{6}\times \mathcal{A}_{q}\times SO\left(
1,1\right) \overset{mcs}{\supsetneq }SO\left( q+1\right) \times \mathcal{A}%
_{q}.  \label{bare-lightlike-Z<>0}
\end{equation}
Thus, under the \textit{``branching''} (\ref{bare-lightlike-Z<>0}) the
irrepr. $\mathbf{R}_{Q}$ of $G_{5}$ in which the electric charges $q_{i}$'s
sit decomposes as follows:
\begin{eqnarray}
\mathbf{R}_{Q} &\rightarrow &\left( \mathbf{1},\mathbf{1}\right)
_{+4}+\left( \mathbf{q+2},\mathbf{1}\right) _{-2}+\left( \mathbf{Spin}\left(
q+2\right) ,\mathbf{Spin}\left( Q_{q}\right) \right) _{+1}  \notag \\
&\rightarrow &\left( \mathbf{1},\mathbf{1}\right) _{I}+\left( \mathbf{q+1},%
\mathbf{1}\right) +\left( \mathbf{1},\mathbf{1}\right) _{II}+\left( \mathbf{%
Spin}\left( q+1\right) ,\mathbf{Spin}\left( Q_{q}\right) \right) .
\label{decomp-1}
\end{eqnarray}
This in turn entails the ``\textit{branching''}
\begin{equation}
q_{i}\longrightarrow \left( q_{\left( \mathbf{1},\mathbf{1}\right)
_{I}},q_{\left( \mathbf{1},\mathbf{1}\right) _{II}},q_{\left( \mathbf{q+1},%
\mathbf{1}\right) },q_{\left( \mathbf{Spin}\left( q+1\right) ,\mathbf{Spin}%
\left( Q_{q}\right) \right) }\right) .  \label{decomp-1-2}
\end{equation}
In the first line of (\ref{decomp-1}) subscripts denote the weight with
respect to $SO\left( 1,1\right) $, whereas in the second line they just
discriminate between the two singlets of $SO\left( q+1\right) \times
\mathcal{A}_{q}$. Also recall that, as given in Table 2, $\mathcal{A}_{q}$
and $Q_{q}$ are absent for $q=8$ and $q=1$.

Therefore, with respect to $SO\left( q+1\right) \times \mathcal{A}_{q}$, one
obtains:

\begin{itemize}
\item  two singlets (note that $\left( \mathbf{1},\mathbf{1}\right) _{I}$ is
a singlet of $SO\left( q+1,1\right) \times \mathcal{A}_{q}$, as well);

\item  one vector $\left( \mathbf{q+1},\mathbf{1}\right) $;

\item  a (double-)spinor $\left( \mathbf{Spin}\left( q+1\right) ,\mathbf{Spin%
}\left( Q_{q}\right) \right) $.
\end{itemize}

The representation decomposition (\ref{decomp-1}) yields that $d^{ijk}$, the
rank-$3$ completely symmetric $G_{5}$-invariant tensor (namely, the unique
singlet in the tensor product $\left( \mathbf{R}_{Q}\right) ^{3}$)
decomposes in a such way that $\left( \mathbf{1},\mathbf{1}\right) _{II}$
and $\left( \mathbf{q+1},\mathbf{1}\right) $ have the same couplings inside $%
\left( \mathbf{R}_{Q}\right) ^{3}$.

Details concerning the various \textit{``magic''} symmetric RSG's are given
further below.

The position which solves (with maximal - compact - symmetry $SO\left(
q+1\right) \times \mathcal{A}_{q}$) the ``small''\textit{\ lightlike} $G_{5}$%
-invariant defining constraints (\ref{lightlike-bare}) in \textit{``bare''}
charges (and in a way consistent with an orbit representative having $Z\neq
0 $) reads as follows:
\begin{equation}
\left\{
\begin{array}{l}
q_{\left( \mathbf{1},\mathbf{1}\right) _{I}}=0; \\
q_{\left( \mathbf{q+1},\mathbf{1}\right) }=0; \\
q_{\left( \mathbf{Spin}\left( q+1\right) ,\mathbf{Spin}\left( Q_{q}\right)
\right) }=0; \\
q_{\left( \mathbf{1},\mathbf{1}\right) _{II}}\neq 0.
\end{array}
\right.  \label{solving-pos-1}
\end{equation}
Since $SO\left( q+1\right) \times \mathcal{A}_{q}$ is the unique group
maximally (and symmetrically) embedded into $G_{6}\times \mathcal{A}%
_{q}\times SO\left( 1,1\right) $ which has $SO\left( q+1\right) \times
\mathcal{A}_{q}$ as (in this case improper) $mcs$, it follows that $SO\left(
q+1\right) \times \mathcal{A}_{q}$ is also the maximal semi-simple symmetry
of $\mathcal{O}_{lightlike,BPS}$, which is thus given by Eq. (\ref
{magic-orbit-lightlike-Z<>0}).

The origin of the non-semi-simple Abelian (namely, translational) factor $%
\mathbb{R}^{\left( spin\left( q+1\right) ,spin\left( Q_{q}\right) \right) }$
in the stabilizer of $\mathcal{O}_{lightlike,BPS}$ will be explained through
the procedure of suitable \.{I}n\"{o}n\"{u}-Wigner contraction performed in
Sub-App. \ref{IW-Contrs.}.

\subsubsection{$\mathcal{O}_{critical,BPS}$}

Eq. (\ref{bare-lightlike-Z<>0}) and subsequent ones are also relevant for
the resolution of the ``small''\textit{\ critical }$G_{5}$-invariant
defining constraints (\ref{critical-bare}) in \textit{``bare''} charges in a
way consistent with an orbit representative having $Z\neq 0\mathcal{\ }$%
(which is the unique possible case; see treatment above). In this case, the
position which solves (with maximal - non-compact - symmetry $G_{6}\times
\mathcal{A}_{q}$) the constraints (\ref{critical-bare}) in \textit{``bare''}
charges reads as follows:
\begin{equation}
\left\{
\begin{array}{l}
q_{\left( \mathbf{1},\mathbf{1}\right) _{II}}=0; \\
q_{\left( \mathbf{q+1},\mathbf{1}\right) }=0; \\
q_{\left( \mathbf{Spin}\left( q+1\right) ,\mathbf{Spin}\left( Q_{q}\right)
\right) }=0; \\
q_{\left( \mathbf{1},\mathbf{1}\right) _{I}}\neq 0.
\end{array}
\right.  \label{solving-pos-2}
\end{equation}
A\textit{t least} for the relevant values $q=8,4,2,1$ it holds that $%
spin\left( q+2\right) =spin\left( q+1\right) $ (recall definition (\ref
{def-1})). Therefore, since
\begin{equation}
\left.
\begin{array}{r}
q_{\left( \mathbf{1},\mathbf{1}\right) _{II}}=0; \\
q_{\left( \mathbf{q+1},\mathbf{1}\right) }=0;
\end{array}
\right\} \Leftrightarrow q_{\left( \mathbf{q+2},\mathbf{1}\right) }=0,
\end{equation}
it follows that the position (\ref{solving-pos-2}) exhibits maximal -
non-compact - symmetry $G_{6}\times \mathcal{A}_{q}$, which then is the
maximal semi-simple symmetry of $\mathcal{O}_{critical,BPS}$, which is thus
given by Eq. (\ref{magic-orbit-critical-Z<>0}).

The origin of $\mathbb{R}^{\left( spin\left( q+2\right) ,spin\left(
Q_{q}\right) \right) }$ in the stabilizer of $\mathcal{O}_{critical,BPS}$
will be explained through the procedure of suitable $SO\left( 1,1\right) $%
-(three-)grading performed in Sub-App. \ref{SO(1,1)-Three-Grading}.

\subsubsection{$\mathcal{O}_{lightlike,nBPS}$}

In order to solve the ``small''\textit{\ lightlike} $G_{5}$-invariant
defining constraints (\ref{lightlike-bare}) in \textit{``bare''} charges in
a way consistent with an orbit representative having $Z=0$, the embedding (%
\ref{bare-starting}) has to be further elaborated as follows:
\begin{equation}
G_{5}\supsetneq _{\max }G_{6}\times \mathcal{A}_{q}\times SO\left(
1,1\right) \supsetneq _{\max }SO\left( q,1\right) \times \mathcal{A}%
_{q}\times SO\left( 1,1\right) \overset{mcs}{\supsetneq }SO\left( q\right)
\times \mathcal{A}_{q}.  \label{mmll-2}
\end{equation}
Thus, under the \textit{``branching''} (\ref{mmll-2}) the irrepr. $\mathbf{R}%
_{Q}$ decomposes as follows :
\begin{eqnarray}
\mathbf{R}_{Q} &\rightarrow &\left( \mathbf{1},\mathbf{1}\right)
_{+4}+\left( \mathbf{q+2},\mathbf{1}\right) _{-2}+\left( \mathbf{Spin}\left(
q+2\right) ,\mathbf{Spin}\left( Q_{q}\right) \right) _{+1}  \notag \\
&\rightarrow &\left( \mathbf{1},\mathbf{1}\right) _{+4}+\left( \mathbf{q+1},%
\mathbf{1}\right) _{-2}+\left( \mathbf{1},\mathbf{1}\right) _{-2}+\left(
\mathbf{Spin}\left( q+1\right) ,\mathbf{Spin}\left( Q_{q}\right) \right)
_{+1}  \notag \\
&\rightarrow &\left( \mathbf{1},\mathbf{1}\right) _{I}+\left( \mathbf{q},%
\mathbf{1}\right) +\left( \mathbf{1},\mathbf{1}\right) _{III}+\left( \mathbf{%
1},\mathbf{1}\right) _{II}+\left( \mathbf{Spin}^{\prime }\left( q\right) ,%
\mathbf{Spin}\left( Q_{q}\right) \right) +\left( \mathbf{Spin}^{\prime
\prime }\left( q\right) ,\mathbf{Spin}\left( Q_{q}\right) \right) ,  \notag
\\
&&  \label{decomp-2}
\end{eqnarray}
where, besides the obvious irrepr. decompositions determining the last line
of (\ref{decomp-2}), one should recall that
\begin{equation}
\left( \mathbf{Spin}\left( q+1\right) ,\mathbf{Spin}\left( Q_{q}\right)
\right) \longrightarrow \left( \mathbf{Spin}^{\prime }\left( q\right) ,%
\mathbf{Spin}\left( Q_{q}\right) \right) +\left( \mathbf{Spin}^{\prime
\prime }\left( q\right) ,\mathbf{Spin}\left( Q_{q}\right) \right) ,
\end{equation}
where the primes discriminate between the two spinor irreprs. of $SO\left(
q\right) \times \mathcal{A}_{q}$. The \textit{``branching''} of electric
charges corresponding to (\ref{decomp-2}) reads
\begin{equation}
q_{i}\longrightarrow \left( q_{\left( \mathbf{1},\mathbf{1}\right)
_{I}},q_{\left( \mathbf{1},\mathbf{1}\right) _{II}},q_{\left( \mathbf{1},%
\mathbf{1}\right) _{III}},q_{\left( \mathbf{q},\mathbf{1}\right) },q_{\left(
\mathbf{Spin}^{\prime }\left( q\right) ,\mathbf{Spin}\left( Q_{q}\right)
\right) },q_{\left( \mathbf{Spin}^{\prime \prime }\left( q\right) ,\mathbf{%
Spin}\left( Q_{q}\right) \right) }\right) .  \label{decomp-2-2}
\end{equation}
In the first and second line of (\ref{decomp-2}) subscripts denote the
weight with respect to $SO\left( 1,1\right) $, whereas in the third line
they just discriminate between the three singlets of $SO\left( q\right)
\times \mathcal{A}_{q}$.

Therefore, with respect to $SO\left( q\right) \times \mathcal{A}_{q}$, one
obtains:

\begin{itemize}
\item  three singlets (notice that $\left( \mathbf{1},\mathbf{1}\right) _{I}$
is also singlet of $SO\left( q,1\right) \times \mathcal{A}_{q}$ and of $%
G_{6}\times \mathcal{A}_{q}$, and that $\left( \mathbf{1},\mathbf{1}\right)
_{II}$ is singlet of $SO\left( q,1\right) \times \mathcal{A}_{q}$,as well);

\item  a vector $\left( \mathbf{q},\mathbf{1}\right) $;

\item  two (double-)spinors $\left( \mathbf{Spin}^{\prime }\left( q\right) ,%
\mathbf{Spin}\left( Q_{q}\right) \right) $ and $\left( \mathbf{Spin}^{\prime
\prime }\left( q\right) ,\mathbf{Spin}\left( Q_{q}\right) \right) $.
\end{itemize}

As a feature peculiar to (\ref{decomp-2}), the vector $\left( \mathbf{q},%
\mathbf{1}\right) $ and the two (double-)spinors $\left( \mathbf{Spin}%
^{\prime }\left( q\right) ,\mathbf{Spin}\left( Q_{q}\right) \right) $ and $%
\left( \mathbf{Spin}^{\prime \prime }\left( q\right) ,\mathbf{Spin}\left(
Q_{q}\right) \right) $ do exhibit a \textit{``triality symmetry''}, realized
differently depending on $q=8,4,2,1$, as given in Sub-App. \ref
{Sol-Constrs-Bare}.

The representation decomposition (\ref{decomp-2}) yields that $d^{ijk}$
decomposes in such a way that the manifest \textit{``triality''} exhibited
by the \textit{``branching''} of $\mathbf{R}_{Q}$ is removed, and the two
(double-)spinors are put on a different footing with respect to the vector.
As a consequence:

\begin{itemize}
\item  $\left( \mathbf{1},\mathbf{1}\right) _{II}$, $\left( \mathbf{1},%
\mathbf{1}\right) _{III}$ and $\left( \mathbf{q},\mathbf{1}\right) $;

\item  $\left( \mathbf{Spin}^{\prime }\left( q\right) ,\mathbf{Spin}\left(
Q_{q}\right) \right) $ and $\left( \mathbf{Spin}^{\prime \prime }\left(
q\right) ,\mathbf{Spin}\left( Q_{q}\right) \right) $
\end{itemize}

separately have the same couplings inside $\left( \mathbf{R}_{Q}\right) ^{3}$%
.

The position which solves (with maximal - compact - symmetry $SO\left(
q\right) \times \mathcal{A}_{q}$) the ``small''\textit{\ lightlike} $G_{5}$%
-invariant defining constraints (\ref{lightlike-bare}) in \textit{``bare''}
charges (and in a way consistent with an orbit representative having $Z=0$)
reads as follows:
\begin{equation}
\left\{
\begin{array}{l}
q_{\left( \mathbf{q},\mathbf{1}\right) }=0; \\
q_{\left( \mathbf{Spin}^{\prime }\left( q\right) ,\mathbf{Spin}\left(
Q_{q}\right) \right) }=0; \\
q_{\left( \mathbf{Spin}^{\prime \prime }\left( q\right) ,\mathbf{Spin}\left(
Q_{q}\right) \right) }=0,
\end{array}
\right.  \label{solving-pos-3}
\end{equation}
with the three singlets $q_{\left( \mathbf{1},\mathbf{1}\right) _{I}}$, $%
q_{\left( \mathbf{1},\mathbf{1}\right) _{II}}$ and $q_{\left( \mathbf{1},%
\mathbf{1}\right) _{III}}$ constrained by
\begin{equation}
q_{\left( \mathbf{1},\mathbf{1}\right) _{I}}\left[
\begin{array}{l}
d_{\left( \mathbf{1},\mathbf{1}\right) _{I}\left( \mathbf{1},\mathbf{1}%
\right) _{II}\left( \mathbf{1},\mathbf{1}\right) _{II}}q_{\left( \mathbf{1},%
\mathbf{1}\right) _{II}}^{2} \\
+2d_{\left( \mathbf{1},\mathbf{1}\right) _{I}\left( \mathbf{1},\mathbf{1}%
\right) _{II}\left( \mathbf{1},\mathbf{1}\right) _{III}}q_{\left( \mathbf{1},%
\mathbf{1}\right) _{II}}q_{\left( \mathbf{1},\mathbf{1}\right) _{III}} \\
+d_{\left( \mathbf{1},\mathbf{1}\right) _{I}\left( \mathbf{1},\mathbf{1}%
\right) _{III}\left( \mathbf{1},\mathbf{1}\right) _{III}}q_{\left( \mathbf{1}%
,\mathbf{1}\right) _{III}}^{2}
\end{array}
\right] =0.  \label{solving-pos-4}
\end{equation}
Notice that in (\ref{solving-pos-3}) the charges related to the vector and
to the two (double-)spinors are on equal footing, thus exhibiting a \textit{%
``triality symmetry''}, as already mentioned above.

Notice that $SO\left( q,1\right) \times \mathcal{A}_{q}$ is the unique group
which is maximally (if one consider also the factor $SO\left( 1,1\right) $)
and symmetrically embedded into $G_{6}\times \mathcal{A}_{q}\times SO\left(
1,1\right) $, and also which has $SO\left( q\right) \times \mathcal{A}_{q}$
as $mcs$. Therefore, it follows that $SO\left( q,1\right) \times \mathcal{A}%
_{q}$ is also the maximal semi-simple symmetry of $\mathcal{O}%
_{lightlike,nBPS}$, which is thus given by Eq. (\ref
{magic-orbit-lightlike-Z=0}).

As mentioned above, the origin of $\mathbb{R}^{\left( spin\left( q+1\right)
,spin\left( Q_{q}\right) \right) }$ in the stabilizer of $\mathcal{O}%
_{lightlike,nBPS}$ will be explained through the procedure of suitable
\.{I}n\"{o}n\"{u}-Wigner contraction performed in Sub-App. \ref{IW-Contrs.}.

\subsubsection{\label{BPS-Bare-Details}Details}

We now explicit some details of the treatment of symmetric ``magic'' RSG.

We start by giving the explicit form of Eqs. (\ref{bare-lightlike-Z<>0}) and
(\ref{decomp-1}) for all $q=8,4,2,1$ classifying symmetric ``magic'' RSG.

\begin{itemize}
\item  $q=8~\left( J_{3}^{\mathbb{O}}\right) $%
\begin{equation}
\begin{array}{l}
E_{6\left( -26\right) }\supsetneq _{\max }SO\left( 9,1\right) \times
SO\left( 1,1\right) \overset{mcs}{\supsetneq }SO\left( 9\right) ; \\
\\
\mathbf{27}\rightarrow \mathbf{1}_{+4}+\mathbf{10}_{-2}+\mathbf{16}%
_{+1}\rightarrow \mathbf{1}_{I}+\mathbf{9}+\mathbf{1}_{II}+\mathbf{16}.
\end{array}
\end{equation}

\item  $q=4$ $\left( J_{3}^{\mathbb{H}}\right) $ ($SO\left( 5,1\right) \sim
SU^{\ast }\left( 4\right) $, $SO\left( 5\right) \sim USp\left( 4\right) $)
\begin{equation}
\begin{array}{l}
SU^{\ast }\left( 6\right) \supsetneq _{\max }SO\left( 5,1\right) \times
SO\left( 3\right) \times SO\left( 1,1\right) \overset{mcs}{\supsetneq }%
SO\left( 5\right) \times SO\left( 3\right) ; \\
\\
\mathbf{15}\rightarrow \left( \mathbf{1},\mathbf{1}\right) _{+4}+\left(
\mathbf{6},\mathbf{1}\right) _{-2}+\left( \mathbf{4},\mathbf{2}\right)
_{+1}\rightarrow \left( \mathbf{1},\mathbf{1}\right) _{I}+\left( \mathbf{5},%
\mathbf{1}\right) +\left( \mathbf{1},\mathbf{1}\right) _{II}+\left( \mathbf{4%
},\mathbf{2}\right) .
\end{array}
\end{equation}

\item  $q=2$ $\left( J_{3}^{\mathbb{C}}\right) $ ($SL\left( 2,\mathbb{C}%
\right) \sim SO\left( 3,1\right) $, $GL\left( 1,\mathbb{C}\right) \sim
SO\left( 2\right) \times SO\left( 1,1\right) $)
\begin{equation}
\begin{array}{l}
SL\left( 3,\mathbb{C}\right) \supsetneq _{\max }SL\left( 2,\mathbb{C}\right)
\times SL\left( 1,\mathbb{C}\right) \times GL\left( 1,\mathbb{C}\right)
\overset{mcs}{\supsetneq }SO\left( 3\right) \times SO\left( 2\right) ; \\
\\
\mathbf{9}\rightarrow \left( \mathbf{1}_{0}\right) _{+4}+\left( \mathbf{3}%
_{0}+\mathbf{1}_{0}\right) _{-2}+\left( \mathbf{2}_{3}+\overline{\mathbf{2}}%
_{-3}\right) _{+1}\rightarrow \left( \mathbf{1}_{0}\right) _{I}+\mathbf{3}%
_{0}+\left( \mathbf{1}_{0}\right) _{II}+\mathbf{2}_{3}+\mathbf{2}_{-3},
\end{array}
\label{d-q=2}
\end{equation}
where the first subscript in the second step and the subscript in the last
step denote charges w.r.t. $SO\left( 2\right) \sim U\left( 1\right) $, as
well as the second subscript in the second step denotes weights w.r.t. $%
SO\left( 1,1\right) $. In order to derive (\ref{d-q=2}), the decompositions
of the irreprs. of $SL\left( 3,\mathbb{C}\right) $ under $SL\left( 2,\mathbb{%
C}\right) \times SL\left( 1,\mathbb{C}\right) \times GL\left( 1,\mathbb{C}%
\right) \sim SL\left( 2,\mathbb{C}\right) \times SO\left( 2\right) \times
SO\left( 1,1\right) $ have been recalled (the charges and weights w.r.t. $%
SO\left( 2\right) $ and $SO\left( 1,1\right) $ are given):
\begin{eqnarray}
\mathbf{3} &\rightarrow &\left( \mathbf{2},\mathbf{1},-\mathbf{1}\right)
+\left( \mathbf{1},-\mathbf{2},\mathbf{2}\right) ;  \label{dd-1} \\
\overline{\mathbf{3}} &\rightarrow &\left( \overline{\mathbf{2}},-\mathbf{1}%
,-\mathbf{1}\right) +\left( \mathbf{1},\mathbf{2},\mathbf{2}\right) ;
\label{dd-2} \\
\mathbf{3}^{\prime } &\rightarrow &\left( \mathbf{2},-\mathbf{1},\mathbf{1}%
\right) +\left( \mathbf{1},\mathbf{2},-\mathbf{2}\right) ;  \label{dd-3} \\
\overline{\mathbf{3}}^{\prime } &\rightarrow &\left( \overline{\mathbf{2}},%
\mathbf{1},\mathbf{1}\right) +\left( \mathbf{1},-\mathbf{2},-\mathbf{2}%
\right) .  \label{dd-4}
\end{eqnarray}
Thus, through (\ref{dd-1}) and (\ref{dd-2}), the irrepr.
\begin{equation}
\mathbf{R}_{q=2}=\mathbf{9}\equiv \mathbf{3}\times \overline{\mathbf{3}}
\end{equation}
branches as given by (\ref{d-q=2}).

\item  $q=1$ $\left( J_{3}^{\mathbb{R}}\right) $ ($SL\left( 2,\mathbb{R}%
\right) \sim SO\left( 2,1\right) $)
\begin{equation}
\begin{array}{l}
SL\left( 3,\mathbb{R}\right) \supsetneq _{\max }SO\left( 2,1\right) \times
SO\left( 1,1\right) \overset{mcs}{\supsetneq }SO\left( 2\right) ; \\
\\
\mathbf{6}^{\prime }\rightarrow \mathbf{1}_{+4}+\mathbf{3}_{-2}+\mathbf{2}%
_{+1}\rightarrow \mathbf{1}_{I}+\mathbf{2}+\mathbf{1}_{II}+\mathbf{2},
\end{array}
\end{equation}
where the the normalizations and conventions of Table 58 of \cite{Slansky}
have been adopted.\medskip
\end{itemize}

Next, we explicit Eqs. (\ref{mmll-2}) and (\ref{decomp-2}) for all $%
q=8,4,2,1 $ classifying symmetric ``magic'' RSG.

\begin{itemize}
\item  $q=8~\left( J_{3}^{\mathbb{O}}\right) $%
\begin{equation}
\begin{array}{l}
E_{6\left( -26\right) }\supsetneq _{\max }SO\left( 9,1\right) \times
SO\left( 1,1\right) \supsetneq _{\max }SO\left( 8,1\right) \times SO\left(
1,1\right) \overset{mcs}{\supsetneq }SO\left( 8\right) ; \\
\\
\mathbf{27}\rightarrow \mathbf{1}_{+4}+\mathbf{10}_{-2}+\mathbf{16}%
_{+1}\rightarrow \mathbf{1}_{+4}+\mathbf{9}_{-2}+\mathbf{1}_{-2}+\mathbf{16}%
_{+1}\rightarrow \mathbf{1}_{I}+\mathbf{8}_{v}+\mathbf{1}_{III}+\mathbf{1}%
_{II}+\mathbf{8}_{s}+\mathbf{8}_{c}.
\end{array}
\end{equation}
The ``triality'' in irreprs. of $SO\left( q\right) $ is here implemented
through the triality of $\left( \mathbf{8}_{v},\mathbf{8}_{s},\mathbf{8}%
_{c}\right) $ of $SO\left( 8\right) $.

\item  $q=4~\left( J_{3}^{\mathbb{H}}\right) $%
\begin{eqnarray}
&&
\begin{array}{c}
SU^{\ast }\left( 6\right) \\
~ \\
~
\end{array}
\begin{array}{l}
\supsetneq _{\max }SO\left( 5,1\right) \times SO\left( 3\right) \times
SO\left( 1,1\right) \\
\supsetneq _{\max }SO\left( 4,1\right) \times SO\left( 3\right) \times
SO\left( 1,1\right) \\
\overset{mcs}{\supsetneq }SO\left( 4\right) \times SO\left( 3\right) \sim
SU\left( 2\right) \times SU\left( 2\right) \times SU\left( 2\right) ;
\end{array}
\\
&&  \notag
\end{eqnarray}
\begin{eqnarray}
\mathbf{15} &\rightarrow &\left( \mathbf{1},\mathbf{1}\right) _{+4}+\left(
\mathbf{6},\mathbf{1}\right) _{-2}+\left( \mathbf{4},\mathbf{2}\right)
_{+1}\rightarrow \left( \mathbf{1},\mathbf{1}\right) _{+4}+\left( \mathbf{5},%
\mathbf{1}\right) _{-2}+\left( \mathbf{1},\mathbf{1}\right) _{-2}+\left(
\mathbf{4},\mathbf{2}\right) _{+1}\rightarrow  \notag \\
&\rightarrow &\left( \mathbf{1},\mathbf{1},\mathbf{1}\right) _{I}+\left(
\mathbf{2},\mathbf{2},\mathbf{1}\right) +\left( \mathbf{1},\mathbf{1},%
\mathbf{1}\right) _{III}+\left( \mathbf{1},\mathbf{1},\mathbf{1}\right)
_{II}+\left( \mathbf{1},\mathbf{2},\mathbf{2}\right) +\left( \mathbf{2},%
\mathbf{1},\mathbf{2}\right) .
\end{eqnarray}
Thus, the ``triality'' in irreprs. of $SO\left( q\right) \times \mathcal{A}%
_{q}$ is implemented for $q=4$ through the triality of\linebreak $\left(
\left( \mathbf{2},\mathbf{2},\mathbf{1}\right) ,\left( \mathbf{2},\mathbf{1},%
\mathbf{2}\right) ,\left( \mathbf{1},\mathbf{2},\mathbf{2}\right) \right) $
of $SU\left( 2\right) \times SU\left( 2\right) \times SU\left( 2\right) $.

\item  $q=2$ $\left( J_{3}^{\mathbb{C}}\right) $%
\begin{eqnarray}
SL\left( 3,\mathbb{C}\right) &\supsetneq &_{\max }SL\left( 2,\mathbb{C}%
\right) \times SL\left( 1,\mathbb{C}\right) \times GL\left( 1,\mathbb{C}%
\right)  \notag \\
&\supsetneq &_{\max }SO\left( 2,1\right) \times SO\left( 2\right) \times
SO\left( 1,1\right) \overset{mcs}{\supsetneq }SO\left( 2\right) \times
SO\left( 2\right) ;
\end{eqnarray}
\begin{eqnarray}
\mathbf{9} &\rightarrow &\left( \mathbf{1}_{0}\right) _{+4}+\left( \mathbf{3}%
_{0}+\mathbf{1}_{0}\right) _{-2}+\left( \mathbf{2}_{3}+\overline{\mathbf{2}}%
_{-3}\right) _{+1}\rightarrow \left( \mathbf{1}_{0}\right) _{+4}+\left(
\mathbf{3}_{0}\right) _{-2}+\left( \mathbf{1}_{0}\right) _{-2}+\left(
\mathbf{2}_{3}\right) _{+1}+\left( \mathbf{2}_{-3}\right) _{+1}  \notag \\
&\rightarrow &\left( \mathbf{1}_{0}\right) _{I}+\mathbf{2}_{0}+\left(
\mathbf{1}_{0}\right) _{III}+\left( \mathbf{1}_{0}\right) _{II}+\mathbf{2}%
_{3}+\mathbf{2}_{-3}.  \notag \\
&&
\end{eqnarray}
Thus, the triality in irreprs. of $SO\left( q\right) \times \mathcal{A}_{q}$
is implemented for $q=2$ through the triality of $\left( \mathbf{2}_{0},%
\mathbf{2}_{3},\mathbf{2}_{-3}\right) $ of $SO\left( 2\right) \times
SO\left( 2\right) $ (notice the different charges w.r.t. $\mathcal{A}%
_{q=2}=SO\left( 2\right) \sim U\left( 1\right) $).

\item  $q=1$ $\left( J_{3}^{\mathbb{R}}\right) $%
\begin{equation}
\begin{array}{l}
SL\left( 3,\mathbb{R}\right) \supsetneq _{\max }SO\left( 2,1\right) \times
SO\left( 1,1\right) \supsetneq _{\max }SO\left( 1,1\right) \times SO\left(
1,1\right) \overset{mcs}{\supsetneq }1; \\
\\
\mathbf{6}^{\prime }\rightarrow \mathbf{1}_{+4}+\mathbf{3}_{-2}+\mathbf{2}%
_{+1}\rightarrow \mathbf{1}_{+4}+\mathbf{2}_{-2}+\mathbf{1}_{-2}+\mathbf{2}%
_{+1}\rightarrow \mathbf{1}_{I}+\mathbf{1}_{II}+\mathbf{1}_{III}+\mathbf{1}%
_{IV}+\mathbf{1}_{V}+\mathbf{1}_{VI},
\end{array}
\label{ddd-q=1}
\end{equation}
where in the first line $1$ denotes the identity element. Notice that there
is no compact symmetry in $\mathcal{O}_{lightlike,nBPS,J_{3}^{\mathbb{R}%
},d=5}$, as also given by the fact that $\mathcal{M}_{lightlike,nBPS,J_{3}^{%
\mathbb{R}},d=5}=SO\left( 1,1\right) \rtimes \mathbb{R}^{2}$ (see Table 3).
Thus, the ``triality'' of irreprs. of $SO\left( q\right) $ in this case
trivially degenerates into a ``sextality'' (six singlets in the r.h.s. of
the second line of (\ref{ddd-q=1})).
\end{itemize}

\subsection{\label{Sol-Constrs-Dressed}\textit{``Dressed''} Charges Basis}

Concerning the resolution of the $G_{5}$-invariant (sets of) constraints in
the basis of \textit{``dressed''} charges, one should notice that for each
of the four \textit{``magic''} symmetric RSG's a unique non-compact, real
form $\widetilde{H}_{5}$ of the compact group $H_{5}\equiv mcs\left(
G_{5}\right) $ exists with maximal symmetric embedding into $G_{5}$ (see
\textit{e.g.} \cite{Gilmore}; also recall Subsect. \ref{Symmetric-RSG} and
Table 1):
\begin{equation}
G_{5}\supsetneq _{\max }\widetilde{H}_{5}.  \label{dressed-starting}
\end{equation}

\subsubsection{$\mathcal{O}_{lightlike,BPS}$}

In order to solve the ``small''\textit{\ lightlike} $G_{5}$-invariant
defining constraints (\ref{lightlike-bare}) in \textit{``dressed''} charges
in a way consistent with an orbit representative with $Z\neq 0$, let us
further embed
\begin{equation}
\widetilde{h}_{5}\equiv mcs\left( \widetilde{H}_{5}\right) =SO\left(
q+1\right) \times \mathcal{A}_{q},  \label{h_5-tilde}
\end{equation}
thus obtaining
\begin{equation}
G_{5}\left( \supsetneq _{\max }\widetilde{H}_{5}\right) \overset{mcs}{%
\supsetneq }SO\left( q+1\right) \times \mathcal{A}_{q},
\label{dressed-lightlike-Z<>0}
\end{equation}
where the brackets denote the auxiliary nature of the embedding. Thus, under
the \textit{``branching''} (\ref{dressed-lightlike-Z<>0}) $\mathbf{R}_{Q}$
decomposes as follows:
\begin{equation}
\mathbf{R}_{Q}\left( \rightarrow \mathbf{1}+\widehat{\mathbf{R}}\right)
\rightarrow \left( \mathbf{1},\mathbf{1}\right) _{I}+\left( \mathbf{q+1},%
\mathbf{1}\right) +\left( \mathbf{Spin}\left( q+1\right) ,\mathbf{Spin}%
\left( Q_{q}\right) \right) +\left( \mathbf{1},\mathbf{1}\right) _{II},
\label{decomp-3}
\end{equation}
where $\widehat{\mathbf{R}}$ is an irrepr. of $\widetilde{H}_{5}$ used as an
intermediate step. Eq. (\ref{decomp-3}) corresponds to the \textit{%
``branching''}
\begin{equation}
\mathcal{Z}\equiv \left( Z,Z_{x}\right) \longrightarrow \left( Z,Z_{\left(
\mathbf{1},\mathbf{1}\right) _{II}},Z_{\left( \mathbf{q+1},\mathbf{1}\right)
},Z_{\left( \mathbf{Spin}\left( q+1\right) ,\mathbf{Spin}\left( Q_{q}\right)
\right) }\right) ,  \label{decomp-3-2}
\end{equation}
where
\begin{equation}
\mathcal{Z}_{\left( \mathbf{1},\mathbf{1}\right) _{I}}\equiv Z  \label{ZZ}
\end{equation}
throughout. Therefore, with respect to $SO\left( q+1\right) \times \mathcal{A%
}_{q}$, one obtains:

\begin{itemize}
\item  two singlets;

\item  one vector $\left( \mathbf{q+1},\mathbf{1}\right) $;

\item  one (double-)spinor $\left( \mathbf{Spin}\left( q+1\right) ,\mathbf{%
Spin}\left( Q_{q}\right) \right) $.
\end{itemize}

The position which solves (with maximal - compact - symmetry $SO\left(
q+1\right) \times \mathcal{A}_{q}$) the ``small''\textit{\ lightlike} $G_{5}$%
-invariant defining constraints (\ref{lightlike-bare}) in \textit{``dressed''%
} charges (and in a way consistent with an orbit representative having $%
Z\neq 0$) reads as follows:
\begin{equation}
\left\{
\begin{array}{l}
Z_{\left( \mathbf{q+1},\mathbf{1}\right) }=0; \\
Z_{\left( \mathbf{Spin}\left( q+1\right) ,\mathbf{Spin}\left( Q_{q}\right)
\right) }=0,
\end{array}
\right.  \label{dressed-solving-pos-1}
\end{equation}
with $Z$ and $Z_{\left( \mathbf{1},\mathbf{1}\right) _{II}}$ constrained by:
\begin{equation}
Z^{3}-\left( \frac{3}{2}\right) ^{2}ZZ_{\left( \mathbf{1},\mathbf{1}\right)
_{II}}^{2}-\left( \frac{3}{2}\right) ^{\frac{3}{2}}T_{\left( \mathbf{1},%
\mathbf{1}\right) _{II}\left( \mathbf{1},\mathbf{1}\right) _{II}\left(
\mathbf{1},\mathbf{1}\right) _{II}}Z_{\left( \mathbf{1},\mathbf{1}\right)
_{II}}^{3}=0.  \label{dressed-solving-pos-1-1}
\end{equation}
Notice that $SO\left( q+1\right) \times \mathcal{A}_{q}$ is the unique group
which is maximally (and symmetrically) embedded into $\widetilde{H}_{5}$ and
which has $SO\left( q+1\right) \times \mathcal{A}_{q}$ as (in this case
improper) $mcs$ (actually, $SO\left( q+1\right) \times \mathcal{A}%
_{q}=mcs\left( \widetilde{H}_{5}\right) $). Therefore, it follows that $%
SO\left( q+1\right) \times \mathcal{A}_{q}$ is also the maximal semi-simple
symmetry of $\mathcal{O}_{lightlike,BPS}$, which is thus given by Eq. (\ref
{magic-orbit-lightlike-Z<>0}).

The explicit form of Eqs. (\ref{dressed-lightlike-Z<>0}) and (\ref{decomp-3}%
) for all $q=8,4,2,1$ classifying symmetric ``magic'' RSG is given below.

\begin{itemize}
\item  $q=8~\left( J_{3}^{\mathbb{O}}\right) $%
\begin{equation}
\begin{array}{l}
E_{6\left( -26\right) }\left( \supsetneq _{\max }F_{4\left( -20\right)
}\right) \overset{mcs}{\supsetneq }SO\left( 9\right) ; \\
\\
\mathbf{27}\left( \rightarrow \mathbf{1}+\mathbf{26}\right) \rightarrow
\mathbf{1}_{I}+\mathbf{9}+\mathbf{16}+\mathbf{1}_{II}.
\end{array}
\end{equation}

\item  $q=4$ $\left( J_{3}^{\mathbb{H}}\right) $%
\begin{equation}
\begin{array}{l}
SU^{\ast }\left( 6\right) \left( \supsetneq _{\max }USp\left( 4,2\right)
\right) \overset{mcs}{\supsetneq }USp\left( 4\right) \times USp\left(
2\right) \sim SO\left( 5\right) \times SO\left( 3\right) ; \\
\\
\mathbf{15}\left( \rightarrow \mathbf{1}+\mathbf{14}\right) \rightarrow
\left( \mathbf{1},\mathbf{1}\right) _{I}+\left( \mathbf{5},\mathbf{1}\right)
+\left( \mathbf{4},\mathbf{2}\right) +\left( \mathbf{1},\mathbf{1}\right)
_{II}.
\end{array}
\end{equation}

\item  $q=2$ $\left( J_{3}^{\mathbb{C}}\right) $%
\begin{equation}
\begin{array}{l}
SL\left( 3,\mathbb{C}\right) \left( \supsetneq _{\max }SU\left( 2,1\right)
\right) \overset{mcs}{\supsetneq }SU\left( 2\right) \times U\left( 1\right)
\sim SO\left( 3\right) \times SO\left( 2\right) ; \\
\\
\mathbf{9}\left( \rightarrow \mathbf{1}+\mathbf{8}\right) \rightarrow \left(
\mathbf{1}_{0}\right) _{I}+\mathbf{2}_{-3}+\mathbf{2}_{3}+\mathbf{3}%
_{0}+\left( \mathbf{1}_{0}\right) _{II}.
\end{array}
\end{equation}

\item  $q=1$ $\left( J_{3}^{\mathbb{R}}\right) $%
\begin{equation}
\begin{array}{l}
SL\left( 3,\mathbb{R}\right) \left( \supsetneq _{\max }SO\left( 2,1\right)
\right) \overset{mcs}{\supsetneq }SO\left( 2\right) ; \\
\\
\mathbf{6}^{\prime }\left( \rightarrow \mathbf{1}+\mathbf{5}\right)
\rightarrow \mathbf{1}_{I}+\mathbf{2}+\mathbf{2}+\mathbf{1}_{II}.
\end{array}
\end{equation}
\medskip
\end{itemize}

As mentioned in the resolution in the basis of \textit{``bare''} (electric)
charges $q_{i}$'s, the origin of $\mathbb{R}^{\left( spin\left( q+1\right)
,spin\left( Q_{q}\right) \right) }$ in the stabilizer of $\mathcal{O}%
_{lightlike,BPS}$ will be explained through the procedure of suitable
\.{I}n\"{o}n\"{u}-Wigner contraction performed in Sub-App. \ref{IW-Contrs.}.

\subsubsection{$\mathcal{O}_{lightlike,nBPS}$}

In order to solve the ``small''\textit{\ lightlike} $G_{5}$-invariant
defining constraints (\ref{lightlike-bare}) in \textit{``dressed''} charges
in a way consistent with an orbit representative having $Z=0$, the embedding
(\ref{dressed-starting}) has to be further elaborated as follows:
\begin{equation}
G_{5}\left( \supsetneq _{\max }\widetilde{H}_{5}\right) \supsetneq _{\max }%
\widehat{h}_{5}\overset{mcs}{\supsetneq }SO\left( q\right) \times \mathcal{A}%
_{q},  \label{dressed-lightlike-Z=0}
\end{equation}
where
\begin{equation}
\widehat{h}_{5}=SO\left( q,1\right) \times \mathcal{A}_{q}  \label{h_5-hat}
\end{equation}
is the unique non-compact form of $\widetilde{h}_{5}$ (defined by (\ref
{h_5-tilde})) to be embedded maximally and symmetrically into $\widetilde{H}%
_{5}$ (see \textit{e.g.} \cite{Gilmore}).

Thus, under the \textit{``branching''} (\ref{dressed-lightlike-Z=0}) $%
\mathbf{R}_{Q}$ decomposes as follows:
\begin{eqnarray}
&&\mathbf{R}_{Q}\left( \rightarrow \mathbf{1}+\widehat{\mathbf{R}}\right)
\notag \\
&\rightarrow &\left( \mathbf{1},\mathbf{1}\right) _{I}+\left( \mathbf{q+1},%
\mathbf{1}\right) +\left( \mathbf{Spin}\left( q+1\right) ,\mathbf{Spin}%
\left( Q_{q}\right) \right) +\left( \mathbf{1},\mathbf{1}\right) _{II}
\notag \\
&\rightarrow &\left( \mathbf{1},\mathbf{1}\right) _{I}+\left( \mathbf{q},%
\mathbf{1}\right) +\left( \mathbf{1},\mathbf{1}\right) _{III}+\left( \mathbf{%
Spin}^{\prime }\left( q\right) ,\mathbf{Spin}\left( Q_{q}\right) \right)
+\left( \mathbf{Spin}^{\prime \prime }\left( q\right) ,\mathbf{Spin}\left(
Q_{q}\right) \right) +\left( \mathbf{1},\mathbf{1}\right) _{II}.  \notag \\
&&  \label{decomp-4}
\end{eqnarray}
Eq. (\ref{decomp-4}) corresponds to the \textit{``branching''} (recall Eq. (%
\ref{ZZ}))
\begin{equation}
\mathcal{Z}\equiv \left( Z,Z_{x}\right) \longrightarrow \left( Z,Z_{\left(
\mathbf{1},\mathbf{1}\right) _{II}},Z_{\left( \mathbf{1},\mathbf{1}\right)
_{III}},Z_{\left( \mathbf{q},\mathbf{1}\right) },Z_{\left( \mathbf{Spin}%
^{\prime }\left( q\right) ,\mathbf{Spin}\left( Q_{q}\right) \right)
},Z_{\left( \mathbf{Spin}^{\prime \prime }\left( q\right) ,\mathbf{Spin}%
\left( Q_{q}\right) \right) }\right) .  \label{decomp-4-2}
\end{equation}
Therefore, with respect to $SO\left( q\right) \times \mathcal{A}_{q}$,
besides $Z$, one obtains:

\begin{itemize}
\item  two singlets (note that $\left( \mathbf{1},\mathbf{1}\right) _{II}$
is a singlet of $SO\left( q,1\right) \times \mathcal{A}_{q}$, as well);

\item  one vector $\left( \mathbf{q},\mathbf{1}\right) $;

\item  two (double-)spinors $\left( \mathbf{Spin}^{\prime }\left( q\right) ,%
\mathbf{Spin}\left( Q_{q}\right) \right) $ and $\left( \mathbf{Spin}^{\prime
\prime }\left( q\right) ,\mathbf{Spin}\left( Q_{q}\right) \right) $.
\end{itemize}

The position which solves (with maximal - compact - symmetry $SO\left(
q\right) \times \mathcal{A}_{q}$) the ``small''\textit{\ lightlike} $G_{5}$%
-invariant defining constraints (\ref{lightlike-dressed}) in \textit{%
``dressed''} charges (and in a way consistent with an orbit representative
having $Z=0$) reads as follows:
\begin{equation}
\left\{
\begin{array}{l}
Z\equiv Z_{\left( \mathbf{1},\mathbf{1}\right) _{I}}=0; \\
Z_{\left( \mathbf{q},\mathbf{1}\right) }=0; \\
q_{\left( \mathbf{Spin}^{\prime }\left( q\right) ,\mathbf{Spin}\left(
Q_{q}\right) \right) }=0; \\
q_{\left( \mathbf{Spin}^{\prime \prime }\left( q\right) ,\mathbf{Spin}\left(
Q_{q}\right) \right) }=0,
\end{array}
\right.  \label{dressed-solving-pos-2}
\end{equation}
with the two singlets $Z_{\left( \mathbf{1},\mathbf{1}\right) _{II}}$ and $%
Z_{\left( \mathbf{1},\mathbf{1}\right) _{III}}$ constrained by
\begin{equation}
T_{\left( \mathbf{1},\mathbf{1}\right) _{II}\left( \mathbf{1},\mathbf{1}%
\right) _{II}\left( \mathbf{1},\mathbf{1}\right) _{II}}Z_{\left( \mathbf{1},%
\mathbf{1}\right) _{II}}^{2}+3T_{\left( \mathbf{1},\mathbf{1}\right)
_{II}\left( \mathbf{1},\mathbf{1}\right) _{III}\left( \mathbf{1},\mathbf{1}%
\right) _{III}}Z_{\left( \mathbf{1},\mathbf{1}\right) _{III}}^{2}=0.
\label{dressed-solving-pos-2-2}
\end{equation}

Besides $SO\left( q+1\right) \times \mathcal{A}_{q}$, the only other group
which is maximally (and symmetrically) embedded into $\widetilde{H}_{5}$ and
which has $SO\left( q\right) \times \mathcal{A}_{q}$ as $\left( m\right) cs$%
, is $SO\left( q,1\right) \times \mathcal{A}_{q}$. Therefore, $SO\left(
q,1\right) \times \mathcal{A}_{q}$ is also the maximal semi-simple symmetry
of $\mathcal{O}_{lightlike,BPS}$, which is thus given by Eq. (\ref
{magic-orbit-lightlike-Z=0}).

The explicit form of Eqs. (\ref{dressed-lightlike-Z=0})-(\ref{h_5-hat}) and (%
\ref{decomp-4}) for all $q=8,4,2,1$ classifying symmetric ``magic'' RSG is
given below.

\begin{itemize}
\item  $q=8~\left( J_{3}^{\mathbb{O}}\right) $%
\begin{equation}
\begin{array}{l}
E_{6\left( -26\right) }\left( \supsetneq _{\max }F_{4\left( -20\right)
}\right) \supsetneq _{\max }SO\left( 8,1\right) \overset{mcs}{\supsetneq }%
SO\left( 8\right) ; \\
\\
\mathbf{27}\left( \rightarrow \mathbf{1}+\mathbf{26}\right) \rightarrow
\mathbf{1}_{I}+\mathbf{9}+\mathbf{16}+\mathbf{1}_{II}\rightarrow \mathbf{1}%
_{I}+\mathbf{8}_{v}+\mathbf{1}_{III}+\mathbf{1}_{II}+\mathbf{8}_{s}+\mathbf{8%
}_{c}.
\end{array}
\end{equation}

\item  $q=4~\left( J_{3}^{\mathbb{H}}\right) $ ($USp\left( 2,2\right) \sim
SO\left( 5,1\right) $, $USp\left( 2\right) \sim SU\left( 2\right) $)
\begin{equation}
SU^{\ast }\left( 6\right) \left( \supsetneq _{\max }USp\left( 4,2\right)
\right) \supsetneq _{\max }USp\left( 2,2\right) \times USp\left( 2\right)
\overset{mcs}{\supsetneq }USp\left( 2\right) \times USp\left( 2\right)
\times USp\left( 2\right) ;
\end{equation}
\begin{eqnarray}
\mathbf{15}\left( \rightarrow \mathbf{1}+\mathbf{14}\right) &\rightarrow
&\left( \mathbf{1},\mathbf{1}\right) _{I}+\left( \mathbf{5},\mathbf{1}%
\right) +\left( \mathbf{4},\mathbf{2}\right) +\left( \mathbf{1},\mathbf{1}%
\right) _{II}\rightarrow  \notag \\
&\rightarrow &\left( \mathbf{1},\mathbf{1},\mathbf{1}\right) _{I}+\left(
\mathbf{1},\mathbf{1},\mathbf{1}\right) _{III}+\left( \mathbf{2},\mathbf{2},%
\mathbf{1}\right) +\left( \mathbf{2},\mathbf{1},\mathbf{2}\right) +\left(
\mathbf{1},\mathbf{2},\mathbf{2}\right) +\left( \mathbf{1},\mathbf{1},%
\mathbf{1}\right) _{II}.
\end{eqnarray}

\item  $q=2$ $\left( J_{3}^{\mathbb{C}}\right) $%
\begin{equation}
\begin{array}{l}
SL\left( 3,\mathbb{C}\right) \left( \supsetneq _{\max }SU\left( 2,1\right)
\right) \supsetneq _{\max }SU\left( 1,1\right) \times U\left( 1\right)
\overset{mcs}{\supsetneq }U\left( 1\right) \times U\left( 1\right) ; \\
\\
\mathbf{9}\left( \rightarrow \mathbf{1}+\mathbf{8}\right) \rightarrow \left(
\mathbf{1}_{0}\right) _{I}+\mathbf{2}_{3}+\mathbf{2}_{-3}+\mathbf{3}%
_{0}+\left( \mathbf{1}_{0}\right) _{II}\rightarrow \left( \mathbf{1}%
_{0}\right) _{I}+\mathbf{2}_{0}+\mathbf{2}_{3}+\mathbf{2}_{-3}+\left(
\mathbf{1}_{0}\right) _{III}+\left( \mathbf{1}_{0}\right) _{II}.
\end{array}
\end{equation}

\item  $q=1$ $\left( J_{3}^{\mathbb{R}}\right) $%
\begin{equation}
\begin{array}{l}
SL\left( 3,\mathbb{R}\right) \left( \supsetneq _{\max }SO\left( 2,1\right)
\right) \supsetneq _{\max }SO\left( 1,1\right) \overset{mcs}{\supsetneq }1;
\\
\\
\mathbf{6}^{\prime }\left( \rightarrow \mathbf{1}+\mathbf{5}\right)
\rightarrow \mathbf{1}_{I}+\mathbf{2}+\mathbf{1}_{II}+\mathbf{2}\rightarrow
\mathbf{1}_{I}+\mathbf{1}_{II}+\mathbf{1}_{III}+\mathbf{1}_{IV}+\mathbf{1}%
_{V}+\mathbf{1}_{VI},
\end{array}
\label{dddd-q=1}
\end{equation}
where $1$ denotes the identity element.\medskip
\end{itemize}

The origin of $\mathbb{R}^{\left( spin\left( q+1\right) ,spin\left(
Q_{q}\right) \right) }$ in the stabilizer of $\mathcal{O}_{lightlike,BPS}$
will be explained through the procedure of suitable \.{I}n\"{o}n\"{u}-Wigner
contraction performed in Sub-App. \ref{IW-Contrs.}.

\section{\label{Group-Theory-Small-Orbits}Equivalent Derivations}

In this Appendix, we determine the general form of ``small'' charge orbits
of symmetric \textit{``magic''} RSG (see Eqs. (\ref
{magic-orbit-lightlike-Z<>0}), (\ref{magic-orbit-lightlike-Z=0}) and (\ref
{magic-orbit-critical-Z<>0})) through suitable group theoretical procedures,
namely:

\begin{itemize}
\item  \.{I}n\"{o}n\"{u}-Wigner contractions, for ``small'' \textit{lightlike%
} orbits, Sub-App. \ref{IW-Contrs.}.

\item  $SO\left( 1,1\right) $-three-grading, for ``small'' \textit{critical}
orbit, Sub-App. \ref{SO(1,1)-Three-Grading}.
\end{itemize}

Such procedures will clarify the origin of the non-semi-simple Abelian
(namely, translational) factor (recall Eq. (\ref{O-small-gen-struct}),
definitions (\ref{def-1})-(\ref{def-2}), and see Eq. (\ref{T-T-3}) below)
\begin{equation}
\mathcal{T}=\mathbb{R}^{\left( spin\left( q+1\right) ,spin\left(
Q_{q}\right) \right) }
\end{equation}
in all three classes (\textit{lightlike} BPS, \textit{lightlike} non-BPS,
and \textit{critical }BPS) of ``small''\textit{\ }orbits (for each relevant $%
q=8,4,2,1$).

\subsection{\label{IW-Contrs.}\.{I}n\"{o}n\"{u}-Wigner Contractions}

\subsubsection{$\mathcal{O}_{lightlike,BPS}$}

In order to deal with $\mathcal{O}_{lightlike,BPS}$, we start from the group
embedding (\ref{dressed-lightlike-Z<>0}). This determines the following
decompositions of irreprs. ($\mathbf{Adj}$ and $\mathbf{Fund}$ respectively
denoting the adjoint and fundamental irrepr.):
\begin{equation}
\mathbf{Adj}\left( G_{5}\right) \rightarrow \mathbf{Adj}\left( \widetilde{H}%
_{5}\right) +\mathbf{Fund}\left( \widetilde{H}_{5}\right) ,  \label{m1}
\end{equation}
and further
\begin{eqnarray}
\mathbf{Adj}\left( \widetilde{H}_{5}\right) &\rightarrow &\left( \mathbf{Adj}%
\left( SO\left( q+1\right) \right) ,\mathbf{1}\right) +\left( \mathbf{1},%
\mathbf{Adj}\left( \mathcal{A}_{q}\right) \right) +\left( \mathbf{Spin}%
\left( q+1\right) ,\mathbf{Spin}\left( Q_{q}\right) \right) _{I};  \label{m2}
\\
&&  \notag \\
\mathbf{Fund}\left( \widetilde{H}_{5}\right) &\rightarrow &\left( \mathbf{1},%
\mathbf{1}\right) +\left( \mathbf{q+1},\mathbf{1}\right) +\left( \mathbf{Spin%
}\left( q+1\right) ,\mathbf{Spin}\left( Q_{q}\right) \right) _{II},
\label{m3}
\end{eqnarray}
where trivially $\mathbf{Adj}\left( SO\left( q+1\right) \right) =\frac{%
\mathbf{q}\left( \mathbf{q+1}\right) }{\mathbf{2}}$. Eqs. (\ref{m1})-(\ref
{m3}) thus imply
\begin{eqnarray}
\mathbf{Adj}\left( G_{5}\right) &\rightarrow &\left( \mathbf{Spin}\left(
q+1\right) ,\mathbf{Spin}\left( Q_{q}\right) \right) _{I}+  \notag \\
&&+\left( \mathbf{Adj}\left( SO\left( q+1\right) \right) ,\mathbf{1}\right)
+\left( 1,\mathbf{Adj}\left( \mathcal{A}_{q}\right) \right) +  \notag \\
&&+\left( \mathbf{1},\mathbf{1}\right) +\left( \mathbf{q+1},\mathbf{1}%
\right) +\left( \mathbf{Spin}\left( q+1\right) ,\mathbf{Spin}\left(
Q_{q}\right) \right) _{II}.  \label{m4}
\end{eqnarray}

The decomposition of the branching (\ref{m2}) yields to
\begin{equation}
\underset{\frak{g}_{\widetilde{H}_{5}}}{\underbrace{\mathbf{Adj}\left(
\widetilde{H}_{5}\right) }}~\rightarrow ~\underset{\frak{h}_{\widetilde{H}%
_{5}}}{\underbrace{\left( \mathbf{Adj}\left( SO\left( q+1\right) \right) ,%
\mathbf{1}\right) +\left( 1,\mathbf{Adj}\left( \mathcal{A}_{q}\right)
\right) }}\underset{
\begin{array}{c}
\oplus _{s}
\end{array}
}{+}\underset{\frak{k}_{\widetilde{H}_{5}}}{\underbrace{\left( \mathbf{Spin}%
\left( q+1\right) ,\mathbf{Spin}\left( Q_{q}\right) \right) _{I}}}.
\label{Cartan}
\end{equation}
The coset (recall Eq. (\ref{nBPS-Z<>0-large-moduli-space}))
\begin{equation}
\frac{\widetilde{H}_{5}}{mcs\left( \widetilde{H}_{5}\right) }=\frac{%
\widetilde{H}_{5}}{SO\left( q+1\right) \times \mathcal{A}_{q}}=\mathcal{M}%
_{nBPS,large}  \label{nnn-1}
\end{equation}
is symmetric, with real dimension, Euclidean signature and character
respectively (see \textit{e.g.} \cite{Gilmore,Helgason}; here ``$c$'' and ``$%
nc$'' respectively stand for ``\textit{compact}'' and ``\textit{non-compact}%
''):
\begin{equation}
\begin{array}{l}
dim_{\mathbb{R}}=2q; \\
\left( c,nc\right) =\left( 0,2q\right) ; \\
\chi \equiv c-nc=-2q.
\end{array}
\label{nnn-2}
\end{equation}
By definition, the symmetricity of $\mathcal{M}_{nBPS,large}$ implies that
\begin{equation}
\begin{array}{l}
\left[ \frak{h}_{\widetilde{H}_{5}},\frak{h}_{\widetilde{H}_{5}}\right] =%
\frak{h}_{\widetilde{H}_{5}}; \\
\left[ \frak{h}_{\widetilde{H}_{5}},\frak{k}_{\widetilde{H}_{5}}\right] =%
\frak{k}_{\widetilde{H}_{5}}; \\
\left[ \frak{k}_{\widetilde{H}_{5}},\frak{k}_{\widetilde{H}_{5}}\right] =%
\frak{h}_{\widetilde{H}_{5}}.
\end{array}
\label{simmetricity-rels}
\end{equation}
The \textit{``decoupling''} of $\frak{h}_{\widetilde{H}_{5}}$, with
subsequent transformation of the irrepr. $\left( \mathbf{Spin}\left(
q+1\right) ,\mathbf{Spin}\left( Q_{q}\right) \right) _{I}$ of $SO\left(
q+1\right) \times \mathcal{A}_{q}$ into the non-semi-simple, Abelian
(namely, translational) part of the stabilizer of $\mathcal{O}%
_{lightlike,BPS}$ is achieved by performing a uniform rescaling of the
generators of $\frak{k}_{\widetilde{H}_{5}}$:
\begin{equation}
\frak{k}_{\widetilde{H}_{5}}\longrightarrow \lambda \frak{k}_{\widetilde{H}%
_{5}},~\lambda \in \mathbb{R}_{0}^{+},  \label{IW-c-1}
\end{equation}
and then by letting $\lambda \rightarrow \infty $. This amounts to
performing an \.{I}n\"{o}n\"{u}-Wigner (IW) contraction \cite{IW-1,IW-2} on $%
\frak{k}_{\widetilde{H}_{5}}$. Thus (recall Eqs. (\ref
{magic-orbit-lightlike-Z<>0}) and (\ref{rrr-1})):
\begin{eqnarray}
IW\left( \mathcal{O}_{nBPS,large}=\frac{G_{5}}{\widetilde{H}_{5}}\right)
\overset{\left( \ref{dressed-lightlike-Z<>0}\right) }{\longrightarrow }%
\mathcal{O}_{lightlike,BPS} &=&\frac{G_{5}}{\left( SO\left( q+1\right)
\times \mathcal{A}_{q}\right) \rtimes \mathbb{R}^{\left( spin\left(
q+1\right) ,spin\left( Q_{q}\right) \right) }};  \notag \\
&&  \label{IW-c-2} \\
\mathcal{T}_{lightlike,BPS} &\equiv &\mathbb{R}^{\left( spin\left(
q+1\right) ,spin\left( Q_{q}\right) \right) }.  \label{T-1}
\end{eqnarray}

Thus, $\mathcal{T}_{lightlike,BPS}$ given by (\ref{T-1}) is the $\frak{k}_{%
\widetilde{H}_{5}}$-part of the decomposition (\ref{Cartan}) of the Lie
algebra $\frak{g}_{\widetilde{H}_{5}}$ of $\widetilde{H}_{5}$ with respect
to $mcs\left( \widetilde{H}_{5}\right) =SO\left( q+1\right) \times \mathcal{A%
}_{q}$, which then gets \textit{``decoupled''} from $\frak{g}_{\widetilde{H}%
_{5}}$ and Abelianized through the IW contraction procedure (\ref{IW-c-1})-(%
\ref{IW-c-2}).

\subsubsection{$\mathcal{O}_{lightlike,nBPS}$}

On the other hand, the treatment of $\mathcal{O}_{lightlike,nBPS}$ requires
to start from the embedding (\ref{dressed-lightlike-Z=0}) (actually, without
the last step involving $SO\left( q\right) \times \mathcal{A}_{q}=mcs\left(
\widehat{h}_{5}\right) $; recall Eq. (\ref{h_5-hat})):
\begin{equation}
G_{5}\supsetneq _{\max }\widetilde{H}_{5}\supsetneq _{\max }\widehat{h}%
_{5}=SO\left( q,1\right) \times \mathcal{A}_{q}.
\end{equation}
The subsequent decompositions of $\mathbf{Adj}\left( G_{5}\right) $, $%
\mathbf{Adj}\left( \widetilde{H}_{5}\right) $ and $\mathbf{Fund}\left(
\widetilde{H}_{5}\right) $ are given by Eqs. (\ref{m1}), (\ref{m2}) and (\ref
{m3}), respectively, thus yielding the same decomposition as in (\ref{m4}).
Consequently, the decomposition of the branching (\ref{m2}) yields the same
result as in (\ref{Cartan}).

The coset (recall Eq. (\ref{nBPS-Z<>0-large-moduli-space}))
\begin{equation}
\frac{\widetilde{H}_{5}}{\widehat{h}_{5}}=\frac{\widetilde{H}_{5}}{SO\left(
q,1\right) \times \mathcal{A}_{q}}~  \label{nnn-3}
\end{equation}
is symmetric, with real dimension, Euclidean signature and character
respectively:
\begin{equation}
\begin{array}{l}
dim_{\mathbb{R}}=2q; \\
\left( c,nc\right) =\left( q,q\right) ; \\
\chi \equiv c-nc=0.
\end{array}
\label{nnn-4}
\end{equation}
By definition, the symmetricity of $\frac{\widetilde{H}_{5}}{\widehat{h}_{5}}
$ implies the same relations as in (\ref{simmetricity-rels}).

Thus, the \textit{``decoupling''} of $\frak{h}_{\widetilde{H}_{5}}$, with
subsequent transformation of the irrepr. \linebreak $\left( \mathbf{Spin}%
\left( q+1\right) ,\mathbf{Spin}\left( Q_{q}\right) \right) _{I}$ of $%
SO\left( q,1\right) \times \mathcal{A}_{q}$ into the non-semi-simple,
Abelian (namely, translational) part of the stabilizer of $\mathcal{O}%
_{lightlike,nBPS}$ is achieved by performing a uniform rescaling of the
generators of $\frak{k}_{\widetilde{H}_{5}}$ as given by Eq. (\ref{IW-c-1}),
and then by letting $\lambda \rightarrow \infty $. This amounts to
performing an IW contraction \cite{IW-1,IW-2} on $\frak{k}_{\widetilde{H}%
_{5}}$. Therefore, one obtains (recall Eqs. (\ref{magic-orbit-lightlike-Z=0}%
) and (\ref{rrr-2})):
\begin{eqnarray}
IW\left( \mathcal{O}_{nBPS,large}\right) \overset{\left( \ref
{dressed-lightlike-Z=0}\right) }{\longrightarrow }\mathcal{O}%
_{lightlike,nBPS} &=&\frac{G_{5}}{\left( SO\left( q,1\right) \times \mathcal{%
A}_{q}\right) \rtimes \mathbb{R}^{\left( spin\left( q+1\right) ,spin\left(
Q_{q}\right) \right) }};  \notag \\
&&  \label{IW-c-3} \\
\mathcal{T}_{lightlike,nBPS} &=&\mathcal{T}_{lightlike,BPS}=\mathbb{R}%
^{\left( spin\left( q+1\right) ,spin\left( Q_{q}\right) \right) }.
\label{T-2}
\end{eqnarray}

Thus, $\mathcal{T}_{lightlike,nBPS}$ given by (\ref{T-2}) is the $\frak{k}_{%
\widetilde{H}_{5}}$-part of the decomposition (\ref{Cartan}) of the Lie
algebra $\frak{g}_{\widetilde{H}_{5}}$ of $\widetilde{H}_{5}$ with respect
to $\widehat{h}_{5}=SO\left( q,1\right) \times \mathcal{A}_{q}$, which then
gets \textit{``decoupled''} from $\frak{g}_{\widetilde{H}_{5}}$ and
Abelianized through the IW contraction procedure (see Eqs. (\ref{IW-c-1})
and (\ref{IW-c-3})).\medskip

Note that the IW contraction does not change the dimension of the starting
orbit. Indeed $\mathcal{O}_{lightlike,BPS}$, obtained through the IW
contraction of $\mathcal{O}_{nBPS,large}$ along the branching (\ref
{dressed-lightlike-Z<>0}), has the same real dimension of $\mathcal{O}%
_{nBPS,large}$ itself. Analogously, also $\mathcal{O}_{lightlike,nBPS}$,
obtained through the IW contraction of $\mathcal{O}_{nBPS,large}$ along the
branching (\ref{dressed-lightlike-Z=0}), has the same real dimension of $%
\mathcal{O}_{nBPS,large}$ itself.

\subsubsection{\label{IW-Details}Details}

Below, we explicit in order, besides (\ref{m1})-(\ref{m3}), the relevant
formul\ae\ of the derivations given above, namely Eqs. (\ref{nnn-1}), (\ref
{nnn-2}), (\ref{IW-c-2}), and (\ref{nnn-3}), (\ref{nnn-4}), (\ref{IW-c-3}),
for all $q=8,4,2,1$ classifying symmetric ``magic'' RSG.

\begin{itemize}
\item  $q=8~\left( J_{3}^{\mathbb{O}}\right) $%
\begin{equation}
\begin{array}{l}
\mathbf{78}\rightarrow \mathbf{26}+\mathbf{52}; \\
\mathbf{52}\rightarrow \mathbf{36}+\mathbf{16}_{I}; \\
\mathbf{26}\rightarrow \mathbf{1}+\mathbf{9}+\mathbf{16}_{II};
\end{array}
\end{equation}
\begin{eqnarray}
&&
\begin{array}{l}
\frac{\widetilde{H}_{5}}{mcs\left( \widetilde{H}_{5}\right) }=\left. \frac{%
\widetilde{H}_{5}}{SO\left( q+1\right) \times \mathcal{A}_{q}}\right| _{q=8}=%
\mathcal{M}_{nBPS,large,J_{3}^{\mathbb{O}},d=5}=\frac{F_{4\left( -20\right) }%
}{SO\left( 9\right) }; \\
\\
dim_{\mathbb{R}}=16;~\left( c,nc\right) =\left( 0,16\right) ;~\chi =-16; \\
\\
IW\left( \mathcal{O}_{nBPS,large,J_{3}^{\mathbb{O}}}=\frac{E_{6\left(
-26\right) }}{F_{4\left( -20\right) }}\right) \overset{\left( \ref
{dressed-lightlike-Z<>0}\right) }{\longrightarrow }\mathcal{O}%
_{lightlike,BPS,J_{3}^{\mathbb{O}}}=\frac{E_{6\left( -26\right) }}{SO\left(
9\right) \rtimes \mathbb{R}^{16}};
\end{array}
\\
&&  \notag \\
&&  \notag \\
&&
\begin{array}{l}
\frac{\widetilde{H}_{5}}{\widehat{h}_{5}}=\left. \frac{\widetilde{H}_{5}}{%
SO\left( q,1\right) \times \mathcal{A}_{q}}\right| _{q=8}~=\frac{F_{4\left(
-20\right) }}{SO\left( 8,1\right) }; \\
\\
dim_{\mathbb{R}}=16;~\left( c,nc\right) =\left( 8,8\right) ;~\chi =0; \\
\\
IW\left( \mathcal{O}_{nBPS,large,J_{3}^{\mathbb{O}}}\right) \overset{\left(
\ref{dressed-lightlike-Z=0}\right) }{\longrightarrow }\mathcal{O}%
_{lightlike,nBPS,J_{3}^{\mathbb{O}}}=\frac{E_{6\left( -26\right) }}{SO\left(
8,1\right) \rtimes \mathbb{R}^{16}}.
\end{array}
\end{eqnarray}

\item  $q=4~\left( J_{3}^{\mathbb{H}}\right) $%
\begin{equation}
\begin{array}{l}
\mathbf{35}\rightarrow \mathbf{14}+\mathbf{21}; \\
\mathbf{21}\rightarrow \left( \mathbf{4},\mathbf{2}\right) _{I}+\left(
\mathbf{10},\mathbf{1}\right) +\left( \mathbf{1},\mathbf{3}\right) ; \\
\mathbf{14}\rightarrow \left( \mathbf{1},\mathbf{1}\right) +\left( \mathbf{5}%
,\mathbf{1}\right) +\left( \mathbf{4},\mathbf{2}\right) _{II};
\end{array}
\end{equation}
\begin{eqnarray}
&&
\begin{array}{l}
\frac{\widetilde{H}_{5}}{mcs\left( \widetilde{H}_{5}\right) }=\left. \frac{%
\widetilde{H}_{5}}{SO\left( q+1\right) \times \mathcal{A}_{q}}\right| _{q=4}=%
\mathcal{M}_{nBPS,large,J_{3}^{\mathbb{H}},d=5}=\frac{USp\left( 4,2\right) }{%
USp\left( 4\right) \times USp\left( 2\right) }; \\
\\
dim_{\mathbb{R}}=8;~\left( c,nc\right) =\left( 0,8\right) ;~\chi =-8; \\
\\
IW\left( \mathcal{O}_{nBPS,large,J_{3}^{\mathbb{H}}}=\frac{SU^{\ast }\left(
6\right) }{USp\left( 4,2\right) }\right) \overset{\left( \ref
{dressed-lightlike-Z<>0}\right) }{\longrightarrow }\mathcal{O}%
_{lightlike,BPS,J_{3}^{\mathbb{H}}}=\frac{SU^{\ast }\left( 6\right) }{\left(
SO\left( 5\right) \times SO\left( 3\right) \right) \rtimes \mathbb{R}%
^{\left( 4,2\right) }};
\end{array}
\\
&&  \notag \\
&&  \notag \\
&&
\begin{array}{l}
\frac{\widetilde{H}_{5}}{\widehat{h}_{5}}=\left. \frac{\widetilde{H}_{5}}{%
SO\left( q,1\right) \times \mathcal{A}_{q}}\right| _{q=4}~=\frac{USp\left(
4,2\right) }{USp\left( 2,2\right) \times USp\left( 2\right) }; \\
\\
dim_{\mathbb{R}}=8;~\left( c,nc\right) =\left( 4,4\right) ;~\chi =0; \\
\\
IW\left( \mathcal{O}_{nBPS,large,J_{3}^{\mathbb{H}}}\right) \overset{\left(
\ref{dressed-lightlike-Z=0}\right) }{\longrightarrow }\mathcal{O}%
_{lightlike,nBPS,J_{3}^{\mathbb{H}}}=\frac{SU^{\ast }\left( 6\right) }{%
\left( SO\left( 4,1\right) \times SO\left( 3\right) \right) \rtimes \mathbb{R%
}^{\left( 4,2\right) }}.
\end{array}
\end{eqnarray}

\item  $q=2$ $\left( J_{3}^{\mathbb{C}}\right) $. Notice that in this case
Eq. (\ref{m1}) gets modified into
\begin{equation}
\begin{array}{l}
\mathbf{Adj}\left( G_{5}\right) \rightarrow \mathbf{Adj}\left( \widetilde{H}%
_{5}\right) +\mathbf{Adj}\left( \widetilde{H}_{5}\right) ; \\
\\
\mathbf{16}\rightarrow \mathbf{8}+\mathbf{8}; \\
\mathbf{8}\rightarrow \mathbf{3}_{0}+\mathbf{1}_{0}+\mathbf{2}_{3}+\mathbf{2}%
_{-3}.
\end{array}
\end{equation}
Everything fits also because for $q=2$ it holds that
\begin{equation}
\begin{array}{l}
\left( \mathbf{q+1},\mathbf{1}\right) =\left( \mathbf{Adj}\left( SO\left(
q+1\right) \right) ,\mathbf{1}\right) =\mathbf{3}_{0}; \\
\\
\left( \mathbf{1},\mathbf{Adj}\left( \mathcal{A}_{q}\right) \right) =\left(
\mathbf{1},\mathbf{1}\right) =\mathbf{1}_{0}.
\end{array}
\end{equation}
\begin{eqnarray}
&&
\begin{array}{l}
\frac{\widetilde{H}_{5}}{mcs\left( \widetilde{H}_{5}\right) }=\left. \frac{%
\widetilde{H}_{5}}{SO\left( q+1\right) \times \mathcal{A}_{q}}\right| _{q=2}=%
\mathcal{M}_{nBPS,large,J_{3}^{\mathbb{C}},d=5}=\frac{SU\left( 2,1\right) }{%
SU\left( 2\right) \times U\left( 1\right) }; \\
\\
dim_{\mathbb{R}}=4;~\left( c,nc\right) =\left( 0,4\right) ;~\chi =-4; \\
\\
IW\left( \mathcal{O}_{nBPS,large,J_{3}^{\mathbb{C}}}=\frac{SL\left( 3,%
\mathbb{C}\right) }{SU\left( 2,1\right) }\right) \overset{\left( \ref
{dressed-lightlike-Z<>0}\right) }{\longrightarrow }\mathcal{O}%
_{lightlike,BPS,J_{3}^{\mathbb{C}}}=\frac{SL\left( 3,\mathbb{C}\right) }{%
\left( SO\left( 3\right) \times SO\left( 2\right) \right) \rtimes \mathbb{R}%
^{\left( 2,2\right) }}.
\end{array}
\\
&&  \notag \\
&&  \notag \\
&&
\begin{array}{l}
\frac{\widetilde{H}_{5}}{\widehat{h}_{5}}=\left. \frac{\widetilde{H}_{5}}{%
SO\left( q,1\right) \times \mathcal{A}_{q}}\right| _{q=2}~=\frac{SU\left(
2,1\right) }{SU\left( 1,1\right) \times U\left( 1\right) }; \\
\\
dim_{\mathbb{R}}=4;~\left( c,nc\right) =\left( 2,2\right) ;~\chi =0; \\
\\
IW\left( \mathcal{O}_{nBPS,large,J_{3}^{\mathbb{C}}}\right) \overset{\left(
\ref{dressed-lightlike-Z=0}\right) }{\longrightarrow }\mathcal{O}%
_{lightlike,nBPS,J_{3}^{\mathbb{C}}}=\frac{SL\left( 3,\mathbb{C}\right) }{%
\left( SO\left( 2,1\right) \times SO\left( 2\right) \right) \rtimes \mathbb{R%
}^{\left( 2,2\right) }}.
\end{array}
\end{eqnarray}

\item  $q=1$ $\left( J_{3}^{\mathbb{R}}\right) $. Notice that in this case
Eq. (\ref{m1}) gets modified into
\begin{equation}
\begin{array}{l}
\mathbf{Adj}\left( G_{5}\right) \rightarrow \mathbf{Adj}\left( \widetilde{H}%
_{5}\right) +\mathbf{Spin}_{s=2}\left( \widetilde{H}_{5}\right) ; \\
\\
\mathbf{8}\rightarrow \mathbf{3}+\mathbf{5}; \\
\mathbf{3}\rightarrow \mathbf{1}_{II}+\mathbf{2}_{I}; \\
\mathbf{5}\rightarrow \mathbf{1}_{I}+\mathbf{2}_{III}+\mathbf{2}_{II};
\end{array}
\end{equation}
Everything fits also because for $q=1$ it holds that
\begin{equation}
\begin{array}{l}
\left( \mathbf{q+1},\mathbf{1}\right) =\left( \mathbf{Adj}\left( SO\left(
q+1\right) \right) ,\mathbf{1}\right) =\mathbf{2}; \\
\\
\left( \mathbf{1},\mathbf{Adj}\left( \mathcal{A}_{q}\right) \right) =\left(
\mathbf{1},\mathbf{1}\right) =\mathbf{1}.
\end{array}
\end{equation}
\begin{eqnarray}
&&
\begin{array}{l}
\frac{\widetilde{H}_{5}}{mcs\left( \widetilde{H}_{5}\right) }=\left. \frac{%
\widetilde{H}_{5}}{SO\left( q+1\right) \times \mathcal{A}_{q}}\right| _{q=1}=%
\mathcal{M}_{nBPS,large,J_{3}^{\mathbb{R}},d=5}=\frac{SO\left( 2,1\right) }{%
SO\left( 2\right) }\sim \frac{SU\left( 1,1\right) }{U\left( 1\right) }; \\
\\
dim_{\mathbb{R}}=2;~\left( c,nc\right) =\left( 0,2\right) ;~\chi =-2; \\
\\
IW\left( \mathcal{O}_{nBPS,large,J_{3}^{\mathbb{R}}}=\frac{SL\left( 3,%
\mathbb{R}\right) }{SO\left( 2,1\right) }\right) \overset{\left( \ref
{dressed-lightlike-Z<>0}\right) }{\longrightarrow }\mathcal{O}%
_{lightlike,BPS,J_{3}^{\mathbb{R}}}=\frac{SL\left( 3,\mathbb{R}\right) }{%
SO\left( 2\right) \rtimes \mathbb{R}^{2}}.
\end{array}
\\
&&  \notag \\
&&  \notag \\
&&
\begin{array}{l}
\frac{\widetilde{H}_{5}}{\widehat{h}_{5}}=\left. \frac{\widetilde{H}_{5}}{%
SO\left( q,1\right) \times \mathcal{A}_{q}}\right| _{q=1}~=\frac{SO\left(
2,1\right) }{SO\left( 1,1\right) }; \\
\\
dim_{\mathbb{R}}=2;~\left( c,nc\right) =\left( 1,1\right) ;~\chi =0; \\
\\
IW\left( \mathcal{O}_{nBPS,large,J_{3}^{\mathbb{R}}}\right) \overset{\left(
\ref{dressed-lightlike-Z=0}\right) }{\longrightarrow }\mathcal{O}%
_{lightlike,nBPS,J_{3}^{\mathbb{R}}}=\frac{SL\left( 3,\mathbb{R}\right) }{%
\left( SO\left( 1,1\right) \right) \rtimes \mathbb{R}^{2}}.
\end{array}
\end{eqnarray}
\end{itemize}

\subsection{\label{SO(1,1)-Three-Grading}$SO\left( 1,1\right) $-Three
Grading and $\mathcal{O}_{critical,BPS}$}

In order to deal with $\mathcal{O}_{critical,BPS}$, we start from the group
embedding (\ref{bare-starting}). As pointed out above, this is the unique
maximal embedding (\textit{at least} among the symmetric ones; see \textit{%
e.g.} \cite{Gilmore}) into $G_{5}$ to exhibit a commuting factor $SO\left(
1,1\right) $.

Therefore, the Lie algebra $\frak{g}_{G_{5}}$ of $G_{5}$ admits a
three-grading with respect to the Lie algebra $\mathbb{R}$ of $SO\left(
1,1\right) $ as follows:
\begin{equation}
\frak{g}_{G_{5}}=\mathcal{W}^{+3}\oplus _{s}\mathcal{W}^{0}\oplus _{s}%
\mathcal{W}^{-3},  \label{tg-2}
\end{equation}
where as above the subscripts denote the weights with respect to $SO\left(
1,1\right) $ itself. At the level of \textit{``branching''} of $\mathbf{Adj}%
\left( G_{5}\right) $, the $SO\left( 1,1\right) $-three-grading reads as
follows:
\begin{eqnarray}
\mathbf{Adj}\left( G_{5}\right) &\rightarrow &\left( \mathbf{1},\mathbf{1}%
\right) _{0}+\left( \mathbf{Adj}\left( G_{6}\right) ,\mathbf{1}\right)
_{0}+\left( \mathbf{1},\mathbf{Adj}\left( \mathcal{A}_{q}\right) \right)
_{0}+  \notag \\
&&+\left( \mathbf{Spin}\left( q+2\right) ,\mathbf{Spin}\left( Q_{q}\right)
\right) _{-3}+  \notag \\
&&+\left( \mathbf{Spin}^{\prime }\left( q+2\right) ,\mathbf{Spin}\left(
Q_{q}\right) \right) _{+3}.  \label{tg-1}
\end{eqnarray}
Thus, the decomposition (\ref{tg-1}) yields the following identification of
the graded terms in (\ref{tg-2}):
\begin{equation}
\begin{array}{c}
\mathcal{W}^{0}\equiv \\
~ \\
~
\end{array}
\begin{array}{c}
\left( \mathbf{1},\mathbf{1}\right) _{0} \\
\downarrow \text{{\tiny exp}} \\
SO(1,1)
\end{array}
\begin{array}{c}
+\left( \mathbf{Adj}\left( G_{6}\right) ,\mathbf{1}\right) _{0} \\
\downarrow \text{{\tiny exp}} \\
G_{6}
\end{array}
\begin{array}{c}
+\left( \mathbf{1},\mathbf{Adj}\left( \mathcal{A}_{q}\right) \right) _{0};
\\
\downarrow \text{{\tiny exp}} \\
\mathcal{A}_{q}
\end{array}
\label{tgg-1}
\end{equation}
\begin{eqnarray}
\mathcal{W}^{+3} &\equiv &\left( \mathbf{Spin}^{\prime }\left( q+2\right) ,%
\mathbf{Spin}\left( Q_{q}\right) \right) _{+3};  \label{tgg-2} \\
\mathcal{W}^{-3} &\equiv &\left( \mathbf{Spin}\left( q+2\right) ,\mathbf{Spin%
}\left( Q_{q}\right) \right) _{-3},  \label{tgg-3}
\end{eqnarray}
with ``$exp$'' denoting the exponential mapping.

Thus, $\mathcal{O}_{critical,BPS}$ is obtained by cosetting $G_{5}$ with the
$+3$ (or equivalently $-3$)-graded extension of $\mathcal{W}^{0}-\left(
\mathbf{1},\mathbf{1}\right) _{0}$, namely:
\begin{equation}
\mathcal{O}_{critical,BPS}=\frac{G_{5}}{\mathcal{N}_{+3(-3)}},  \label{eve-1}
\end{equation}
where
\begin{eqnarray}
\mathcal{N}_{+3} &\equiv &\exp \left[ \left( \mathcal{W}^{0}-\left( \mathbf{1%
},\mathbf{1}\right) _{0}\right) \oplus _{s}\mathcal{W}^{+3}\right]  \notag \\
&=&\exp \left[ \left( \left( \mathbf{Adj}\left( G_{6}\right) ,\mathbf{1}%
\right) _{0}+\left( \mathbf{1},\mathbf{Adj}\left( \mathcal{A}_{q}\right)
\right) _{0}\right) \oplus _{s}\left( \mathbf{Spin}^{\prime }\left(
q+2\right) ,\mathbf{Spin}\left( Q_{q}\right) \right) _{+3}\right]  \notag \\
&=&\left( G_{6}\times \mathcal{A}_{q}\right) \rtimes \mathbb{R}^{\left(
spin\left( q+2\right) ,spin\left( Q_{q}\right) \right) };  \label{N+3} \\
&&  \notag \\
\mathcal{N}_{-3} &\equiv &\exp \left[ \left( \mathcal{W}^{0}-\left( \mathbf{1%
},\mathbf{1}\right) _{0}\right) \oplus _{s}\mathcal{W}^{-3}\right]  \notag \\
&=&\exp \left[ \left( \left( \mathbf{Adj}\left( G_{6}\right) ,\mathbf{1}%
\right) _{0}+\left( \mathbf{1},\mathbf{Adj}\left( \mathcal{A}_{q}\right)
\right) _{0}\right) \oplus _{s}\left( \mathbf{Spin}\left( q+2\right) ,%
\mathbf{Spin}\left( Q_{q}\right) \right) _{-3}\right]  \notag \\
&=&\left( G_{6}\times \mathcal{A}_{q}\right) \rtimes \mathbb{R}^{\left(
spin\left( q+2\right) ,spin\left( Q_{q}\right) \right) }.  \label{N-3}
\end{eqnarray}
Thus, it holds that Eqs. (\ref{eve-1}) and (\ref{N+3}) (or equivalently Eqs.
(\ref{eve-1}) and (\ref{N-3})) are consistent with the general form of $%
\mathcal{O}_{critical,BPS}$ given by Eq. (\ref{magic-orbit-critical-Z<>0}).

Therefore, in the stabilizer of $\mathcal{O}_{critical,BPS}$, the factor
\begin{equation}
\mathcal{T}_{critical,BPS}=\mathbb{R}^{\left( spin\left( q+2\right)
,spin\left( Q_{q}\right) \right) }=\mathbb{R}^{\left( spin\left( q+1\right)
,spin\left( Q_{q}\right) \right) }  \label{T-T-2}
\end{equation}
is given by the exponential mapping of the Abelian subalgebra of $\frak{g}%
_{G_{5}}$ contained into the $+3$ (or equivalently $-3$)-graded extension of
$\mathcal{W}^{0}-\left( \mathbf{1},\mathbf{1}\right) _{0}$ through the $%
SO\left( 1,1\right) $-three grading (\ref{tg-2}), corresponding to the
irrepr. $\left( \mathbf{Spin}^{\prime }\left( q+2\right) ,\mathbf{Spin}%
\left( Q_{q}\right) \right) _{+3}$ (or equivalently $\left( \mathbf{Spin}%
\left( q+2\right) ,\mathbf{Spin}\left( Q_{q}\right) \right) _{-3}$) of $%
G_{6}\times \mathcal{A}_{q}(\times SO\left( 1,1\right) )$.\bigskip

The results obtained in Sub-Apps. \ref{IW-Contrs.} and \ref
{SO(1,1)-Three-Grading} (and reported in Tables 3 and 4) allows one to
conclude that all ``small'' charge orbits of symmetric \textit{``magic''}
RSG (classified by $q=8,4,2,1$) share the same non-semi-simple, Abelian
(namely, translational) part of the stabilizer. Namely, Eqs. (\ref{T-2}) and
(\ref{T-T-2}) yield to:
\begin{equation}
\mathcal{T}_{lightlike,BPS}=\mathcal{T}_{lightlike,nBPS}=\mathcal{T}%
_{critical,BPS}=\mathbb{R}^{\left( spin\left( q+1\right) ,spin\left(
Q_{q}\right) \right) }.  \label{T-T-3}
\end{equation}

\subsubsection{\label{Grading-Details}Details}

Below, we explicit Eqs. (\ref{tg-1})-(\ref{tgg-3}) for all $q=8,4,2,1$
classifying symmetric ``magic'' RSG.

\begin{itemize}
\item  $q=8$ $\left( J_{3}^{\mathbb{O}}\right) $%
\begin{equation}
\mathbf{78}\rightarrow \overset{\mathcal{W}^{0}}{\overbrace{\mathbf{1}_{0}+%
\mathbf{45}_{0}}}+\overset{\mathcal{W}^{-3}}{\overbrace{\mathbf{16}_{-3}}}+%
\overset{\mathcal{W}^{+3}}{\overbrace{\mathbf{16}_{+3}^{\prime }}}.
\end{equation}

\item  $q=4$ $\left( J_{3}^{\mathbb{H}}\right) $
\begin{equation}
\mathbf{35}\rightarrow \overset{\mathcal{W}^{0}}{\overbrace{\left( \mathbf{1}%
,\mathbf{1}\right) _{0}+\left( \mathbf{15},\mathbf{1}\right) _{0}+\left(
\mathbf{1},\mathbf{3}\right) _{0}}}+\overset{\mathcal{W}^{-3}}{\overbrace{%
\left( \mathbf{4},\mathbf{2}\right) _{-3}}}+\overset{\mathcal{W}^{+3}}{%
\overbrace{\left( \mathbf{4,2}\right) _{+3}}}.
\end{equation}

\item  $q=2$ $\left( J_{3}^{\mathbb{C}}\right) $. In this case it should be
recalled that
\begin{equation}
\mathbf{Adj}\left( SL\left( 3,\mathbb{C}\right) \right) =\mathbf{16}\equiv
\mathbf{3}\times \mathbf{3}^{\prime }+\overline{\mathbf{3}}\times \overline{%
\mathbf{3}}^{\prime }-2~\text{singlets}.
\end{equation}
Thus, by recalling Eqs. (\ref{dd-4})-(\ref{dd-4}), one can compute that
under $SL\left( 3,\mathbb{C}\right) \supsetneq _{\max }SL\left( 2,\mathbb{C}%
\right) \times SL\left( 1,\mathbb{C}\right) \times GL\left( 1,\mathbb{C}%
\right) $:
\begin{equation}
\begin{array}{l}
\mathbf{3}\times \mathbf{3}^{\prime }\rightarrow \left( \mathbf{3}%
_{0}\right) _{0}+\left( \mathbf{1}_{0}\right) _{0}+\left( \mathbf{2}%
_{3}\right) _{-3}+\left( \overline{\mathbf{2}}_{-3}\right) _{3}+\left(
\mathbf{1}_{0}\right) _{0}; \\
\overline{\mathbf{3}}\times \overline{\mathbf{3}}^{\prime }\rightarrow
\left( \mathbf{3}_{0}\right) _{0}+\left( \mathbf{1}_{0}\right) _{0}+\left(
\overline{\mathbf{2}}_{-3}\right) _{-3}+\left( \mathbf{2}_{3}\right)
_{3}+\left( \mathbf{1}_{0}\right) _{0}.
\end{array}
\end{equation}
Therefore:
\begin{equation}
\mathbf{Adj}\left( SL\left( 3,\mathbb{C}\right) \right) =\mathbf{16}%
\rightarrow \overset{\mathcal{W}^{0}}{\overbrace{2\left( \mathbf{3}%
_{0}\right) _{0}+2\left( \mathbf{1}_{0}\right) _{0}}}+\overset{\mathcal{W}%
^{-3}}{\overbrace{\left( \mathbf{2}_{3}\right) _{-3}+\left( \overline{%
\mathbf{2}}_{-3}\right) _{-3}}}+\overset{\mathcal{W}^{+3}}{\overbrace{\left(
\mathbf{2}_{3}\right) _{+3}+\left( \overline{\mathbf{2}}_{-3}\right) _{+3}}}.
\end{equation}

\item  $q=1$ $\left( J_{3}^{\mathbb{R}}\right) $
\begin{equation}
\mathbf{8}\rightarrow \overset{\mathcal{W}^{0}}{\overbrace{\mathbf{1}_{0}+%
\mathbf{3}_{0}}}+\overset{\mathcal{W}^{-3}}{\overbrace{\mathbf{2}_{-3}}}+%
\overset{\mathcal{W}^{+3}}{\overbrace{\mathbf{2}_{+3}}}.
\end{equation}
\end{itemize}

\end{appendix}

\end{document}